\documentclass[letterpaper,11pt,leqno]{article}
\usepackage{paper}
\usepackage{lscape}
\usepackage{comment}
\usepackage{float}
\usepackage{placeins}  
\usepackage{microtype}
\hypersetup{pdftitle={Geopolitical Barriers}}
\newcommand{\bib}{paper.bib}

\makeatletter
\renewcommand\paragraph{\@startsection{paragraph}{4}{\z@}%
  {1.25ex \@plus1ex \@minus.2ex}%
  {-1em}%
  {\normalfont\normalsize\bfseries}}
\makeatother

\begin{document}

\title{Geopolitical Barriers to Globalization}

\author{
  Tianyu Fan\\
  \textit{Yale University}
  \and
  Mai Wo\\
  \textit{Yale University} 
  \and
  Wei Xiang\\
  \textit{University of Michigan} 
  \thanks{Tianyu Fan: tianyu.fan@yale.edu. Mai Wo: mai.wo@yale.edu. Wei Xiang: xwe@umich.edu. We especially thank Stephen Terry for his constructive feedback. We thank Costas Arkolakis, R\"udiger Bachmann, Chris Clayton, Javier Cravino, Sam Kortum, John Leahy, Fernando Leibovici, Andrei Levchenko, Ana Maria Santacreu, Matthew Schwartzman, Sebastian Sotelo, Linda Tesar, and Erik Voeten for their comments. We thank Imryoung Jeong for her research assistance. Bilateral geopolitical alignment scores and codebook are available at \url{https://github.com/tianyufan-econ/global-geopolitics}. } 
}

 
\date{\today \\[0.5em]
\href{https://www.tianyu-fan.com/files/FWX_Geopolitical_Barriers.pdf}{(Click here for the most recent version)}}


\begin{titlepage}
\maketitle

We show that since the mid-1990s, the trade-promoting effects of tariff liberalization have been increasingly offset by deteriorating geopolitical alignment, slowing trade globalization after 2007. To quantify this barrier, we use large language models to compile 833,485 geopolitical events across 193 countries, 1950--2024, and construct a bilateral geopolitical alignment score. Using local projections, we estimate that a one-standard-deviation permanent improvement in alignment raises bilateral trade by 22 percent in the long run. In an Armington framework, tariff reductions raised 2021 global trade by about 7.5 percent, while geopolitical deterioration reduced it by about 5.3 percent, with uneven welfare effects.

\vspace{0.5cm}


\end{titlepage}
\onehalfspacing                      
\section{Introduction}\label{s:introduction}

\begin{figure}[ht]
    \centering
    \begin{subfigure}[b]{0.48\textwidth}
        \centering
        \includegraphics[width=\linewidth]{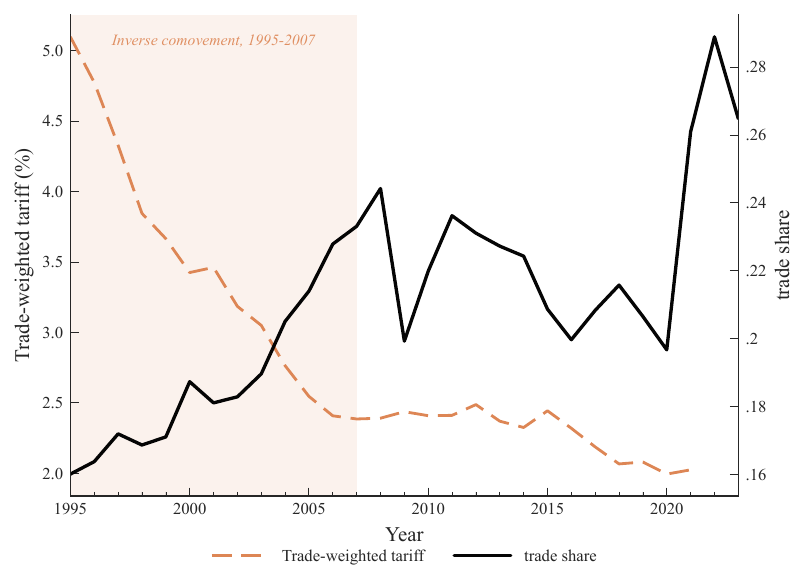}
        \caption{Tariffs and trade share}
    \end{subfigure}
    \hfill
    \begin{subfigure}[b]{0.48\textwidth}
        \centering
        \includegraphics[width=\linewidth]{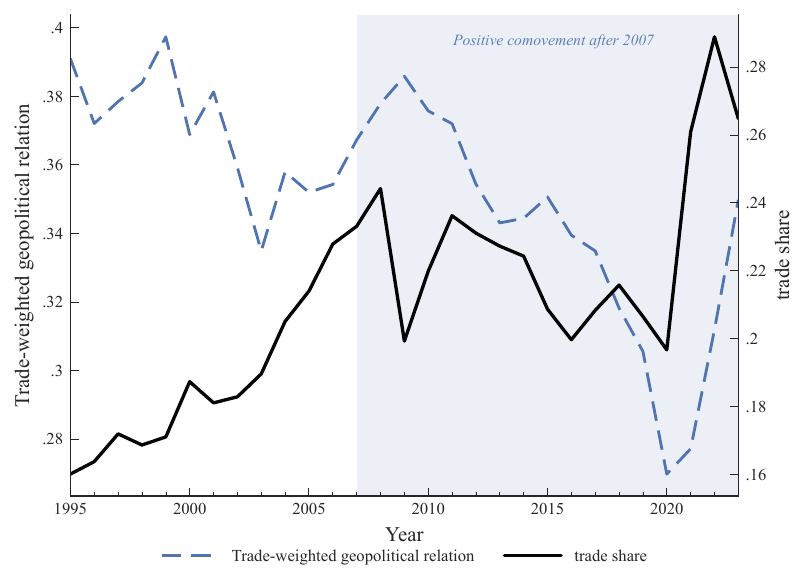}
        \caption{Geopolitical alignment and trade share}
    \end{subfigure}
    \caption{Aggregate Trade Shares, Tariffs, and Geopolitical Alignment}
    \label{fig:motivation_trade_geo}
    \note{\emph{Notes:} The black line in both panels plots global trade as a share of world GDP. Panel A plots the trade-weighted average bilateral tariff, and Panel B plots the trade-weighted average bilateral geopolitical alignment measure constructed in the paper. The orange shaded region in Panel A, 1995--2007, highlights the period in which trade shares moved inversely with tariffs. The blue shaded region in Panel B, 2007--2023, highlights the period in which trade shares moved positively with the geopolitical alignment measure constructed in Section~\ref{measure_alignment}.}
\end{figure}

From the mid-1990s to 2007, global trade rose as tariffs fell. After 2007, that link weakened: tariffs stayed low, regional trade agreements continued to accumulate, and the WTO persisted, yet trade integration stalled. This paper shows that a second force increasingly shaped trade globalization: deteriorating bilateral geopolitical alignment. Trade liberalization continued to expand trade, but geopolitical fragmentation increasingly offset its effects (Figure~\ref{fig:motivation_trade_geo}).

To evaluate this geopolitical barrier, we first construct a bilateral event-based measure of geopolitical alignment covering all 193 UN member states from 1950 to 2024. Second, we estimate the effect of geopolitical alignment on trade using local projections within a gravity framework. Third, we embed the estimated trade effects in an Armington framework to assess the implied aggregate trade and welfare consequences of geopolitical change. 

We begin by constructing a bilateral measure of geopolitical alignment that captures the evolution of strategic and diplomatic relations between country pairs. Using the event-based measurement framework in \citet{fan2026measuringgeopoliticalalignmenteconomic}, we employ Gemini 2.5 Pro, a large language model with web-search capability, to collect major political events for all bilateral pairs among the 193 UN member states from 1950 to 2024. The resulting database contains 833,485 bilateral geopolitical events covering 18,528 country pairs. We classify these events using the Conflict and Mediation Event Observations (CAMEO) framework \citep{schrodt2012cameo}, map them into Goldstein scores from $-10$ to $+10$ \citep{goldstein1992}, and aggregate them into a dynamic bilateral alignment measure bounded between $-1$ and $+1$. This measure is bilateral, continuous, and time-varying. Compared with widely used proxies such as UN voting similarity \citep{Bailey2017-po}, it is designed to capture bilateral political relations rather than similarity in multilateral positioning, and it provides the within-pair variation needed for panel gravity estimation. 

Before turning to dynamic estimation, we use the data to establish the basic trade pattern that motivates the panel analysis. We begin with four bilateral case studies: U.S.--Russia and China--Japan for major-power relations, and Greece--T\"{u}rkiye and Iran--Saudi Arabia for regional rivalries. In each case, we compare residualized trade and residualized geopolitical alignment, removing country-year shocks and time-invariant dyad factors so that the comparison isolates bilateral movements over time. Periods of stronger alignment coincide with stronger bilateral trade, while periods of deterioration coincide with weaker trade. The same positive relationship appears broadly across countries, separately during the Cold War and after 1991, suggesting that the pattern is not specific to a single geopolitical regime.

We then place geopolitical alignment in a standard gravity specification to compare its explanatory power with conventional bilateral trade determinants. Because zero trade flows raise selection concerns in log-linear specifications, we use as our main benchmark the 32 major economies, defined as countries that ever ranked among the world's top 20 economies by GDP since 1960. Bilateral flows within this group are almost always positive, and these economies account for the bulk of global trade. Among major economies, a one-standard-deviation increase in alignment is associated with 11.9 log points higher bilateral trade. Geopolitical alignment explains a share of bilateral trade variation comparable to linguistic distance.



We next estimate dynamic trade responses to changes in bilateral geopolitical alignment using local projections with country-pair, origin-year, and destination-year fixed effects, so identifying variation comes from within-pair changes in relations. We use the same 32-major-economy benchmark for the dynamic analysis, where bilateral flows are nearly always positive. A unit improvement in geopolitical alignment, about four standard deviations, raises bilateral trade by about 28 log points at its peak. A one-standard-deviation change, 0.26 units, is close to the deterioration in U.S.--China bilateral alignment around the 2018 trade war. The response has no detectable pre-trend, builds over several years, and is hump-shaped. Decomposing the response into transitory and permanent components shows that even temporary diplomatic improvements generate trade gains that persist for five to seven years, while a one-standard-deviation permanent improvement in alignment implies roughly 22 percent higher trade in the long run.

The dynamic trade effect is similar in the full country-pair sample, across pre- and post-1995 subsamples, and under alternative constructions of the alignment measure; the estimates barely change when observable bilateral controls are added. Inverse hyperbolic sine specifications produce comparable results in both the 32-major-economy benchmark and the full country-pair sample. The effect also holds when we split the alignment measure into economic and non-economic events, and into cooperative and conflictual events. Two diagnostics further support the causal interpretation. Alignment changes around unexpected leader transitions generate trade responses similar to the baseline, and a transportation-cost LP-IV exercise provides little evidence that trade shocks cause subsequent geopolitical alignment.


We also study heterogeneity and mechanisms. The effect is stronger for country pairs with greater executive discretion and lower institutional-quality scores, weaker within formal alliances, and positive across a wide range of sectors, with the largest responses in manufacturing. On mechanisms, deteriorating geopolitical alignment increases the use of sanctions and other restrictive trade measures, while having little systematic effect on tariffs. Sanctions account for roughly 18 percent of the trade response and tariffs for at most 2.7 percent; the remaining share is not directly decomposed. We provide suggestive evidence on one residual channel: country pairs with stronger geopolitical alignment exhibit more favorable bilateral attitudes, and these attitudes are positively associated with trade.

Finally, we translate the reduced-form estimates into aggregate magnitudes using a 78-country Armington framework. In the calibrated model, tariff liberalization raised 2021 global trade by approximately 7.5 percent relative to a counterfactual with 1995 tariffs, while geopolitical deterioration reduced global trade by approximately 5.3 percent relative to a counterfactual with 1995 alignment. Geopolitical deterioration therefore offset roughly 70 percent of the trade expansion generated by tariff reduction. A complementary decomposition shows that about 80 percent of the geopolitical contribution comes from non-trade events. The qualitative offset pattern is robust across alternative specifications. Changes in bilateral alignment involving the United States, Russia, or China together account for just over half of the aggregate geopolitical effect, with the remainder reflecting dispersed realignment across other economies.

Geopolitical realignment generates substantially more heterogeneous country-level welfare effects than tariff liberalization. Tariff liberalization produces modest average gains, with 50 of 78 countries gaining. Geopolitical realignment, by contrast, has a negative average welfare effect, generates losses for 48 of 78 countries, and produces a cross-country standard deviation roughly three times larger than that of tariff liberalization. In the model, geopolitical change is therefore a quantitatively meaningful source of welfare dispersion.

\paragraph{Literature}
Our paper contributes to three strands of literature: the measurement of geopolitical alignment, the empirical analysis of how geopolitics shapes trade, and the estimation of trade barriers in gravity frameworks.

\noindent \emph{Measuring Global Geopolitical Alignment}. 
We use the event-based measurement framework introduced in \citet{fan2026measuringgeopoliticalalignmenteconomic}, which constructs geopolitical alignment measures from political events for 24 major economies and studies their country-level implications for economic growth. We extend the event compilation to universal country-pair coverage and exploit the resulting bilateral variation to study international trade. Most prior work relies on UN General Assembly voting similarity \citep{Signorino1999-yb,Bailey2017-po}. However, UN voting primarily captures multilateral positioning rather than bilateral relations and exhibits limited within-dyad time variation \citep{Broner2025-fy,Voeten2026review}. Other approaches also face limitations in studying the geopolitics-trade nexus. Discrete measures capture only particular relationship types, such as strategic rivalry \citep{thompson2001identifying}, alliances \citep{Gibler2008-xr}, sanctions \citep{felbermayr2020global,Felbermayr2021-ws}, or treaties \citep{broner2025hegemonic}, and therefore miss the broader spectrum of bilateral interaction. Aggregate measures of geopolitical risk or fragmentation \citep{caldara2022measuring,fernandez2024we} are useful for macro analysis but do not provide the bilateral variation needed for trade. Our event-based measure addresses these limitations by constructing a continuous bilateral score from $-1$ to $+1$ that captures both the timing and intensity of geopolitical relations.

We also contribute to the literature on machine-learning-based measurement. Whereas standard text-based approaches typically infer political risk from keyword frequencies \citep{Baker2016-ni,Hassan2019-rf,caldara2022measuring}, we use an LLM to identify, classify, and score major bilateral geopolitical events using contextual information \citep{clayton2025geoeconomic,Dell2025-qj,Fang2025-ca}. Our approach uses LLMs not to summarize text, but to construct a structured measure of bilateral geopolitical relations.

\noindent \emph{Geopolitics and Trade}. 
We engage with the growing literature on how geopolitics shapes trade patterns. Early work dates back to \citet{hirschman1945national}. Recent theoretical work has advanced rapidly \citep{Alesina2000-pn,couttenier2024gravity,becko2024strategic,clayton2025theory,clayton2025framework, clayton2026great, mayer2025fragmentation, meyer2025hegemonic}, and \citet{mohr2025geoeconomics} provide a recent review. Empirical analysis, however, remains limited by measurement challenges. Our paper provides evidence on how geopolitical changes translate into trade reallocation and welfare effects.

The closest related empirical paper is \citet{gopinath2025changing}, which studies trade flows within and across geopolitical blocs defined using UN voting similarity and documents recent evidence of fragmentation. While their bloc-based analysis uncovers important aggregate patterns, our bilateral event-based measure allows us to estimate dynamic trade elasticities, trace adjustment paths, and quantify aggregate and distributional welfare effects in the Armington framework. Our paper also relates to recent quantitative work on fragmentation, which has measured rising trade barriers through bilateral trade-cost changes for specific country pairs \citep{chen2022non,bonadio2025playing,cai2025quantifying} and country image \citep{chang2022good}. Our measure differs by providing continuous bilateral coverage of all country pairs over seven decades, which allows us to estimate dynamic reduced-form trade elasticities under our identifying assumptions rather than inferring fragmentation from observed trade-cost shifts.

More broadly, our paper connects to a literature on international politics and trade \citep{grossman1994protection,morrow1998political,mansfield2000free,Martin2008-vf}, to studies of the channels linking geopolitics to trade \citep{korovkin2023conflict,kleinman2024international,liu2025international}, and to work on how geopolitical tensions shape innovation \citep{alfaro2025trade,flynn2025foreign}, asset allocation \citep{pellegrino2025Capital}, the economic impact of war \citep{federle2026price}, and economic growth \citep{fan2026measuringgeopoliticalalignmenteconomic}. \citet{goldberg2025changing} review how geopolitical shifts are reshaping trade and development.

\noindent \emph{Gravity Estimation and Trade Barriers}. 
Finally, we contribute to the literature on gravity models and trade barriers \citep{anderson2003gravity,anderson2004trade,head2014gravity}. Our use of local projections within a gravity framework is closest to \citet{boehm2023long}. Whereas much of the gravity literature emphasizes observable trade costs, including tariffs \citep{Baier2007-xv,Caliendo2015-xv}, distance \citep{Disdier2008-ri}, language \citep{melitz2008language}, regulatory barriers \citep{Looi-Kee2009-ej}, and shipping time \citep{hummels2013time}, we show that geopolitical alignment is an important, and often omitted, component of bilateral trade frictions.

\paragraph{Roadmap}
The remainder of the paper is organized as follows. Section~\ref{measure_alignment} introduces the measure of global geopolitical alignment. Section~\ref{s:geo_trade_patterns} documents descriptive patterns linking alignment to trade. Section~\ref{s:trade_elas_geo} estimates dynamic trade elasticities via local projections. Section~\ref{s:het_mechanisms} examines heterogeneity and mechanisms. Section~\ref{s:quantitative} embeds the estimated dynamic effects in an Armington framework to quantify aggregate trade and welfare consequences. Section~\ref{s:conclusion} concludes.

\section{Measuring Global Geopolitical Alignment}
\label{measure_alignment}

This section describes the construction of our bilateral measure of geopolitical alignment. The measure is designed to capture the state of the strategic and diplomatic relationship between two countries, as reflected in whether their bilateral political interactions are cooperative or conflictual. We use the event-based measurement framework developed by \citet{fan2026measuringgeopoliticalalignmenteconomic}, extending its event compilation from 24 major economies to universal country-pair coverage. The resulting database contains 833,485 events across 18,528 bilateral pairs and provides the core measure used in the dyadic trade analysis that follows.

For exposition, we describe the construction in three steps. First, we construct a global database of major bilateral political events. Second, we classify each event and map it into a numerical score. Third, we aggregate event-level scores into an annual country-pair measure of geopolitical alignment.\footnote{In implementation, the prompt performs the first two tasks jointly: it identifies relevant bilateral political events, classifies them using the CAMEO framework, and assigns Goldstein scores within a single workflow.}

\subsection{Step 1: Event Database Construction}

We use Gemini 2.5 Pro, a large language model with web search capability, to construct a database of major bilateral political events for all undirected country-pair-year observations from 1950 to 2024.\footnote{Web search augments the model's pretraining knowledge and helps standardize source verification, especially for less prominent dyad-years.} The prompt prioritizes four types of sources: official government materials, international organizations, major news archives, and academic or diplomatic archives. For each country-pair-year observation, the model verifies the relevant political entities, accounting for state succession when necessary, and identifies the major political events involving the two countries in that year. When an event involves more than two countries, we decompose it into the relevant bilateral relationships. Because the model processes each country-pair-year independently, multilateral episodes such as coalition sanctions or international treaties enter separately for each involved pair, with scores reflecting each dyad's specific role in the episode. The resulting database contains 833,485 political events. Appendix~\ref{app:llm_prompt} reports the full prompt and implementation details.

Our approach differs from existing global event databases such as the Global Database of Events, Language, and Tone (GDELT) \citep{leetaru2013gdelt} and the Integrated Crisis Early Warning System (ICEWS) \citep{boschee2015icews} in two respects. First, we focus on major bilateral political events that are most likely to alter the underlying relationship between two countries, rather than attempting to capture the full universe of international interactions. This narrower focus reduces noise from routine or low-salience events and yields a cleaner record of consequential bilateral political developments. Second, our data span 1950--2024, extending historical coverage well beyond that of GDELT and ICEWS and aligning more closely with the trade panel used in the empirical analysis.

Table~\ref{tab:us_china_2024} provides an illustration of the specific events recorded for a given country pair, using U.S.--China interactions in 2024 as an example. The seven events include diplomatic exchanges, military dialogue, tariffs, sanctions, export controls, and security tensions. Together they capture the mix of crisis-management diplomacy and strategic rivalry that characterized the bilateral relationship in that year.

\begin{table}[!ht]
\centering
\caption{Major U.S.--China Bilateral Events in 2024: Recorded Events and Scores}
\label{tab:us_china_2024}
\footnotesize
\resizebox{\linewidth}{!}{
\begin{tabular}{@{}p{3.0cm}p{6.0cm}p{2.7cm}p{1.5cm}p{1.5cm}@{}}
\toprule
\textbf{Event Name} & \textbf{Event Description} & \textbf{CAMEO Class.} & \textbf{Econ. Type} & \textbf{Goldstein Score} \\
\midrule
US-China Defense Chiefs Meet in Singapore &
Defense Secretary Austin and Minister Dong Jun met at the Shangri-La Dialogue, agreeing to resume military communications and crisis-management discussions &
Verbal Cooperation (04-044) &
Not econ. &
$+4.5$ \\
\addlinespace[0.3em]
Resumption of US-China MMCA Talks &
U.S. and Chinese defense officials resumed Military Maritime Consultative Agreement (MMCA) talks in Honolulu after a two-year suspension, reopening a channel to manage military incidents &
Verbal Cooperation (04-040) &
Not econ. &
$+4.0$ \\
\addlinespace[0.3em]
Blinken Visit to China &
Secretary Blinken visited Shanghai and Beijing in April, meeting senior Chinese officials to discuss Russia, trade frictions, and regional security &
Verbal Cooperation (04-042) &
Not econ. &
$+3.0$ \\
\addlinespace[0.3em]
US Tariff Increases on Chinese Goods &
The Biden administration raised tariffs on \$18 billion of Chinese imports, including electric vehicles, solar cells, semiconductors, and medical products &
Material Conflict (16-163) &
Tariff &
$-6.5$ \\
\addlinespace[0.3em]
US Sanctions Chinese Companies Supporting Russia &
The U.S. sanctioned Chinese and Hong Kong firms supplying inputs to Russia's military-industrial base, linking bilateral tensions to the Ukraine war &
Material Conflict (17-172) &
Sanction &
$-7.0$ \\
\addlinespace[0.3em]
US Adds Chinese Firms to Entity List &
The Commerce Department added 42 Chinese companies to the Entity List for supporting Russia's defense industrial base, intensifying technological pressure &
Material Conflict (17-172) &
Sanction &
$-7.0$ \\
\addlinespace[0.3em]
Mainland China Military Drills Around Taiwan &
Mainland China conducted large-scale PLA exercises around Taiwan in May, raising cross-Strait tensions and bilateral tensions with the United States &
Material Conflict (15-150) &
Not econ. &
$-8.0$ \\
\bottomrule
\end{tabular}
}
\note{\emph{Notes:} This table shows U.S.--China bilateral events recorded for 2024. Numbers in parentheses are CAMEO subcategory codes. Econ.\ Type classifies the instrument (tariff, sanction, or non-economic). Goldstein scores range from $-10$ (maximum conflict) to $+10$ (maximum cooperation). See Appendix~\ref{app:llm_prompt} for the full prompt and scoring guidelines.}
\end{table}

\subsection{Step 2: Mapping Events into Scores}

Although event identification and scoring are implemented jointly, it is useful to separate the scoring logic conceptually. We classify each event using the Conflict and Mediation Event Observations (CAMEO) framework \citep{schrodt2012cameo} and assign it a Goldstein score \citep{goldstein1992}. CAMEO is a hierarchical taxonomy of political interactions organized by whether actions are cooperative or conflictual and whether they are verbal or material. It contains 20 top-level categories and 143 subcategories, ranging from diplomatic engagement and aid provision to sanctions and military force. The Goldstein scale maps these event types into numerical weights from $-10$ to $+10$, with lower values indicating more conflictual behavior and higher values indicating more cooperative behavior. We use the standard CAMEO-to-Goldstein mapping as a benchmark and provide it to Gemini as a reference, while allowing limited contextual adjustments for event severity and historical setting.\footnote{This preserves comparability with existing event datasets while allowing the scoring to reflect the context of particular bilateral episodes.} The U.S.--China example in Table~\ref{tab:us_china_2024} illustrates the mapping from event descriptions to CAMEO categories and Goldstein scores: diplomatic engagement and military communication are coded as verbal cooperation, while tariffs, sanctions, export controls, and security tensions are coded as material conflict.

Table~\ref{tab:event_summary} reports summary statistics for bilateral geopolitical events across three phases of the postwar international system: the Cold War, the post-Cold War globalization period, and the recent period of geopolitical fragmentation. We classify events with positive Goldstein scores as cooperative and those with negative Goldstein scores as conflictual. The Cold War exhibits the highest conflict share, the lowest mean Goldstein score, and the greatest dispersion. The globalization period is more cooperative on average, while the fragmentation period partly reverses that pattern, with a higher conflict share and a lower mean score. Event counts also rise after 2010 despite the shorter time span, indicating more recorded bilateral interactions in recent decades but less cooperative average relations than during the globalization period.

\begin{table}[ht]
\centering
\caption{Summary of Bilateral Geopolitical Events by Era, 1950--2024}
\label{tab:event_summary}
\resizebox{\linewidth}{!}{
\begin{tabular}{lcccccc}
\toprule
& \multicolumn{4}{c}{Event Counts} & \multicolumn{2}{c}{Goldstein Score} \\
\cmidrule(lr){2-5} \cmidrule(lr){6-7}
Period & Cooperation & Conflict & Total & Conflict (\%) & Mean & SD \\
\midrule
Cold War (1950--1990) & 206,857 & 58,770 & 265,627 & 22.1 & 2.33 & 5.04 \\
Globalization (1991--2009) & 220,840 & 37,958 & 258,798 & 14.7 & 3.92 & 4.11 \\
Fragmentation (2010--2024) & 251,060 & 58,000 & 309,060 & 18.8 & 3.70 & 3.68 \\
\midrule
Full Period & 678,757 & 154,728 & 833,485 & 18.6 & 3.38 & 4.33 \\
\bottomrule
\end{tabular}
}
\note{\emph{Notes:} This table reports summary statistics for the 833,485 bilateral geopolitical events in our sample, grouped into three eras. Events are classified as cooperative if their Goldstein score is positive and conflictual if their Goldstein score is negative. The column ``Conflict (\%)'' reports the share of conflictual events in total events within each era. Goldstein scores range from $-10$ (maximum conflict) to $+10$ (maximum cooperation).}
\end{table}

Because our baseline measure is constructed using Gemini 2.5 Pro, we assess whether the resulting scores are stable across alternative LLMs and repeated runs. Appendix~\ref{app:model_robustness} reports cross-model and within-model robustness checks. Cross-model correlations on randomly selected dyad-years yield a pooled $\rho = 0.88$, while within-model reruns yield $\rho = 0.93$.

\subsection{Step 3: Aggregating Events into Bilateral Relationship Scores}
\label{ss:measuring_alignment}

Bilateral geopolitical alignment should characterize the persistent state of the relationship between two countries rather than a mere summary of events within a single year. Major incidents durably reshape subsequent relations, whereas isolated events typically exert only transient influence. We therefore aggregate event-level scores into a persistent relationship stock using the recursive construction as in \citet{fan2026measuringgeopoliticalalignmenteconomic}: The stock is updated by current events, with the update weight equal to their share in the effective cumulative event number, which is a discounted sum with decay rate $\lambda$. This approach captures both the persistence of bilateral relationships and mitigates the noise inherent in annual score averages for dyads with sparse events.

\paragraph{Construction of Geopolitical Alignment Scores.}
For each country pair $(o,d)$ and year $t$, let $\{s^n_{od,t}\}_{n=1}^{\tilde N_{od,t}}$ denote the Goldstein scores of the $\tilde N_{od,t}$ recorded bilateral events. We first define the annual event score as the average Goldstein score, normalized to the interval $[-1,1]$:
\begin{equation}
\tilde{S}_{od,t}=\frac{1}{\tilde N_{od,t}}\sum_{n=1}^{\tilde N_{od,t}} s^n_{od,t}/10.
\end{equation}

We then define the geopolitical alignment score recursively as
\begin{align}
S_{od,t} &= (1-\phi_{od,t}) S_{od,t-1} + \phi_{od,t}\tilde{S}_{od,t}, \label{eq:dynamic_score}\\
\phi_{od,t} &= \tilde N_{od,t}/N_{od,t}, \qquad
N_{od,t} = (1-\lambda)N_{od,t-1} + \tilde N_{od,t}, \nonumber
\end{align}
where $N_{od,t}$ is the effective cumulative event count and $\phi_{od,t}$ is the updating weight. The score changes gradually when event flow is sparse and responds more strongly when many events are recorded in a given year. The parameter $\lambda$ determines how much weight is placed on current events versus past accumulated events: a higher $\lambda$ puts more weight on the current year by discounting past events more quickly. We set $\lambda = 0.3$, which corresponds to a half-life of about two years for $N_{od,t}$ and implies that the alignment score $S_{od,t}$ evolves with moderate persistence. Section~\ref{ss:robustness} shows the main results are robust to alternative values of $\lambda$.\footnote{When no events are observed ($\tilde{N}_{od,t}=0$), Eq.~\eqref{eq:dynamic_score} would yield $S_{od,t}=S_{od,t-1}$ since $\phi_{od,t}=0$. Instead, we assume that in periods with no new events, both the effective event count and the alignment score decay at rate $\lambda$: $N_{od,t}=(1-\lambda)N_{od,t-1}$ and $S_{od,t}=(1-\lambda)S_{od,t-1}$. This captures the idea that, absent new interactions, previously accumulated relationships gradually fade over time. For the initial year of each bilateral pair, $S_{od,t_0}=\tilde{S}_{od,t_0}$ and $N_{od,t_0}=\tilde{N}_{od,t_0}$.} By construction, $S_{od,t}\in[-1,1]$, and is symmetric: $S_{od,t}=S_{do,t}$. Higher values indicate stronger geopolitical alignment.

\begin{table}[htb]
\centering
\caption{Summary Statistics of Bilateral Geopolitical Alignment Scores}
\label{tab:geo_summary}
\resizebox{\linewidth}{!}{
\begin{tabular}{lcccccc}
\toprule
& \multicolumn{2}{c}{All Countries} & \multicolumn{2}{c}{Major-Major Pairs} & \multicolumn{2}{c}{Dynamics (All)} \\
\cmidrule(lr){2-3} \cmidrule(lr){4-5} \cmidrule(lr){6-7}
Period & Mean & Std Dev & Mean & Std Dev & \% Decline & \% Improve \\
\midrule
Cold War (1950--1990) & 0.087 & 0.224 & 0.233 & 0.292 & 18.2 & 13.7 \\
Globalization (1991--2009) & 0.150 & 0.231 & 0.317 & 0.254 & 19.3 & 14.5 \\
Fragmentation (2010--2024) & 0.202 & 0.230 & 0.321 & 0.257 & 22.9 & 16.6 \\
\midrule
\multicolumn{7}{l}{\textit{Variance Decomposition (All Countries):}} \\
\quad Between-dyad: 28.8\% & \multicolumn{5}{r}{Within-dyad: 71.2\%} \\
\bottomrule
\end{tabular}
}
\note{\emph{Notes:} This table reports summary statistics for bilateral geopolitical alignment scores covering 18,528 country pairs from 1950 to 2024. Scores range from $-1$ to $+1$ and are constructed using equation~\eqref{eq:dynamic_score} with $\lambda=0.3$. The Dynamics columns report the share of dyad-years with annual score changes below $-0.05$ and above $+0.05$, respectively.}
\end{table}

Table~\ref{tab:geo_summary} reports summary statistics. Mean alignment rises over time, from 0.087 in the Cold War to 0.150 during globalization and 0.202 in the fragmentation period, while the all-country cross-sectional standard deviation remains close to 0.23. Defining deteriorating dyad-years as those with annual changes in alignment below $-0.05$ and improving dyad-years as those above $+0.05$, both tails become more active after 2010: the deteriorating share rises from 18.2 percent to 22.9 percent, while the improving share rises from 13.7 percent to 16.6 percent. Dyads among the 32 major economies, defined as countries that ever ranked among the top 20 by GDP since 1960, have higher mean alignment throughout, although the gap narrows over time.\footnote{The 32 major economies are: Argentina, Australia, Austria, Belgium, Brazil, Canada, China, Denmark, France, Germany, India, Indonesia, Iran, Iraq, Italy, Japan, Mexico, Netherlands, Nigeria, Philippines, Poland, Russia, Saudi Arabia, South Africa, South Korea, Spain, Sweden, Switzerland, T\"{u}rkiye, United Kingdom, United States, and Venezuela. Appendix Figure~\ref{fig:trade_major_decomp} shows that these economies account for the bulk of global trade.} The variance decomposition shows that 71.2 percent of total variation is within pair rather than between pairs, leaving substantial within-pair variation for gravity estimation. The deterioration highlighted in Figure~\ref{fig:motivation_trade_geo} and the larger post-2010 decline share in Table~\ref{tab:geo_summary} therefore capture rising tensions among economically central dyads rather than a decline in the unweighted average relationship across all country pairs.

\subsection{Case Study: United States and China}

\begin{figure}[ht]
    \centering
    \caption{Geopolitical Alignment Score Between the United States and China, 1950--2024}
    \includegraphics[width=\linewidth]{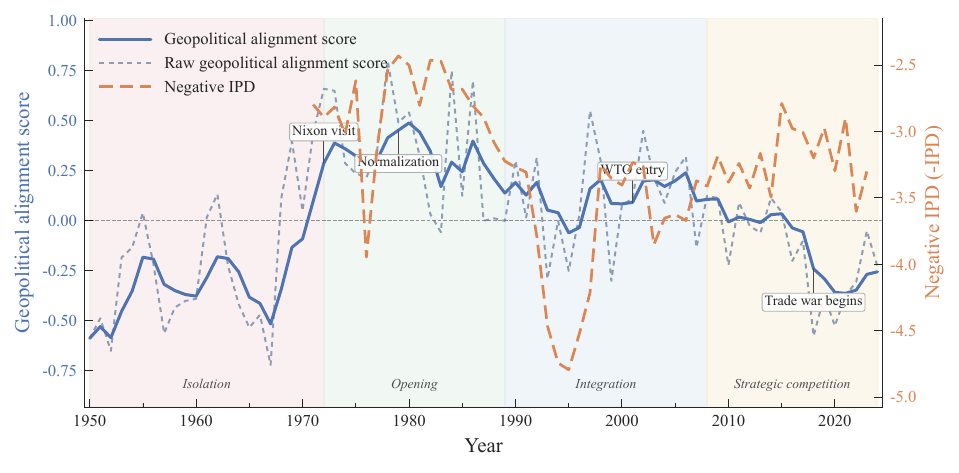}
    \label{fig:geo_score_USA_CHN}
    \note{\emph{Notes:} The solid blue line reports the geopolitical alignment score. The light blue dashed line reports the raw annual event score. The orange dashed line reports negative Ideal Point Distance (IPD) from United Nations General Assembly (UNGA) voting data. Shaded regions denote major phases of U.S.--China relations.}
\end{figure}

Figure~\ref{fig:geo_score_USA_CHN} illustrates the geopolitical alignment measure between the United States and China. The measure captures the diplomatic opening following Nixon’s 1972 visit and the normalization in 1979, the period of cooperation associated with commercial integration in the 1990s and China’s WTO entry in 2001, and the recent deterioration after 2018 amid the trade war and rising tensions over Taiwan. Appendix~\ref{app:validation} provides additional evidence for other dyads.

To relate the measure to existing proxies for geopolitical alignment, we compare it with negative Ideal Point Distance (IPD) from UNGA voting data \citep{Bailey2017-po}. IPD is widely used in the literature but captures alignment in multilateral voting positions rather than bilateral political interactions. The comparison is therefore informative about whether the event-based series adds bilateral information beyond standard voting-based measures. In the U.S.--China case, the two measures are correlated, since China's distance from U.S.-led positions on multilateral issues partly reflects bilateral tensions. However, negative IPD shows a pronounced dip in the mid-1990s and only a muted response to the post-2018 deterioration. These patterns suggest that IPD may be less well suited to capturing bilateral geopolitical dynamics than the event-based measure, consistent with its construction from multilateral voting rather than bilateral interactions.

\subsection{Bilateral and Multilateral Dimensions of Geopolitical Alignment}
\label{ss:bilateral_multilateral}

Because UNGA voting similarity is the most widely used proxy for geopolitical alignment in the literature, we compare it with our event-based measure. The comparison clarifies a key distinction: UNGA voting captures countries' positions in a multilateral voting space, whereas our measure captures bilateral political relations through country-pair events.

\subsubsection{UNGA Voting and Multilateral Positioning}

\begin{figure}[hbt]
    \centering
    \caption{Bilateral and Multilateral Dimensions of Geopolitical Alignment}
    \begin{subfigure}[b]{0.48\textwidth}
        \includegraphics[width=\textwidth, height=0.32\textheight, keepaspectratio]{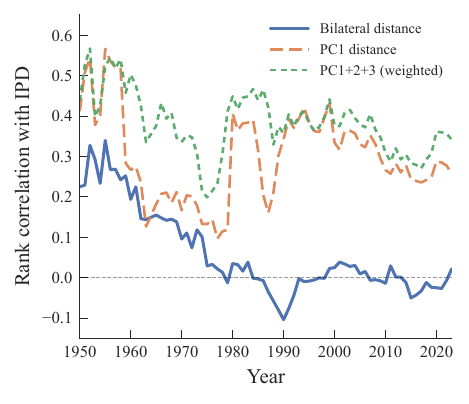}
        \caption{Bilateral alignment vs.\ UNGA voting similarity}
    \end{subfigure}
    \hfill
    \begin{subfigure}[b]{0.48\textwidth}
        \includegraphics[width=\textwidth, height=0.32\textheight, keepaspectratio]{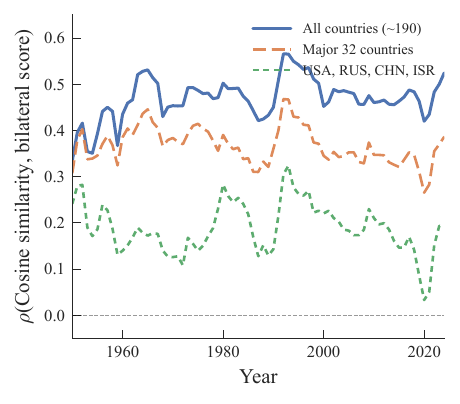}
        \caption{Multilateral position vs.\ bilateral score}
    \end{subfigure}
    \label{fig:bilateral_multilateral}
    \note{\emph{Notes:} Panel A reports the annual Spearman correlation between UN ideal-point distance (IPD) and three bilateral distance measures based on our data. Panel B reports the annual Spearman correlation between cosine similarity of third-country stance vectors and the bilateral score under alternative reference sets.}
\end{figure}

Figure~\ref{fig:bilateral_multilateral}A compares the two measures across all country pairs. We report the year-by-year cross-sectional Spearman rank correlation between three bilateral distance measures (constructed from our data) and the UN ideal-point distance (IPD), with all measures defined so that higher values indicate greater dissimilarity. The correlation between the raw bilateral distance (negative geopolitical alignment) and IPD is low after the 1960s, averaging only 0.06. The near-zero average correlation across all dyads indicates that UNGA voting primarily reflects multilateral positioning rather than bilateral relations \citep{Voeten2026review}.

To sharpen this comparison, we extract principal components from the annual bilateral score matrix.\footnote{For each year $t$, we construct the symmetric $N_t \times N_t$ matrix of bilateral scores $\mathbf{S}_t = [S_{ij,t}]$ and extract principal components. Let $p_{ik,t}$ denote the PC$k$ score for country $i$ and $\sigma^2_{k,t}$ the share of total variance explained by the $k$-th component. The PC$k$ distance for dyad $(i,j)$ is $|p_{ik,t} - p_{jk,t}|$. The variance-weighted PC1+2+3 distance is $d^{\text{w3}}_{ij,t} = \sum_{k=1}^{3} \omega_{k,t} |p_{ik,t} - p_{jk,t}|$, where $\omega_{k,t} = \sigma^2_{k,t} / \sum_{\ell=1}^{3} \sigma^2_{\ell,t}$.} Principal component distances correlate more strongly with IPD than the raw bilateral distance does: the mean correlation rises to 0.30 using the first principal component and to 0.38 using a variance-weighted distance based on the first three components. This pattern indicates that UNGA voting captures the broad multilateral structure embedded in bilateral relations, while most of the dyad-specific variation in our measure is orthogonal to voting behavior.

Appendix Figure~\ref{fig:ipd_trade} reinforces this distinction at the aggregate level. UNGA voting similarity continued to improve after 2007, even as trade globalization stagnated. Our event-based measure, by contrast, records bilateral deterioration over the same period (Figure~\ref{fig:motivation_trade_geo}b). The two measures thus track different dimensions of the international system: one tied to multilateral positioning, the other to bilateral political relations.

\subsubsection{Irreducibly Bilateral Variation}

A related approach in the literature summarizes bilateral alignment using countries' relations with a small number of reference powers \citep{bonadio2025playing, gopinath2025changing, liu2025international}. To assess how much bilateral variation this approach recovers, we compute cosine similarity in countries' third-country stance vectors and compare it with the direct bilateral score $S_{od,t}$. Figure~\ref{fig:bilateral_multilateral}B reports the year-by-year cross-sectional correlation under alternative reference sets.

Even using the full set of countries as references, the correlation stabilizes at only 0.45 to 0.55, implying an $R^2$ of roughly 0.25. Most bilateral variation is therefore not recoverable from countries' broader geopolitical orientation. Reducing the reference set to four geopolitical poles weakens the correlation to 0.10--0.30, accounting for a much smaller fraction of bilateral geopolitical alignment. Bilateral relations cannot be summarized well by countries' positions relative to a few hegemons.

These results point to two distinct dimensions of the international system: a multilateral dimension, which UNGA voting and low-dimensional projections partially capture, and a bilateral dimension, shaped by dyad-specific diplomatic, security, and economic interactions. For trade, where frictions operate through country-pair channels, this bilateral dimension is essential.

\section{Geopolitical Alignment and Trade Patterns}
\label{s:geo_trade_patterns}

Before proceeding to the dynamic estimation, we first document a positive comovement between bilateral trade and geopolitical alignment, both on specific country pairs and in the cross section. We begin with bilateral case studies and then estimate a standard gravity specification. The results are consistent: periods of closer alignment are associated with higher trade flows, both for specific dyads and in cross-sectional averages.

\subsection{Case Studies and Descriptive Patterns} \label{ss:case_studies}

\begin{figure}[ht]
    \centering
    \caption{Geopolitics and Trade: Selected Bilateral Dyads}
    \includegraphics[width=\textwidth]{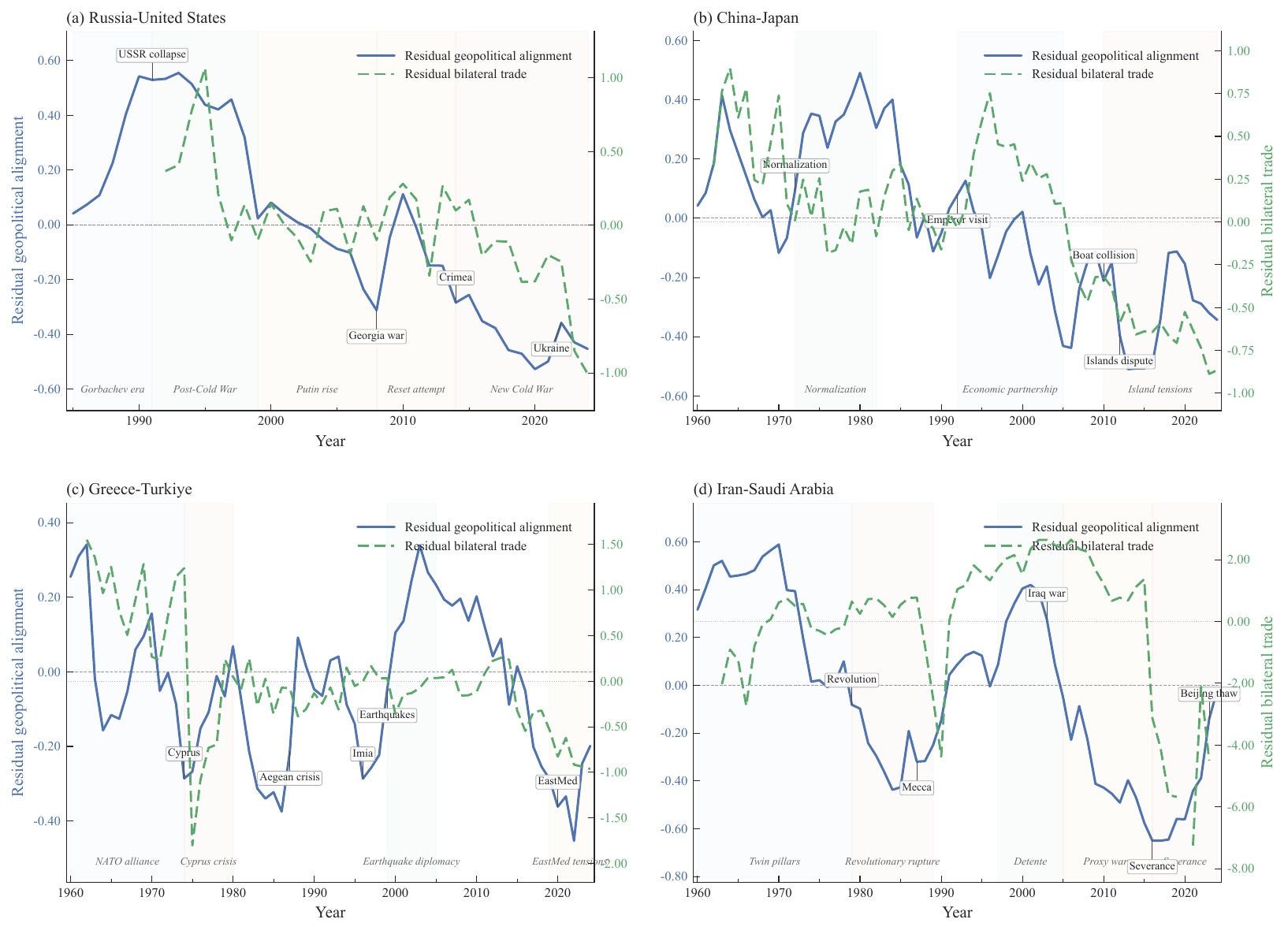}
    \label{fig:geo_trade_cases}
    \note{\emph{Notes:} Each panel plots residualized geopolitical alignment (blue, left axis) against residualized bilateral trade (green dashed, right axis). Geopolitical alignment scores are residualized on two country-year fixed effects and dyad fixed effects. Trade flows (log values from COMTRADE) are residualized on origin-year, destination-year, and origin-destination fixed effects. Shaded regions indicate major diplomatic periods.}
\end{figure}

We begin with bilateral case studies that illustrate how geopolitical alignment and trade comove across different political settings. Figure~\ref{fig:geo_trade_cases} plots residualized geopolitical alignment against residualized bilateral trade for four dyads spanning major-power relations and regional rivalries. Both series are residualized to isolate bilateral variation from global and country-specific trends: geopolitical scores are residualized on two country-year and dyad fixed effects, and trade flows (log COMTRADE values) on origin-year, destination-year, and origin-destination fixed effects. In each case, periods of stronger geopolitical alignment are associated with stronger bilateral trade, while periods of deterioration coincide with weaker trade.

Panels (a) and (b) examine major-power dyads. The Russia--U.S.\ relationship shows a post-Soviet improvement in alignment coinciding with stronger trade, followed by renewed deterioration beginning in the late 1990s and intensifying after 2014, which coincides with sharply weaker bilateral trade. The China--Japan relationship displays a similar pattern: the improvement in relations during the 1990s is associated with stronger trade, while the deterioration after 2010 amid territorial disputes coincides with declining trade residuals.

Panels (c) and (d) extend the analysis to regional rivalries outside the major-power sample. In the Greece--T\"{u}rkiye dyad, episodes of political conflict coincide with weaker trade, while periods of diplomatic improvement coincide with stronger trade. In the Iran--Saudi Arabia dyad, the improvement in relations around the late 1990s is associated with stronger bilateral trade, while the deterioration after the mid-2000s and especially after 2016 coincides with weaker trade.

These cases are illustrative rather than exhaustive. Appendices~\ref{app:add_cases_usa} and~\ref{app:add_cases_asia} present additional dyads and show similar patterns across a broader set of bilateral relationships.

\begin{figure}[ht]
    \centering
    \caption{Geopolitical Alignment and Trade Intensity: Major-Economy Dyads}
    \begin{subfigure}[b]{0.49\textwidth}
        \includegraphics[width=\textwidth]{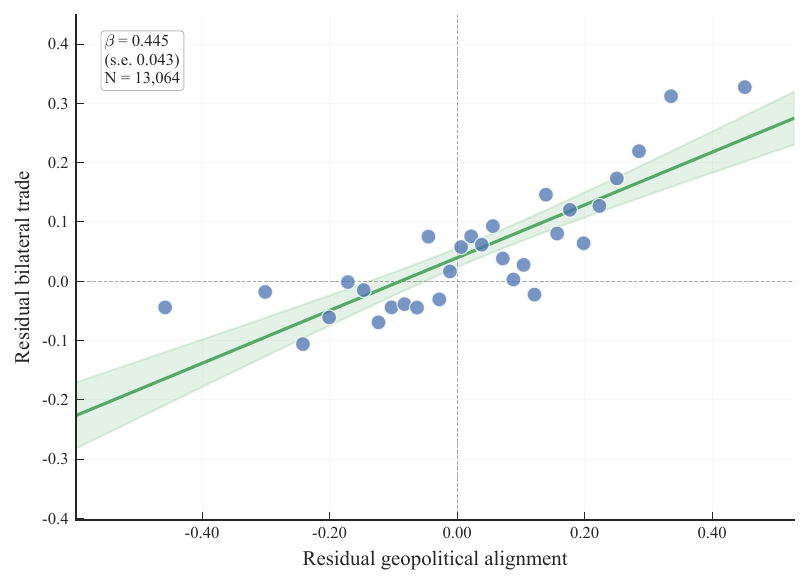}
        \caption{Cold War Era (Pre-1991)}
    \end{subfigure}
    \hfill
    \begin{subfigure}[b]{0.49\textwidth}
        \includegraphics[width=\textwidth]{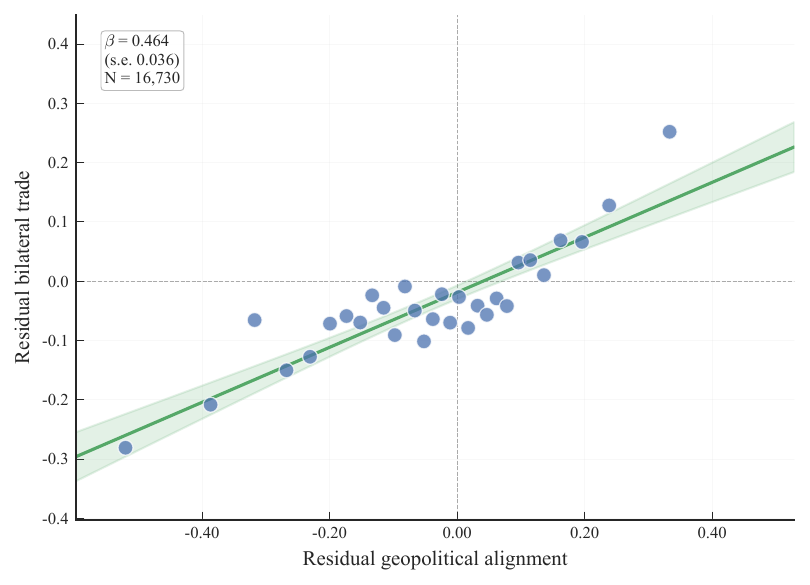}
        \caption{Post-Cold War (1991--2024)}
    \end{subfigure}
    \label{fig:binscatter_major}
    \note{\emph{Notes:} Binscatter plots of residualized log trade flows against residualized geopolitical alignment for dyads among the 32 major economies (defined in Section~\ref{ss:measuring_alignment}). Each point represents the mean within 30 quantile-based bins. Fitted lines from bivariate regressions are shown with 95\% confidence intervals. Trade and geopolitical alignment scores are residualized as in Figure~\ref{fig:geo_trade_cases}.}
\end{figure}

To extend the case-study evidence to a broader sample, Figure~\ref{fig:binscatter_major} plots residualized bilateral trade against residualized geopolitical alignment across all major-economy dyads, separately for the Cold War and post-1991 periods. Since both variables are residualized, the figure uses within-dyad variation net of country-year shocks and time-invariant bilateral factors. The positive association holds in both time periods: a one-unit increase in residual alignment is associated with a 45 log point increase in residual trade in the Cold War era ($\beta = 0.445$, s.e.\ $= 0.043$, $N = 13{,}064$) and a 46 log point increase after 1991 ($\beta = 0.464$, s.e.\ $= 0.036$, $N = 16{,}730$). The stability of these coefficients suggests that the within-dyad positive correlation is not driven by a particular geopolitical regime.

\subsection{Cross-Country Gravity Estimates} \label{ss:cross_country_gravity}

To quantify the effect of geopolitical alignment on trade and benchmark it against standard gravity covariates, we estimate the following specification:
\begin{equation}
\label{eq:gravity}
\ln X_{od,t} = \alpha S_{od,t} + \boldsymbol{\beta}'\mathbf{Z}_{od} + \delta_{ot} + \delta_{dt} + \varepsilon_{od,t},
\end{equation}
where \(X_{od,t}\) denotes bilateral trade from origin \(o\) to destination \(d\) in year \(t\), \(S_{od,t}\) is our measure of geopolitical alignment, and \(\mathbf{Z}_{od}\) includes standard time-invariant gravity controls: geographic distance, contiguity, and linguistic distance. Origin-year and destination-year fixed effects absorb all country-specific time-varying determinants of trade, including GDP, multilateral resistance, and unilateral trade policies, so that the coefficient captures cross-sectional variation in bilateral relationships. We omit country-pair fixed effects in the baseline specification in order to estimate the coefficients on the standard gravity controls.

We estimate this specification on two samples. Our benchmark consists of the 32 major economies introduced in Section~\ref{ss:measuring_alignment}. Trade among these economies accounts for 59--71\% of world trade over our sample period (Appendix Figure~\ref{fig:trade_major_decomp}), despite representing fewer than 3\% of potential country pairs; bilateral flows within this group are almost always strictly positive, so the log-linear specification raises limited concerns about selection from zero flows. The full sample includes all country pairs.

\begin{table}[ht]
\footnotesize
\centering
\caption{Geopolitical Alignment and Bilateral Trade: Cross-Sectional Gravity Estimates}
\label{tab:static}
\begin{tabular*}{\textwidth}{@{\extracolsep{\fill}}lcccccccc@{}}
\hline \hline
 & \multicolumn{4}{c}{Major Countries} & \multicolumn{4}{c}{All Countries} \\
\cmidrule(lr){2-5}\cmidrule(lr){6-9}
Dependent Variable: log Trade Value & (1) & (2) & (3) & (4) & (5) & (6) & (7) & (8)\\
\hline
Geopolitical Alignment & 0.729  & 0.458  &  & 0.595  & 2.571 & 1.096 & & 0.357\\
 & (0.121)  & (0.106)  &  & (0.077) & (0.043)  & (0.027) & & (0.017) \\
UNGA Voting Similarity: & & & -0.342 & & & & -0.084 & \\
$\quad$ Negative Ideal Point Distance & & & (0.045) & & & & (0.013) & \\
Geographic Distance & & -0.813 & -0.947 & & & -1.448 & -1.562 & \\
& & (0.052) & (0.051) & & & (0.016) & (0.017) \\
Neighbor & & 0.226 & 0.257 & & & 0.843 & 0.765 & \\
& & (0.171) & (0.157) & & & (0.079) & (0.083) & \\
Linguistic Distance & & -1.099 & -1.705 & &  & -1.701 & -1.905 & \\
& & (0.330) & (0.301) & & & (0.075) & (0.078) & \\
Mean Dep. Var. & 12.52 & 12.52 & 12.72 & 12.52 & 7.23 & 7.23 & 7.26 & 7.23 \\
Observations & 58,948 & 58,948 & 50,946 & 58,948 & 1,087,543 & 1,087,543 & 974,780 & 1,087,543 \\
\hline
Origin $\times$ Year FE & Yes & Yes & Yes & Yes & Yes & Yes & Yes & Yes\\
Destination $\times$ Year FE & Yes & Yes & Yes & Yes & Yes & Yes & Yes & Yes\\
Origin $\times$ Destination FE & No & No & No & Yes & No & No & No & Yes\\
\hline \hline
\noalign{\vskip -0.8em}
\end{tabular*}
\note{\emph{Notes:} The unit of observation is an origin-destination country pair in a year. Columns 1--4 report results for country pairs among 32 major countries, while columns 5--8 include all country pairs. Standard errors are clustered at the country-pair level.}
\end{table}

Table~\ref{tab:static} reports the baseline estimates. The coefficient on geopolitical alignment is positive and precisely estimated in all specifications. Among major economies, a one-standard-deviation increase in alignment (0.26 units, comparable in magnitude to the U.S.--China deterioration around the 2018 trade war) is associated with 19.0 log points higher bilateral trade in the bivariate specification (column 1) and 11.9 $(= 0.26 \times 0.458)$ log points after controlling for geography, contiguity, and linguistic distance (column 2). The association is larger in the full-country trade sample, where the standard deviation of alignment is 0.23 units: a one-standard-deviation increase is associated with 25.2 log points higher bilateral trade with gravity controls (column 6), and the bivariate coefficient in column 5 is larger still.

Table~\ref{tab:variance_decomp_app} reports a variance decomposition that quantifies the share of bilateral trade variation explained by different gravity variables. Geopolitical alignment is comparable in magnitude to standard trade barriers such as language: among major countries, geopolitical alignment explains 2.2\% of the variation in bilateral trade, compared with 2.8\% for linguistic distance.

To benchmark our measure against the standard approach in the literature, columns 3 and 7 replace geopolitical alignment with UNGA voting similarity (negative Ideal Point Distance). The coefficient is negative, implying that countries with more similar voting patterns tend to trade \textit{less}. This counterintuitive sign is consistent with the evidence in Section~\ref{ss:bilateral_multilateral} that UNGA voting captures multilateral positioning rather than bilateral relations. India--Pakistan and Greece--T\"{u}rkiye illustrate the disconnect: both dyads exhibit high voting similarity despite persistent bilateral tensions.

\paragraph{Robustness}
The pattern is robust along several dimensions. First, adding country-pair fixed effects in columns 4 and 8 preserves a positive and precisely estimated coefficient on geopolitical alignment, although it is somewhat attenuated in the full sample. With country-pair fixed effects, the sign of UNGA voting similarity is positive (Table~\ref{tab:static_odfe}). Second, Poisson pseudo-maximum likelihood (PPML) and inverse hyperbolic sine (IHS) specifications that account for zero trade flows yield similar results (Appendix Table~\ref{tab:static_ppml}). Third, the results are unchanged when we use BACI or IMF trade data in place of UN Comtrade (Table~\ref{tab:static_baci_imf}). Fourth, alternative geopolitical alignment measures, such as raw scores, alternative discount factors $\lambda$, and moving-average constructions, yield similar coefficients (Table~\ref{tab:static_other_geo}). Fifth, repeated cross-sectional estimates by year show that the coefficient is large during the Cold War, declines in the post--Cold War period, and rises again after 2018 (Figure~\ref{fig:static_by_year_combined}), indicating that the cross-sectional association between geopolitics and trade varies across time periods while remaining economically important throughout.


\section{Dynamic Trade Effects of Geopolitical Alignment} \label{s:trade_elas_geo}
Section~\ref{s:geo_trade_patterns} documented the cross-sectional association between alignment and trade; we now estimate dynamic responses. We use local projection methods to trace how changes in bilateral alignment propagate to trade over time, providing the reduced-form elasticities that support the quantitative analysis in Section~\ref{s:quantitative}. We include all two-way fixed effects, thereby relying exclusively on within-dyad variation. The absence of pre-trends, the robustness exercises in Section~\ref{ss:robustness}, and two identification diagnostics in Section~\ref{ss:iv_checks} help mitigate endogeneity concerns.

\subsection{Empirical Specification} \label{ss:empirical_spec}

Rather than modeling the origins of bilateral geopolitical shifts, we treat $S_{od,t}$ as given and estimate its dynamic effects on bilateral trade. The event-based measure captures bilateral alignment through major political events, such as sanctions, export controls, diplomatic breakdowns, security incidents, summits, treaty signings, and their accumulated relationship consequences.\footnote{When bilateral trade reorients through any of these channels, the reorientation is part of the estimand rather than a separate confounder.} The object of interest is the effect of broad bilateral geopolitical realignment as captured by observable political events, not the effect of a narrowly defined political gesture holding all other bilateral changes fixed.

Formally, following \citet{Jorda2025-pi}, we define the impulse response function (IRF) as
\begin{equation}
\mathcal{R}_{S \to X}(h) \equiv \mathbb{E}\!\left[\ln X_{od,t+h} \mid S_{od,t}=s+1;\,\mathbf{X}_{od,t}\right] - \mathbb{E}\!\left[\ln X_{od,t+h} \mid S_{od,t}=s;\,\mathbf{X}_{od,t}\right]
\label{eq:irf_trade}
\end{equation}
for nonnegative horizons $h=0,\ldots,20$, where $\mathbf{X}_{od,t} \equiv \{\ln X_{od,t-\ell},\,S_{od,t-\ell}\}_{\ell=1}^{L}\cup\{\delta_{od},\delta_{ot},\delta_{dt}\}$ collects lagged trade, lagged alignment, and the two-way fixed effects. The IRF traces the counterfactual difference in bilateral trade at horizon $h$ between a one-unit increase in current alignment and the baseline, holding the conditioning set fixed. We also estimate the same projection for $h=-8,\ldots,-1$ as pre-trend diagnostics.

To estimate $\mathcal{R}_{S\to X}(h)$, we use local projections \citep{Jorda2005-te,Jorda2025-pi}. The estimating equation follows the structure of \citet{boehm2023long}:
\begin{equation}
\ln X_{od,t+h} = \beta_h S_{od,t} + \sum_{\ell=1}^{L} \gamma_{h,\ell} \ln X_{od,t-\ell} + \sum_{\ell=1}^{L} \beta_{h,\ell} S_{od,t-\ell} + \delta_{od} + \delta_{ot} + \delta_{dt} + \varepsilon_{od,t+h}.
\label{eq:lp_trade}
\end{equation}
for horizons $h \in \{-8, -7, \ldots, 19, 20\}$. For $h \geq 0$, $\beta_h$ estimates the dynamic trade response to a unit change in geopolitical alignment at time $t$; for $h < 0$, the same coefficients implement the pre-trend diagnostics described above. The specification controls for the autocorrelation in both variables through $L = 3$ lags.\footnote{Appendix Figure~\ref{fig:dynamic_lag} reports sensitivity to alternative lag specifications. Using $L = 5$ yields virtually identical impulse responses, indicating that the baseline dynamics are not sensitive to the three-lag choice.} Local projections trace both the contemporaneous impact and the adjustment path of trade following changes in alignment, and accommodate the autocorrelation structure of both variables.\footnote{Local projections offer three advantages over VAR methods \citep{Jorda2005-te, Montiel-Olea2021-oo, Plagborg-Moller2021-hi}. First, they remain robust to mis-specification of the underlying dynamic process. Second, they naturally accommodate our high-dimensional fixed-effects structure. Third, estimating pre-treatment coefficients for $h < 0$ provides a diagnostic for pre-trends, anticipation, and some forms of reverse causality.}

Our identification strategy exploits within-dyad temporal variation through all two-way fixed effects. Country-pair fixed effects ($\delta_{od}$) absorb all time-invariant bilateral determinants: geographic distance, colonial history, common language, and other standard gravity variables. Origin-year ($\delta_{ot}$) and destination-year ($\delta_{dt}$) fixed effects control for all country-specific time-varying factors, including GDP, multilateral resistance terms, unilateral trade policies, and global value chain participation. Identification therefore derives from differential changes in bilateral geopolitical alignment, isolating the relationship-specific component from global trends and country-level shocks.

\begin{assumption}[Identification]
\label{assumption_trade}
Conditional on the fixed effects and lagged controls, current geopolitical alignment is uncorrelated with future innovations in bilateral trade:
\[
\mathbb{E}\!\left[\varepsilon_{od,t+h}\mid S_{od,t},\,\mathbf{X}_{od,t}\right]=0.
\]
\end{assumption}

Under Assumption~\ref{assumption_trade}, $\beta_h = \mathcal{R}_{S\to X}(h)$: the LP coefficient identifies the impulse response function in equation~\eqref{eq:irf_trade}. Two threats warrant discussion: reverse causality, and omitted time-varying bilateral confounders.

Reverse causality would arise if bilateral trade shocks move measured alignment. At $h=0$, the concern is direct: a contemporaneous trade shock can push $S_{od,t}$, producing correlation between current alignment and current trade that does not reflect alignment driving trade. The concern also extends to $h>0$ through a two-step channel: a time-$t$ trade shock moves $S_{od,t}$, and the same shock persists into future trade $X_{od,t+h}$ via ordinary trade dynamics. The LP coefficient $\beta_h$ then picks up this joint correlation even when alignment has no causal effect on trade. At $h<0$, the LP provides a pre-trend diagnostic: non-zero coefficients at negative horizons would signal that current alignment is correlated with past trade innovations. The absence of such pre-trends and the three lags of trade in the specification narrow this concern; Section~\ref{ss:iv_checks} probes the reverse direction further using the transportation-cost LP-IV and examines forward-direction confounding using the event-window diagnostic.

The second threat is omitted bilateral shocks: time-varying bilateral factors that affect both trade and measured alignment but are neither encoded in the score nor absorbed by the fixed effects. Three concerns deserve note. First, trade-policy events, including FTAs, sanctions, and tariff changes, may enter $S_{od,t}$ as scored events while also affecting trade through direct policy channels. Under the broad-realignment estimand, these policy responses are part of the treatment. At the same time, the non-economic decomposition in Section~\ref{ss:robust_decomposition} and the aggregate decomposition in Section~\ref{ss:channel_attribution} show that the baseline response is not driven solely by direct trade-policy events; Section~\ref{ss:mechanisms} separately documents tariffs, sanctions, and other restrictive measures. Second, events involving multiple countries or third-country linkages may generate correlated shocks across pairs. A key feature of the measure is that such events are recorded at the dyad level, with pair-specific scores assigned during construction, which narrows the concern by capturing each pair's role in the event rather than imposing a common shock on all involved countries. Country-year fixed effects further absorb each country's overall external posture, leaving identifying variation from the pair-specific component of these events. Third, some bilateral shocks may remain unobserved, including covert commercial arrangements without public signatures and sector-specific shocks that later generate diplomatic responses. The measure's broad event coverage narrows this concern, but does not eliminate it.

Across these concerns, augmenting the specification with observable bilateral controls (aid, sanctions, migration, FDI, country risk, and applied tariffs) leaves the point estimates nearly unchanged (Appendix Figure~\ref{fig:dynamic_controls}). Section~\ref{ss:robustness} shows that the response persists across country samples, Cold War and post--Cold War decades, and event-type subsets, which makes a single omitted bilateral shock less plausible. The event-window diagnostic in Section~\ref{ss:iv_checks} further examines alignment variation around leadership transitions, whose timing is partially exogenous to contemporaneous trade.

\subsection{Baseline Results} \label{ss:dynamic_trade}

We retain the major-economy benchmark sample defined in Section~\ref{ss:measuring_alignment}, both for the econometric reason that local projections require positive trade flows at each horizon\footnote{Among the 32 major economies, fewer than 5\% of potential bilateral flows are zero, compared with over 50\% in the full sample. Unlike static gravity \citep{silva2006log}, local projections do not have a standard PPML counterpart that accommodates the dynamic structure with leads, lags, and high-dimensional fixed effects.} and the economic reason that these economies drive the bulk of global trade patterns. Appendix Figure~\ref{fig:dynamic_ihs} shows that our results are robust to using the inverse hyperbolic sine of trade as the dependent variable, which helps mitigate concerns arising from zero trade values.

\begin{figure}[ht]
    \centering
    \caption{Dynamic Effect of Geopolitical Alignment on Trade}
    \begin{subfigure}[b]{0.49\textwidth}
        \caption{Persistence of Geopolitical Alignment}
        \includegraphics[width=\textwidth]{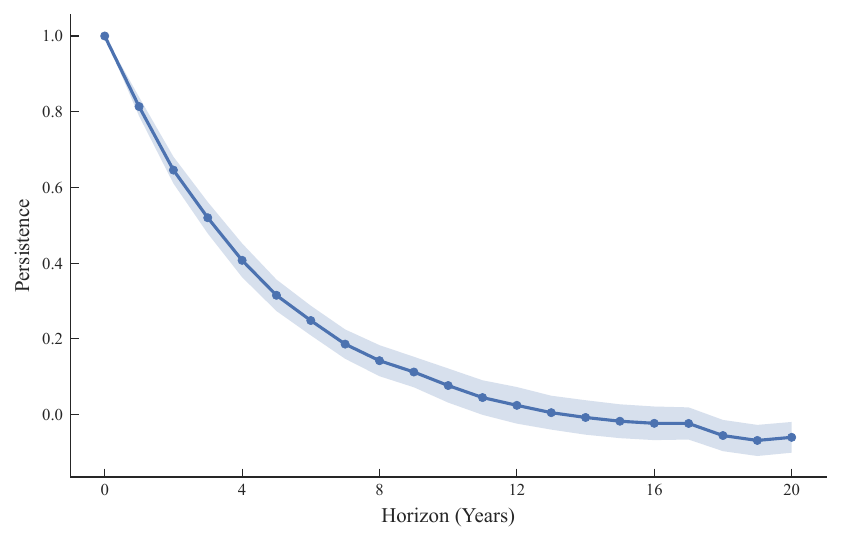}
    \end{subfigure}
    \hfill
    \begin{subfigure}[b]{0.49\textwidth}
        \caption{Response of Trade to Geopolitical Alignment}
        \includegraphics[width=\textwidth]{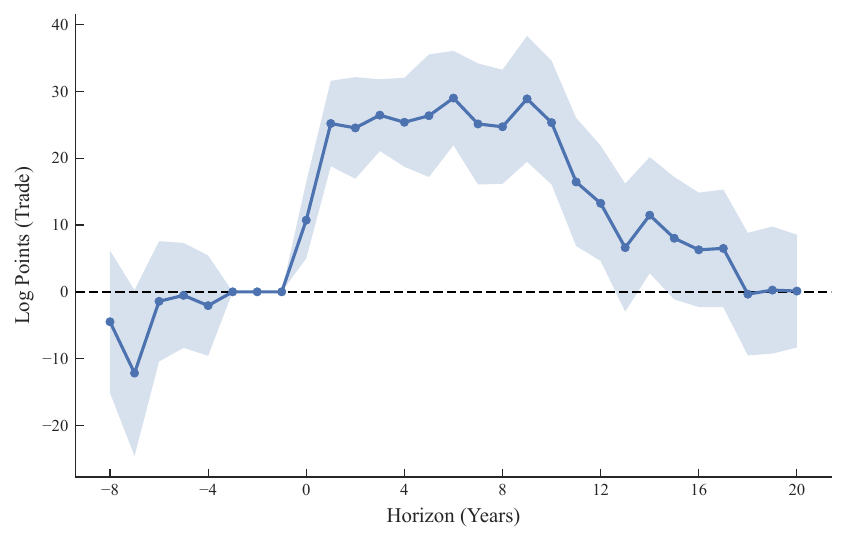}
    \end{subfigure}
    \label{fig:dynamic}
    \note{\emph{Notes:} Panel A reports estimates $\{\phi_h\}$ from: $S_{od,t+h} = \phi_h S_{od,t}+ \sum_{\ell=1}^{3} \phi_{h,\ell} S_{od,t-\ell} +\delta_{ot} +\delta_{dt}+\delta_{od}+ \varepsilon_{od,t+h}$. Panel B shows estimates $\{\beta_h\}$ from $\ln X_{od,t+h} = \beta_h S_{od,t} + \sum_{\ell=1}^{3} \gamma_{h,\ell} \ln X_{od,t-\ell} + \sum_{\ell=1}^{3} \beta_{h,\ell} S_{od,t-\ell} + \delta_{od} + \delta_{ot} + \delta_{dt} + \varepsilon_{od,t+h}$. The sample includes country pairs among 32 major economies. Both panels report estimated coefficients with 95\% confidence intervals based on Driscoll--Kraay standard errors.}
\end{figure}

Figure~\ref{fig:dynamic} displays two central results: the persistence of geopolitical shocks and their dynamic impact on bilateral trade.

\paragraph{Dynamics of Geopolitical Alignment and Trade}

Panel~A shows that shocks to geopolitical alignment are persistent but mean-reverting. Following a one-unit improvement in alignment, roughly half of the effect remains after three years, and it largely dissipates within a decade. This persistence drives the cumulative trade response. Panel~B highlights three features of that response. First, the coefficients at horizons $-8$ to $-4$ are small and statistically indistinguishable from zero, providing no evidence of pre-trends and supporting the identifying assumption.\footnote{Because the specification includes three lags of trade as controls, the coefficients at horizons $h = -3, -2, -1$ are mechanically equal to zero.} Second, the effect builds up rapidly: the coefficient rises from about 10 log points on impact to roughly 25 log points within a year and peaks at about 29 log points around year six. This buildup indicates that geopolitical shocks are followed by swift adjustment in bilateral trade flows, consistent with the rapid supply chain adjustments documented by \citet{fajgelbaum2024us} in the context of the U.S.--China trade war. Third, the response is hump-shaped, declining after year 10 as the geopolitical impulse dissipates.

Our inference employs Driscoll--Kraay standard errors to account for both serial correlation and cross-sectional dependence arising from common shocks \citep{Driscoll1998-oh}. Appendix Figure~\ref{fig:dynamic_inference} demonstrates robustness to alternative inference methods: clustering at the country-pair level yields nearly identical confidence intervals, while bootstrap inference produces only modestly wider bands.

\paragraph{Transitory versus Permanent Shocks} \label{ss:trans_perm}

The local projection estimates in Figure~\ref{fig:dynamic} combine the effect of an initial geopolitical shock with the persistence of that shock over time. To separate these margins, we decompose the responses into transitory and permanent components following \citet{bilal2024macroeconomic}. Appendix~\ref{app:irf_transitory_persistent} provides details of the decomposition.

\begin{figure}[ht]
    \centering
    \caption{Impulse Responses of Trade to Transitory and Permanent Geopolitical Shocks}
    \begin{subfigure}[b]{0.48\textwidth}
        \caption{Transitory Shock}
        \includegraphics[width=\textwidth]{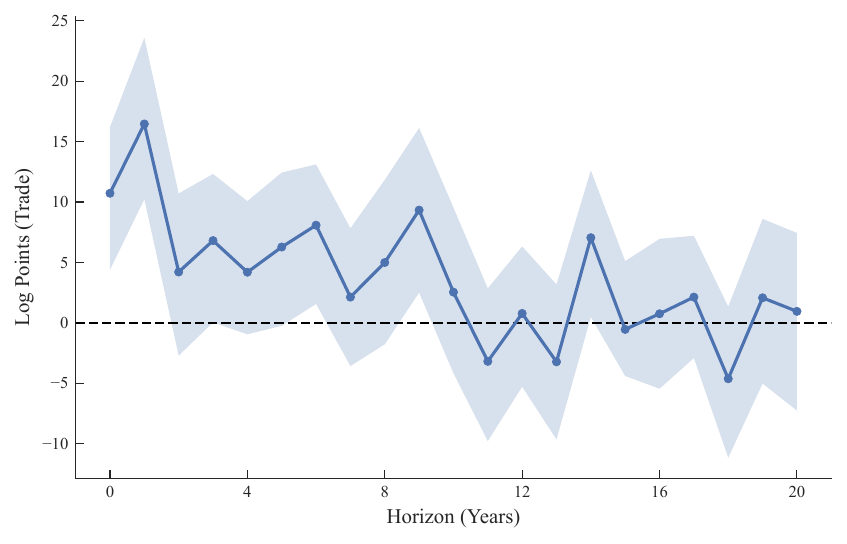}
    \end{subfigure}
    \hfill
    \begin{subfigure}[b]{0.48\textwidth}
        \caption{Permanent Shock}
        \includegraphics[width=\textwidth]{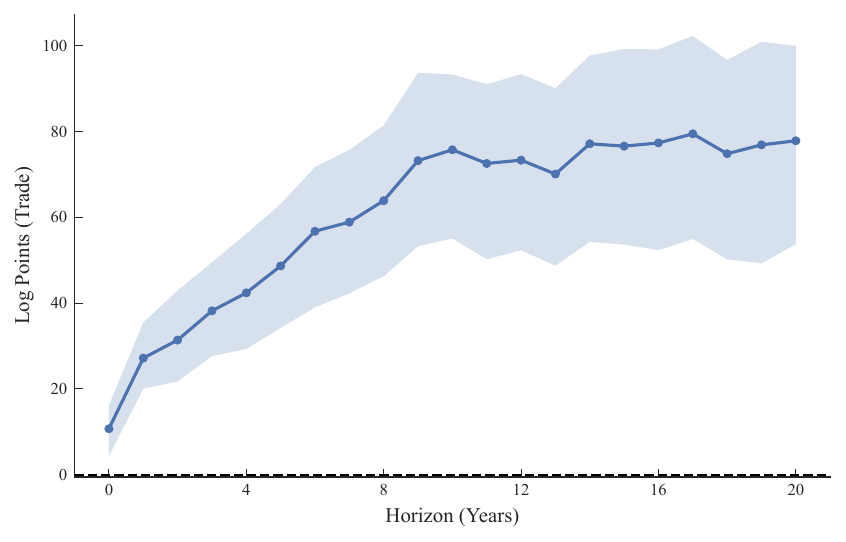}
    \end{subfigure}
    \label{fig:IRF_main}
    \note{\emph{Notes:} Panel A reports the impulse response of log trade to a purely transitory unit shock to geopolitical alignment. Panel B reports the cumulative response to a permanent unit shock. Both panels use the baseline specification with three lags and all two-way fixed effects (country-pair, origin-year, destination-year). The 95\% confidence intervals are based on 200 bootstrap iterations with country-pair block resampling.}
\end{figure}

Figure~\ref{fig:IRF_main} presents the results. In Panel A, a unit transitory improvement in geopolitical alignment at $t=0$ increases trade by approximately 11 log points on impact, peaking at about 17 log points after one year, with the effect persisting for 5--7 years before dissipating. This pattern indicates that even transitory geopolitical shocks generate persistent effects on trade.

Panel B shows that permanent improvements in alignment produce larger long-run effects. Following a permanent one-unit increase in geopolitical alignment, trade rises steadily, reaching approximately 78 log points after 20 years, with the response flattening after about 10 years. Scaling by the standard deviation of alignment (0.26 units), a permanent one-standard-deviation improvement implies a long-run increase in trade of about 20 log points, or roughly 22 percent. This magnitude exceeds the cross-sectional association in Table~\ref{tab:static}, consistent with the LP capturing cumulative adjustment not present in the static specification.

\subsection{Robustness} \label{ss:robustness}

We assess the robustness of our dynamic trade elasticities along four dimensions: country sample and time period; event-type decomposition by category and valence (cooperative vs. conflictual); observable bilateral controls; and alternative measure constructions.

\subsubsection{Sample and Period Variation} \label{ss:robust_sample_period}

\begin{figure}[ht]
    \centering
    \caption{Dynamic Effect of Geopolitical Alignment on Trade: Sample and Period Variation}
    \begin{subfigure}[b]{0.48\textwidth}
        \caption{Sample Variation}\label{fig:dynamic_sample_overlay}
        \includegraphics[width=\textwidth]{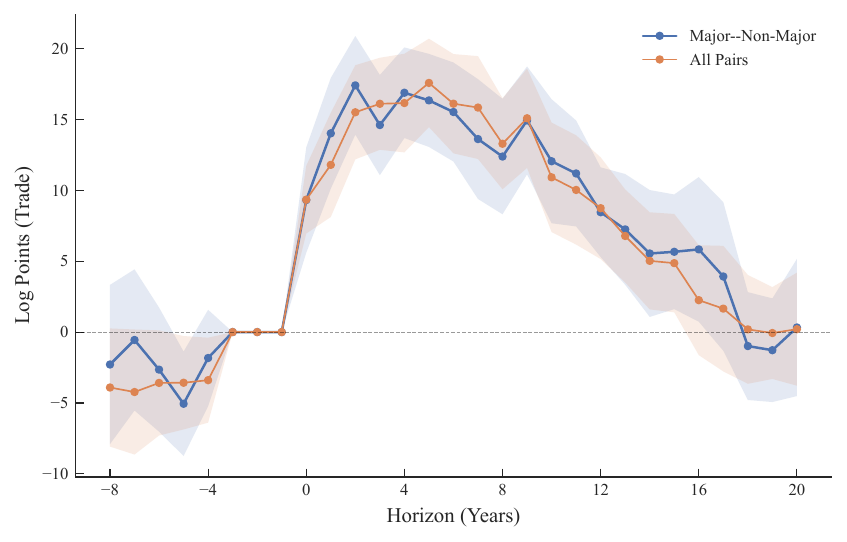}
    \end{subfigure}
    \hfill
    \begin{subfigure}[b]{0.48\textwidth}
        \caption{Period Variation}\label{fig:dynamic_period_overlay}
        \includegraphics[width=\textwidth]{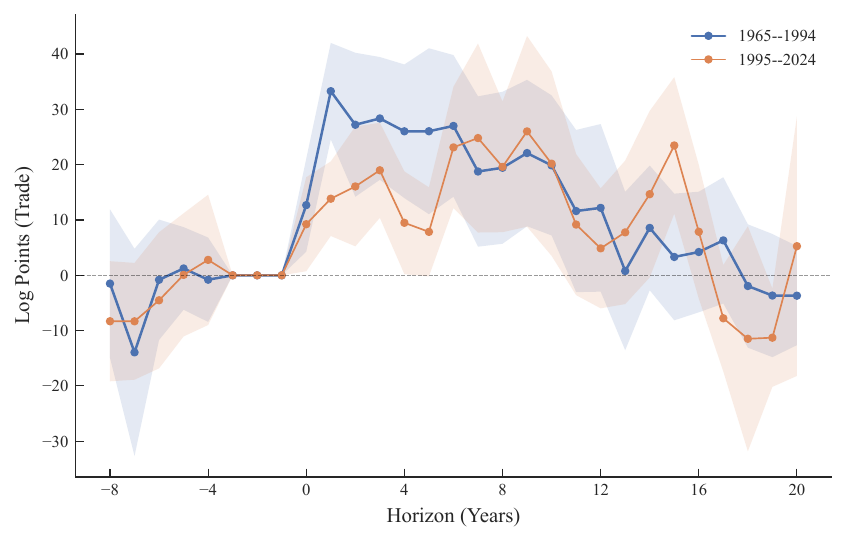}
    \end{subfigure}
    \label{fig:trade_robustness_sample_period}
    \note{\emph{Notes:} Panel~A overlays LP estimates $\{\beta_h\}$ from equation~\eqref{eq:lp_trade} for two samples: country pairs between major and non-major economies, and all country pairs. Panel~B overlays LP estimates for two subsample periods on the 32-major-economy benchmark: 1965--1994 and 1995--2024. All specifications use country-pair, origin-year, and destination-year fixed effects and three lags. Shaded bands show 95\% confidence intervals based on Driscoll--Kraay standard errors.}
\end{figure}

Panels~A and B of Figure~\ref{fig:trade_robustness_sample_period} extend the baseline along two dimensions: country sample and time period.

Panel~A overlays two sample expansions beyond the 32-major-economy benchmark: bilateral trade between major and non-major economies, and the full all-country panel. Both produce positive, statistically significant responses with attenuated peaks, approximately 17 log points for major--non-major pairs and 18 log points for all country pairs, compared with 28 log points in the baseline. The major--non-major response rises more quickly and peaks earlier, while the all-country response builds more gradually and peaks around year five. The pattern is consistent with geopolitical alignment being more consequential and faster acting among large economies, while operating in the same direction across sample definitions.

Panel~B overlays two subsample periods on the major-economy benchmark: 1965--1994 and 1995--2024. The effect is present in both periods but with different adjustment dynamics. In the earlier period, largely coinciding with the Cold War, the trade response is rapid, peaking at approximately 33 log points within two to three years. In the later period, the response is more gradual and hump-shaped, reaching a peak of about 26 log points at horizons 6--9. A natural interpretation is that geopolitical shocks translated more quickly into trade reallocation when trade blocs and formal restrictions were more prominent; later adjustment is slower, consistent with deeper production networks.\footnote{The corresponding results using UNGA voting similarity are less stable across periods: Appendix Figure~\ref{fig:dynamic_time_ipd} shows that IPD-based estimates are statistically significant before 1995 but attenuate thereafter. This instability is consistent with the evidence in Section~\ref{ss:bilateral_multilateral} that voting-based measures capture multilateral positioning rather than the bilateral variation most relevant for trade.}

\subsubsection{Event-Type Decomposition} \label{ss:robust_decomposition}

The baseline alignment score aggregates economic and non-economic events, as well as cooperative and conflictual events, into a single measure. We decompose the score along two dimensions: economic versus non-economic events, and cooperative versus conflictual events. We then re-estimate the dynamic response on each component. The two decompositions address different questions. The first asks whether the baseline response operates through economic interactions alone or also through non-economic diplomatic and security channels. The second examines whether cooperative and conflictual shocks propagate symmetrically. For each decomposition, we compute the average geopolitical alignment score within the event subset, smooth it with $\lambda = 0.3$, and estimate
{\small
\begin{equation}
    \label{eq:lp_two}
    \ln X_{od,t+h} = \sum_{g\in\{\text{A},\text{B}\}}\beta_h^{g} S^{g}_{od,t} + \sum_{\ell=1}^{3} \gamma_{h,\ell} \ln X_{od,t-\ell} + \sum_{g\in\{\text{A},\text{B}\}}\sum_{\ell=1}^{3} \beta^{g}_{h,\ell} S^{g}_{od,t-\ell} + \delta_{od} + \delta_{ot} + \delta_{dt} + \varepsilon_{od,t+h},
\end{equation}
}
where $\{A, B\}$ indexes the two event subsets.

\begin{figure}[ht]
    \centering
    \caption{Dynamic Effect of Geopolitical Alignment on Trade: Event-Type Decomposition}
    \begin{subfigure}[b]{0.49\textwidth}
        \caption{Economic vs.\ Non-Economic Events}\label{fig:dynamic_econ_nonecon}
        \includegraphics[width=\textwidth]{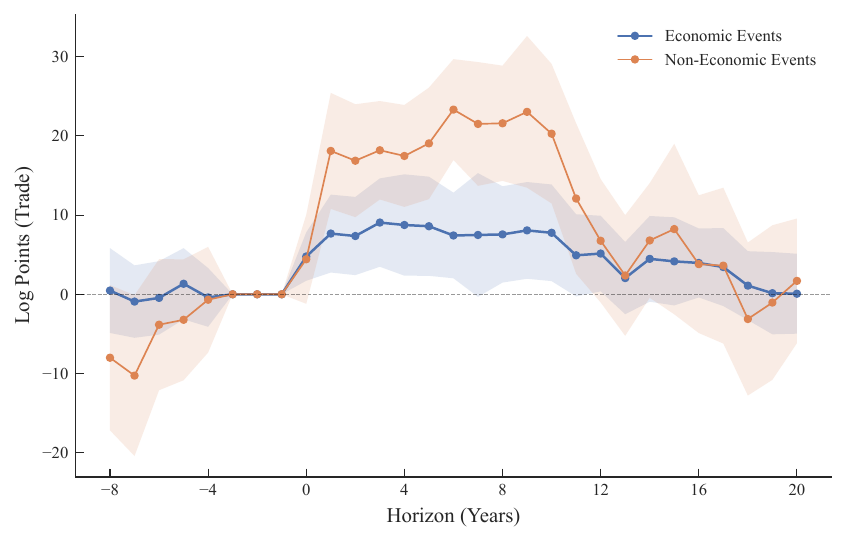}
    \end{subfigure}
    \hfill
    \begin{subfigure}[b]{0.49\textwidth}
        \caption{Cooperative vs.\ Conflictual Events}\label{fig:dynamic_coop_conflict}
        \includegraphics[width=\textwidth]{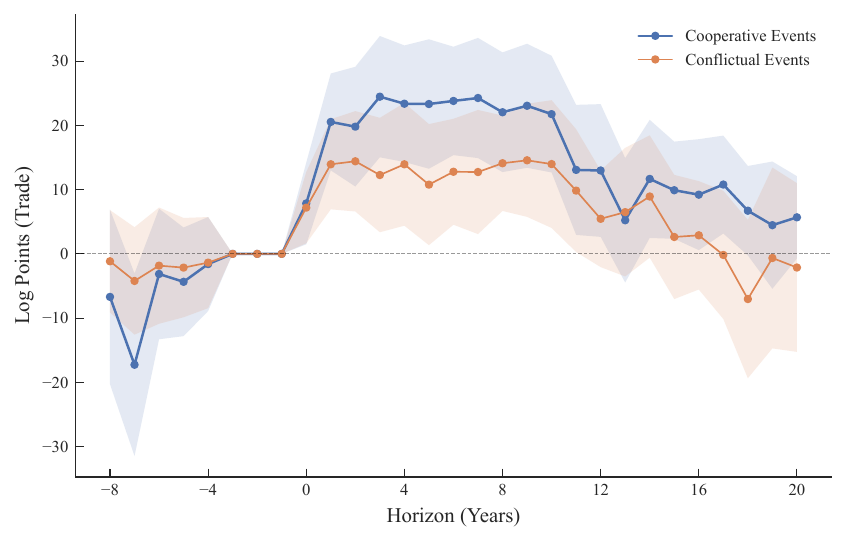}
    \end{subfigure}
    \label{fig:trade_robustness_types}
    \note{\emph{Notes:} Panel~A reports estimates $\{\beta_h^{\text{econ}}, \beta_h^{\text{non-econ}}\}$ from the two-series specification in equation~\eqref{eq:lp_two} with A, B indexing economic and non-economic events. Panel~B reports analogous estimates with A, B indexing cooperative and conflictual events; the conflictual-events coefficients are sign-flipped so that a positive value expresses the trade response to a unit reduction in conflictual interactions. All specifications use country-pair, origin-year, and destination-year fixed effects and three lags and are estimated on 32-major-economy pairs. Shaded bands show 95\% confidence intervals based on Driscoll--Kraay standard errors.}
\end{figure}

Panel~A of Figure~\ref{fig:trade_robustness_types} decomposes alignment by economic and non-economic events. A natural question is whether the baseline response is driven primarily by economic events, including sanctions, tariff actions, and trade agreements, or also by non-economic diplomatic and security interactions.\footnote{The two dynamic scores share related but largely distinct information. In the major-economy sample, their raw correlation is 0.38; after absorbing origin-year, destination-year, and country-pair fixed effects, the residual correlation falls to 0.21, so the two components identify separate effects in the specification above.} Non-economic alignment exhibits effects similar in magnitude and persistence to the aggregate score, while the economic component is weaker. The baseline response therefore reflects more than directly economic events: diplomatic, security, and political interactions carry much of the trade response. The reduced-form effect is therefore not confined to instruments that mechanically alter trade costs.

Panel~B decomposes alignment by event valence, using the CAMEO classification to construct a cooperative-events measure from verbal and material cooperation categories and a conflictual-events measure from verbal and material conflict categories. The conflictual-events coefficients are sign-flipped so that a positive value expresses the trade response to a unit reduction in conflictual interactions, placing both series in the same direction. Cooperative events produce a larger and more persistent response, peaking at approximately 23 log points, while conflictual events produce a smaller response of around 14 log points. The asymmetry suggests that cooperation opens new trading relationships and reduces uncertainty, while conflict disrupts existing flows that may be partially sustained by contractual commitments and sunk investments. Despite the asymmetry, both directions generate significant trade responses, indicating that the baseline pattern is not driven by a particular event subtype.

\subsubsection{Observable Bilateral Controls} \label{ss:robust_controls}

A residual concern with the two-way fixed-effects design is that time-varying bilateral shocks correlated with alignment could drive the trade response. We probe this by augmenting the baseline with three lags of observable bilateral variables. Conditioning on lagged rather than contemporaneous values keeps the controls predetermined relative to the alignment shock at time $t$, avoiding the bad-controls concern that arises when controls are themselves downstream of the shock of interest.

Panel~A of Appendix Figure~\ref{fig:dynamic_controls} adds bilateral aid, sanctions, and migration, preserving the full horizon. Panel~B additionally includes bilateral FDI, country risk, and applied tariffs, which restrict the sample to post-2001. In both specifications, the impulse response tracks its unconstrained baseline closely: the peak coefficient moves by less than half a log point in each case. The baseline trade response is therefore not driven by these observable bilateral channels.

\subsubsection{Alternative Geopolitical Alignment Measures} \label{ss:robust_measure}

We next examine whether the results are sensitive to the construction of the geopolitical alignment measure. We re-estimate the local projections using the unsmoothed average event score, $\tilde{S}_{od,t}$, which captures contemporaneous bilateral interactions without incorporating historical persistence. The results are reported in Figure~\ref{fig:dynamic_nonsmooth}. Although the unsmoothed measure exhibits rapid mean reversion, with approximately 75\% of the initial shock dissipating within one year, the implied trade response remains economically meaningful. The peak elasticity declines to around 10 log points (compared to 28 log points in the baseline), consistent with the more transitory nature of individual events.

When we scale the estimates to reflect a permanent one-unit change in geopolitical alignment, the long-run trade effect converges to our baseline estimate of approximately 78 log points. Because the local projection controls for lags of alignment, identifying variation comes from innovations rather than the accumulated stock, so smoothed and unsmoothed measures yield similar long-run estimates. The convergence extends to alternative smoothing parameters ($\lambda = 0.1$ and $\lambda = 0.5$) and a four-period moving average of the raw score (Figure~\ref{fig:dynamic_robust_specification}), indicating that the relationship between geopolitics and trade is not confined to a particular smoothing choice.

\subsection{Identification Diagnostics} \label{ss:iv_checks}

Two residual threats remain after the baseline design: confounders from bilateral shocks that move trade and alignment together, and reverse causality from trade to alignment. We address each with a targeted diagnostic.

\subsubsection{Exogenous Leader Transition}

The first diagnostic focuses on changes in bilateral alignment around leadership transitions. These episodes are useful because the timing of unexpected deaths in office and close-election turnovers is less likely to be driven by contemporaneous bilateral trade shocks. We use 26 unexpected deaths in office among leaders of major countries \citep{goemans2009introducing} and 56 close-election turnovers with vote margins below 10 percentage points \citep{marx2024elections}, for a total of 82 transition years.\footnote{Deaths in office provide plausibly exogenous timing. Close-election turnovers with narrow margins provide additional identifying variation, but we do not interpret them as a sharp regression-discontinuity design.}

For each dyad-year, let $\mathbf{1}^{\mathrm{LC}}_{od,t}$ denote an indicator equal to one if either country in dyad $(o,d)$ experiences an exogenous leadership transition in year $t$. We define
\[
z^{\mathrm{LC}}_{od,t}
=
\mathbf{1}^{\mathrm{LC}}_{od,t}
\left(S_{od,t+1}-S_{od,t-1}\right).
\]
Thus, $z^{\mathrm{LC}}_{od,t}$ equals the change in bilateral alignment from the year before to the year after the transition for dyad-years exposed to a leadership transition, and equals zero for all other dyad-years. Positive values indicate improved alignment, while negative values indicate deterioration. The magnitude captures the size of the short-window shift in the bilateral relationship.

We estimate two local-projection equations using the same fixed effects and controls as in equation~\eqref{eq:lp_trade}. First, we project future trade on the leadership-transition alignment shift,
\[
\ln X_{od,t+h}
=
\zeta_h z^{\mathrm{LC}}_{od,t}
+ \text{controls}
+ \delta_{od}+\delta_{ot}+\delta_{dt}
+ \varepsilon_{od,t+h}.
\]
Second, we estimate the corresponding first-stage association between the leadership-transition shift and the baseline alignment measure,
\[
S_{od,t}
=
\varphi z^{\mathrm{LC}}_{od,t}
+ \text{controls}
+ \delta_{od}+\delta_{ot}+\delta_{dt}
+ u_{od,t}.
\]
We report $\hat{\zeta}_h/\hat{\varphi}$, which expresses the trade response per unit of alignment change and is therefore on the same scale as the baseline local-projection coefficient. Because $z^{\mathrm{LC}}_{od,t}$ is constructed from the alignment series itself, this ratio is not an instrumental-variables estimate. It is a diagnostic that asks whether similar trade responses appear when alignment changes are concentrated around leadership transitions.

\begin{figure}[htb]
\centering
\caption{Identification Diagnostics}
\begin{subfigure}[b]{0.48\textwidth}
    \caption{Event-Window: Exogenous Leader Transition}
    \includegraphics[width=\textwidth]{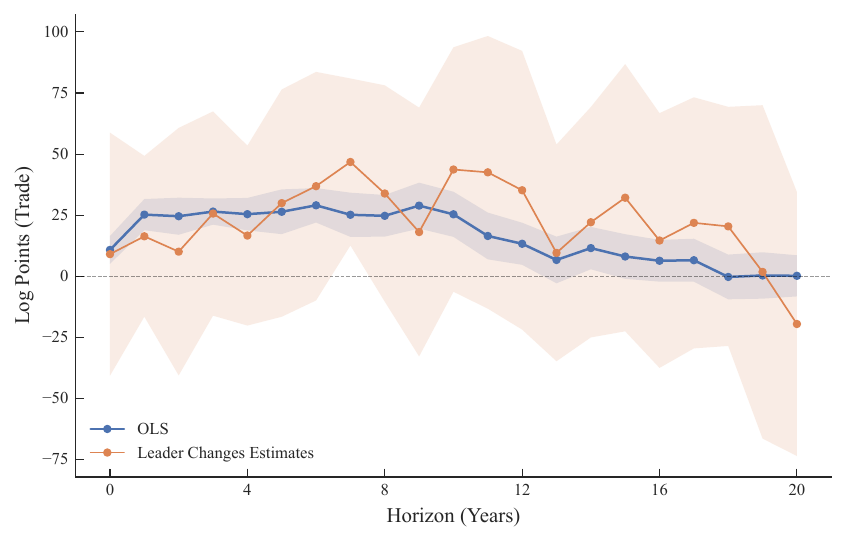}
\end{subfigure}
\hfill
\begin{subfigure}[b]{0.48\textwidth}
    \caption{Reverse-Direction: Transportation-Cost LP-IV}
    \includegraphics[width=\textwidth]{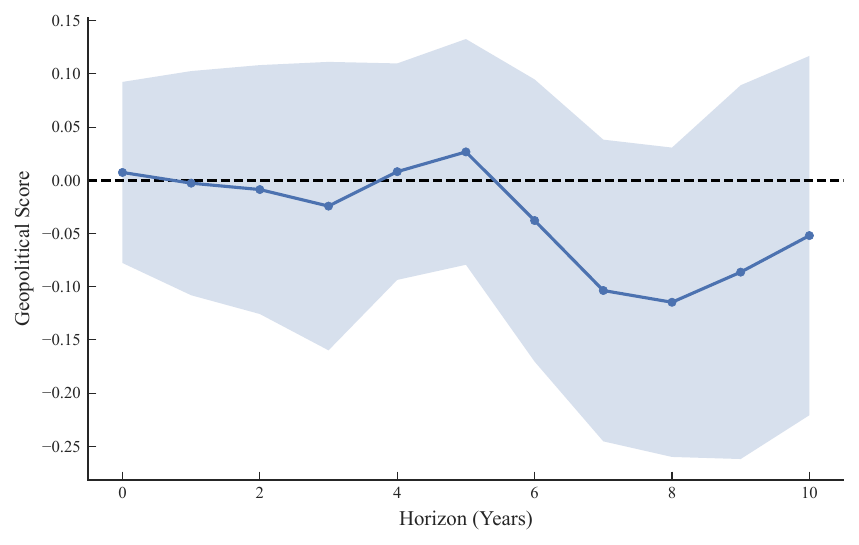}
\end{subfigure}
\label{fig:iv_checks}
\note{\emph{Notes:} This figure reports two identification diagnostics for country pairs among 32 major economies. Panel~A compares the baseline local-projection estimates with event-window estimates based on leadership transitions. The leadership-transition shift is defined as 
$z^{\mathrm{LC}}_{od,t}=\mathbf{1}^{\mathrm{LC}}_{od,t}(S_{od,t+1}-S_{od,t-1})$, 
where $\mathbf{1}^{\mathrm{LC}}_{od,t}$ equals one if either country in dyad $(o,d)$ experiences a leadership transition in year $t$, and zero otherwise. Panel~B reports the LP-IV response of bilateral alignment to trade, instrumenting $\ln X_{od,t}$ with gravity-predicted flows based on time-varying sea and air distance coefficients. All specifications include country-pair, origin-year, and destination-year fixed effects; Panel~B also includes five lags of trade, alignment, and predicted trade. Shaded bands show 95\% confidence intervals based on Driscoll--Kraay standard errors.}
\end{figure}

Panel~A of Figure~\ref{fig:iv_checks} reports the leadership-transition estimates together with the baseline local-projection estimates. The two series are similar in shape and magnitude, although the leadership-transition estimates have wider confidence intervals because they rely on a smaller set of event-years. This similarity indicates that the baseline trade response also appears when alignment changes are concentrated around leadership transitions, where timing is less likely to be driven by contemporaneous bilateral trade shocks.

\subsubsection{Reverse-Direction Diagnostic: Transportation-Cost LP-IV}

The second diagnostic targets the reverse causality: whether transport-cost-induced changes in bilateral trade generate subsequent changes in alignment. Following \citet{feyrer2019trade}, we instrument bilateral trade using predicted flows from a gravity equation with time-varying transportation-cost coefficients:
\begin{equation*}
    \ln X_{od,t}=\beta_{\text{sea},t}\times \ln {\text{seadist}}_{od}+\beta_{\text{air},t}\times \ln {\text{airdist}}_{od}+\delta_{od}+\delta_{ot}+\delta_{dt}+\epsilon_{od,t}.
\end{equation*}
We then estimate the LP-IV response of alignment to trade. Let $Z_{od,t}$ denote the gravity-predicted component of bilateral trade generated by time-varying sea and air distance coefficients. For each horizon $h$, we estimate the first stage and reduced form,
\[
\ln X_{od,t} = \pi Z_{od,t} + \mathbf{W}_{od,t}'\Gamma + u_{od,t},
\qquad
S_{od,t+h} = \rho_h Z_{od,t} + \mathbf{W}_{od,t}'\Lambda_h + v_{od,t+h},
\]
where $\mathbf{W}_{od,t}$ includes the fixed effects and lag controls. Following the LP-IV representation in \citet{Plagborg-Moller2021-hi}, the coefficient is $\hat{\theta}_h=\hat{\rho}_h/\hat{\pi}$. The exclusion restriction requires that gravity-predicted trade affects alignment only through actual trade, conditional on the fixed effects and lag controls.

Panel~B of Figure~\ref{fig:iv_checks} reports the LP-IV response of alignment to trade. The Kleibergen--Paap rk Wald F-statistic is 62.0, indicating a strong first stage. The estimated effects are statistically indistinguishable from zero over the horizon shown, providing little evidence of reverse causality operating through bilateral trade.\footnote{While the literature has extensively examined how economic fundamentals shape geopolitical outcomes, it typically does not focus on bilateral trade per se. Existing studies emphasize broader mechanisms such as economic dependence through global trade networks \citep{kleinman2024international}, the opportunity cost of conflict \citep{mayer2025fragmentation}, and trade asymmetries \citep{liu2025international}.}

Taken together with the absence of pre-trends and the robustness exercises above, the two diagnostics narrow the set of confounders consistent with the baseline. Using exogenous leadership transitions to generate plausibly exogenous variation in geopolitical alignment yields estimates similar to the baseline, while the transportation-cost LP-IV provides little evidence of effects running from trade to alignment. Remaining concerns center on bilateral confounders that jointly affect alignment and trade and may not be fully absorbed by the two-way fixed-effects specification.

\section{Heterogeneity and Mechanisms} \label{s:het_mechanisms}

The local projections in Section~\ref{s:trade_elas_geo} estimate the reduced-form average trade response to changes in bilateral alignment. This section unpacks that response along two dimensions: Section~\ref{ss:heterogeneity} examines how domestic institutions and sectoral composition shape where the effect is strongest, and Section~\ref{ss:mechanisms} traces the policy and non-policy channels through which alignment affects trade.

\subsection{Heterogeneity} \label{ss:heterogeneity}

\paragraph{Institutional Heterogeneity}

Government intervention plays an important role in mediating the effect of geopolitics on trade. Deteriorating bilateral alignment increases the likelihood that countries impose restrictive trade policies, and such policies are more readily implemented where executives face fewer institutional constraints. We therefore expect the effect of geopolitical alignment on trade to be stronger among country pairs with greater executive discretionary power. Using data from the Varieties of Democracy project (V-Dem), we proxy discretionary power with the index of judicial and legislative constraints on the executive and classify 32 major economies into discretionary and non-discretionary types.\footnote{We split the 32 major economies by the sample median of discretionary power. The below-median group includes Australia, Austria, Belgium, Canada, Denmark, France, Germany, India, Italy, Japan, the Netherlands, Sweden, Switzerland, the United Kingdom, and the United States. The above-median group includes Argentina, Brazil, China, Indonesia, Iran, Iraq, Mexico, Nigeria, the Philippines, Poland, Russia, Saudi Arabia, South Africa, South Korea, Spain, T\"{u}rkiye, and Venezuela.} We then estimate:
\begin{equation}
\label{eq:het_institution}
    \ln X_{od,t+h} = \sum_{g=1}^{3} \beta_h^{g}\, S_{od,t} \times \mathbf{1}[(o,d)\in \mathcal{C}_g] + \mathbf{W}_{od,t}'\boldsymbol{\gamma}_h + \delta_{od} + \delta_{ot} + \delta_{dt} + \varepsilon_{od,t+h},
\end{equation}
where $\mathcal{C}_1$, $\mathcal{C}_2$, and $\mathcal{C}_3$ denote country pairs in which both countries have low discretionary power, one country has high discretionary power, and both countries have high discretionary power, respectively.

\begin{figure}[ht]
    \centering
    \caption{Dynamic Trade Responses: Institutional Heterogeneity}
    \begin{subfigure}[b]{0.49\textwidth}
        \caption{Discretionary Power}
        \includegraphics[width=\textwidth]{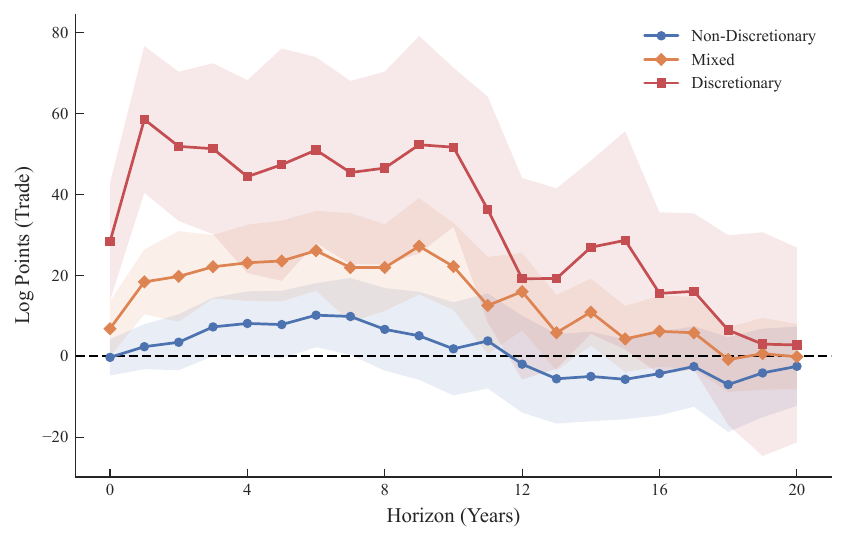}
    \end{subfigure}
    \hfill
    \begin{subfigure}[b]{0.49\textwidth}
        \caption{Overall Institutional Quality}
        \includegraphics[width=\textwidth]{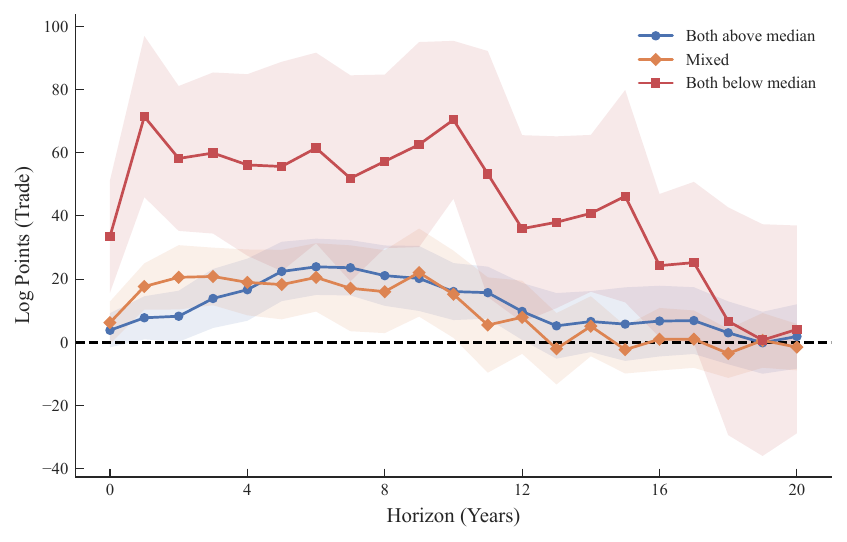}
    \end{subfigure}
    \label{fig:dynamic_het_institution}
    \note{\emph{Notes:} This figure reports estimates $\{\beta_h^{g}\}$ from equation~\eqref{eq:het_institution}, augmented with three lags of $\ln X_{od}$ and group-interacted lags of $S_{od,t-\ell}$ for $\ell = 1, 2, 3$. The sample includes country pairs among 32 major economies. In Panel A, countries are split by the sample median of discretionary power; in Panel B, countries are split by the sample median of the institutional-quality index. Coefficient estimates are shown with 95\% confidence intervals based on Driscoll--Kraay standard errors.}
\end{figure}

Panel A of Figure~\ref{fig:dynamic_het_institution} presents the estimates. The response is strongest among pairs in which both countries are above the sample median of discretionary power and is substantially attenuated among pairs in which both countries are below the median. As a robustness check, we partition countries by overall institutional quality using the Worldwide Governance Indicators from the World Bank and estimate a specification analogous to equation~\eqref{eq:het_institution}.\footnote{We split the 32 major economies by the sample median of the institutional-quality index. The higher-score group includes Australia, Austria, Belgium, Canada, Denmark, France, Germany, Italy, Japan, the Netherlands, Poland, South Korea, Spain, Sweden, Switzerland, the United Kingdom, and the United States. The lower-score group includes Argentina, Brazil, China, India, Indonesia, Iran, Iraq, Mexico, Nigeria, the Philippines, Russia, Saudi Arabia, South Africa, T\"{u}rkiye, and Venezuela.}
Panel B confirms that the effect is largest among pairs in which both countries fall below the sample median of the institutional-quality index, consistent with the discretionary power results.

\paragraph{Formal Alliances}

The effect of geopolitical alignment on trade is attenuated among countries linked through formal alliances. Institutionalized cooperation within alliances may dampen the transmission of geopolitical shocks into trade flows in both directions. Alliance members are less likely to impose restrictive policies in response to temporary bilateral deterioration, as formal coordination and dispute resolution mechanisms help contain conflicts. At the same time, improvements in alignment generate more limited trade expansion within alliances, since member countries already exhibit high baseline integration and low policy uncertainty.

Appendix Figure~\ref{fig:dynamic_alliance} presents the dynamic effects of geopolitical alignment on trade within the European Union and NATO. Consistent with this interpretation, the estimated responses are substantially weaker and statistically insignificant among alliance members, indicating that formal institutional ties reduce the sensitivity of trade to geopolitical shocks.

\paragraph{Sectoral Heterogeneity}

We examine how the effects of geopolitical alignment vary across sectors. Grouping industries by their 2-digit North American Industry Classification System (NAICS) codes, we estimate effects separately for agriculture, energy and mining, and three manufacturing subsectors. Appendix Figure~\ref{fig:trade_sector} presents the results. The effects are positive across all sectors except energy and mining. The largest and most persistent responses occur in manufacturing, particularly in NAICS sector 33 (primary metals, machinery, electronics, and transportation equipment).

We also examine whether our findings are concentrated in sectors that the U.S.\ International Trade Administration classifies as ``critical.'' The effects are positive in both critical and non-critical sectors, with no statistically significant difference between the two (Appendix Figure~\ref{fig:trade_sector} Panel F). The sizable effects in sectors not explicitly targeted by government policy suggest that our results are not primarily driven by directed intervention.

\subsection{Mechanisms} \label{ss:mechanisms}

\paragraph{Trade Policies: Tariffs and Sanctions}

We turn to whether the effect of geopolitical alignment on trade operates through policy instruments such as tariffs and sanctions. A deterioration in bilateral relations may increase the likelihood of restrictive policy measures, which in turn reduce trade. To investigate this channel, we compile data on three types of restrictive policies: tariffs from \citet{Teti2024}, sanctions from the Global Sanctions Database \citep{felbermayr2020global,yalcin2025global},\footnote{The GSDB records both comprehensive and partial trade sanctions, encompassing a broader set of restrictive measures than simple trade embargoes.} and other restrictive trade policies from the Global Trade Alert (GTA) Database. We estimate how geopolitical alignment is associated with the imposition of these policies.

\begin{table}[ht]
\caption{Geopolitical Alignment and Policies}\label{tab:tp}
\begin{center}
\scalebox{0.9}{
\begin{tabular}{lcccccc}
\hline \hline
& (1)      & (2)      & (3)      &  (4)      & (5)   & (6)  \\
\multirow{2}{*}{Dependent Variable:}  & \multirow{2}{*}{log (1+Tariff Rate)} & \multicolumn{3}{c}{$\mathbf{1}$ [Sanction]} & \multicolumn{2}{c}{Restricting Trade Policy} \\
  & & Trade & Financial & Travel & $\geq 1$ Product & $\geq 20$ Products \\
\hline
Geopolitical Alignment   & -0.003 & -0.051 & -0.052 & -0.050 & -4.24 & -2.16 \\
        & (0.003)  & (0.009)  & (0.006)  &  (0.005) & (2.02)   & (0.88) \\ 
Mean Dep. Var. & 0.050 & 0.049 & 0.037 & 0.019 & 21.4 & 3.24\\
Observations   & 24760   & 74400  & 74400  & 74400 & 16864  & 16864 \\
\hline 
Imposer $\times$ Year FE & Yes & Yes & Yes & Yes & Yes & Yes\\
Receiver $\times$ Year FE & Yes & Yes & Yes & Yes & Yes & Yes\\
Imposer $\times$ Receiver FE & Yes & Yes & Yes & Yes & Yes & Yes\\
\hline \hline
\end{tabular}}
\end{center}
\vspace{-5mm}
\note{\emph{Notes:} This table shows the relationship between geopolitical alignment and policy outcomes. The unit of observation is an imposer--receiver country pair in a given year. The dependent variables are the log $(1+\text{tariff rate})$ in column 1, indicators for the presence of trade, financial, and travel sanctions in columns 2--4, and the number of restrictive trade policies in columns 5--6. The sample includes country pairs among 32 major countries. Standard errors are clustered at the country-pair level.}
\end{table}

As shown in Column 1 of Table~\ref{tab:tp}, there is no statistically significant correlation between geopolitical alignment and tariffs. One explanation is that tariff setting largely takes place within multilateral frameworks, limiting the role of bilateral geopolitical relations. In contrast, the likelihood of imposing sanctions rises as geopolitical relations deteriorate. A one-standard-deviation decline in alignment is associated with a 1.4 percentage point increase in the probability of imposing trade, financial, or travel sanctions. Columns 5--6 indicate that worsening relations also increase the number of restrictive trade measures: a one-standard-deviation decline raises the count by about 1.1, with roughly half covering more than 20 products. Taken together, these results suggest that geopolitical alignment shapes the use of restrictive policies, which in turn affect trade flows. Table~\ref{tab:tp_nonecon} shows similar results when we use non-economic geopolitical alignment as the independent variable.

\begin{figure}[ht]
    \centering
    \caption{Dynamic Effect of Geopolitical Alignment on Policies}
    \begin{subfigure}[b]{0.49\textwidth}
        \caption{Tariffs}
        \includegraphics[width=\textwidth]{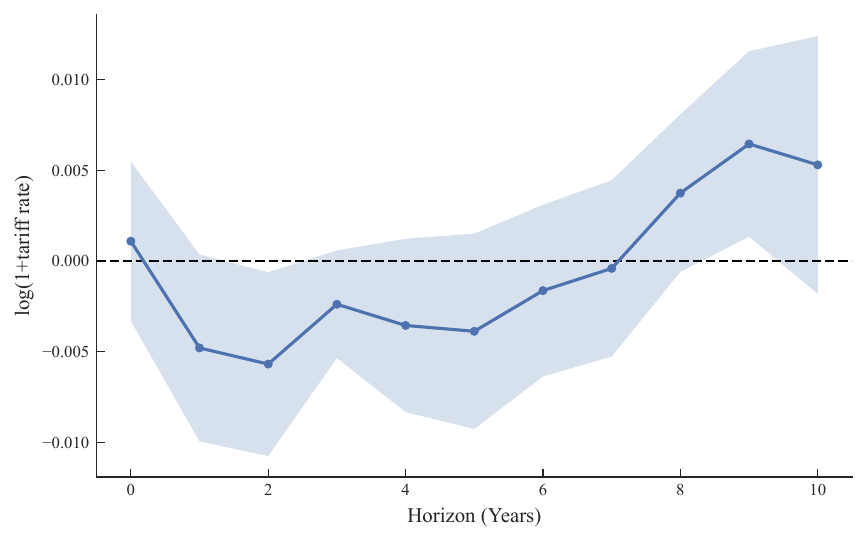}
    \end{subfigure}
    \hfill
    \begin{subfigure}[b]{0.49\textwidth}
        \caption{Sanctions}
        \includegraphics[width=\textwidth]{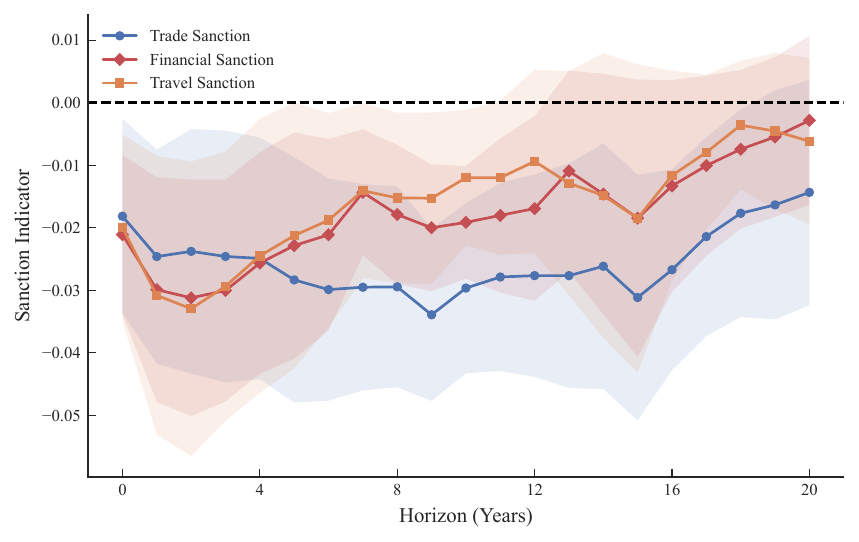}
    \end{subfigure}
    \label{fig:dynamic_tp}
    \note{\emph{Notes:} This figure shows estimates $\{\beta_h\}$ from: $Y_{od,t+h} = \beta_h S_{od,t}+\sum_{\ell=1}^{3} \gamma_{h,\ell} Y_{od,t-\ell} + \sum_{\ell=1}^{3} \beta_{h,\ell} S_{od,t-\ell}+\delta_{ot} +\delta_{dt}+\delta_{od}+ \epsilon_{od,t+h}$, where the dependent variable is $\log(1+\text{tariff rate})$ in Panel A and indicators for trade, financial, or travel sanctions in Panel B. The sample includes country pairs among 32 major economies. Both panels report coefficient estimates with 95\% confidence intervals based on Driscoll--Kraay standard errors.}
\end{figure}

Figure~\ref{fig:dynamic_tp} examines the dynamic responses of tariffs and sanctions to geopolitical shocks. The responses differ sharply across the two instruments. Panel A shows little systematic response of tariffs to bilateral geopolitical shocks, consistent with Table~\ref{tab:tp}. One interpretation is that the GATT/WTO system and its Most Favored Nation commitments limited the scope for bilateral tariff adjustment over this period.\footnote{The tariff data begin in 1995.}

By contrast, Panel B shows that sanctions respond systematically to geopolitical deterioration. A one-standard-deviation decline in alignment increases the probability of sanctions by about 0.7 percentage points on impact, with the effect persisting for up to a decade.

These results identify sanctions as one measurable policy channel: tariffs appear insulated by multilateral rules, while sanctions remain flexible tools of bilateral statecraft. A back-of-the-envelope calculation suggests that the direct tariff channel explains at most 2.7\%\footnote{As shown in Appendix~\ref{app:add_tp}, a one-unit increase in $\log(1+\text{tariff rate})$ reduces log trade by about 3 units. The dynamic LP implies that a one-unit deterioration in geopolitical alignment changes $\log(1+\text{tariff rate})$ by no more than 0.007 in absolute value over the plotted horizon. The direct tariff channel therefore accounts for at most $0.007 \times 3 / 0.78 = 2.7\%$ of the total trade response.} of the total trade response, while sanctions account for roughly 18\%.\footnote{Following the approach in Appendix~\ref{app:irf_transitory_persistent}, we estimate that the long-run impulse responses of a permanent shock to trade, financial, and travel sanctions on trade value are approximately $-0.4$, $-0.5$, and $-0.6$, respectively. A permanent one-unit deterioration in geopolitical alignment increases the probability of trade, financial, and travel sanctions by roughly 12, 9, and 8 percentage points. Combining these estimates implies that the sanctions channel explains roughly $(0.4\times 0.12 + 0.5\times 0.09 + 0.6\times 0.08)/0.78=18\%$ of the total effect of geopolitical alignment on trade. Sanctions may also affect trade indirectly through exchange rate and financial channels \citep{itskhoki2025sanctions}.} The remaining share is not directly decomposed. It may operate through firm uncertainty, informal policy pressure, bilateral attitudes, and other non-tariff channels. We provide suggestive evidence below on one residual channel---bilateral attitudes. Section~\ref{ss:channel_attribution} reports a complementary decomposition at the aggregate level.

\paragraph{Beyond Formal Policies: Bilateral Attitudes}

The residual channels are harder to quantify directly. Geopolitical tensions can increase uncertainty about future tariffs, sanctions, or regulatory disruptions, inducing firms to adjust supply chains, sourcing decisions, and investment plans in advance. Here we focus on a related attitudinal channel: tensions may also shape consumer demand and firm behavior through social attitudes or political sentiment.

\begin{table}[ht]
\caption{Geopolitical Alignment and Social Attitudes}\label{tab:preference}
\begin{center}
\scalebox{0.9}{
\begin{tabular}{lcccccc}
\hline \hline
& (1)      & (2)      & (3)      &  (4)      & (5)   & (6)  \\
Dependent Variable:  & Attitudes & \multicolumn{2}{c}{log Trade} & Attitudes & \multicolumn{2}{c}{log Trade} \\
\cmidrule(lr){2-4}\cmidrule(lr){5-7}
 & \multicolumn{3}{c}{Major Countries} & \multicolumn{3}{c}{All Countries} \\
\hline
Geopolitical Alignment   & 0.147 & 0.427 & 0.309 & 0.140 & 0.254 & 0.144 \\
        & (0.017)  & (0.143)  & (0.149)  &  (0.016) & (0.109)   & (0.112) \\ 
Attitudes   &  &  & 0.763 &  &  & 0.829 \\
        &  &  & (0.341)  &   &   & (0.241) \\ 
Mean Dep. Var. & 0.48 & 15.8 & 15.8 & 0.49 & 14.6 & 14.6\\
Observations   & 1665   & 3048  & 3048 & 2772 & 5138  & 5138 \\
\hline 
Origin $\times$ Year FE & Yes & Yes & Yes & Yes & Yes & Yes\\
Destination $\times$ Year FE & Yes & Yes & Yes & Yes & Yes & Yes\\
Origin $\times$ Destination FE & Yes & Yes & Yes & Yes & Yes & Yes\\
\hline \hline
\end{tabular}}
\end{center}
\vspace{-5mm}
\note{\emph{Notes:} This table shows the relationship between geopolitical alignment, bilateral attitudes, and trade. The unit of observation is an origin--destination country pair in a given year. The dependent variables are the attitude index in columns 1 and 4 and log trade value in columns 2--3 and 5--6. Columns 1--3 report results for country pairs among the 32 major countries, while columns 4--6 include all country pairs. Standard errors are clustered at the country-pair level.}
\end{table}

We study this channel using bilateral attitude data from the Pew Research Center's \textit{Global Attitudes and Trends Survey}, which measures views toward other countries for a limited sample beginning in 2002. We rescale the index to lie between 0 and 1, with higher values indicating more favorable attitudes. Table~\ref{tab:preference} reports the results. Among major economies, country pairs with higher geopolitical alignment exhibit more favorable bilateral attitudes. Columns 2 and 3 show that more favorable attitudes are associated with larger bilateral trade flows, and including attitudes attenuates the coefficient on geopolitical alignment.\footnote{The social attitudes measure is directional, and in the Global Attitudes and Trends Survey most country pairs are observed in only one direction. When estimating the trade regressions, we assign the attitude for each origin--destination pair using the available directional measure; if both directions are observed, we take their average. As a result, the number of observations in columns 2 and 3 is less than double that in column 1.} Similar patterns emerge in the full sample of countries. Taken together, the mechanisms evidence identifies sanctions as a measurable policy channel and provides suggestive evidence on bilateral attitudes, while leaving most of the reduced-form trade response unattributed.

\section{Model and Quantitative Results}
\label{s:quantitative}

The local projections in Section~\ref{s:trade_elas_geo} estimate the dynamic effects of geopolitical alignment on bilateral trade. This section embeds those reduced-form estimates in a single-sector Armington accounting exercise to gauge model-implied aggregate trade and welfare magnitudes. We decompose changes in bilateral trade costs from 1995 to 2021 into tariff, geopolitical, and residual components, holding productivity and endowments at their 1995 values in the baseline counterfactuals. The framework provides a parsimonious aggregation device: it takes the estimated effects as given and maps them into implied trade-cost changes and general-equilibrium responses.

\subsection{Framework}

\textbf{Model.} The economy has $N$ countries indexed by origin $o$ and destination $d$; country $d$ is endowed with labor $\ell_d$ and consumes varieties from all countries with CES preferences,
\begin{equation}
C_d = \left[ \sum_{o=1}^{N} c_{od}^{\frac{\sigma-1}{\sigma}} \right]^{\frac{\sigma}{\sigma-1}},
\end{equation}
where $\sigma>1$ is the elasticity of substitution. Producers in $o$ operate technology $y_o=z_o\ell_o$ with productivity $z_o$ and wage $w_o$. Bilateral trade is subject to multiplicative trade costs $d_{od} = d^{\text{iceberg}}_{od}\,\tilde{\tau}_{od}\,d^{\text{geo}}_{od}$, where $\tilde{\tau}_{od}\equiv 1+\tau_{od}$ is the ad valorem tariff factor and $d^{\text{geo}}_{od}$ reflects geopolitical friction. The expenditure share of $d$ on varieties from $o$ is
\begin{equation}
\pi_{od}=\left(\frac{p_{od}}{P_d}\right)^{1-\sigma}, \quad
p_{od}=\frac{w_o d_{od}}{z_o}, \quad
P_d=\left[\sum_{o} p_{od}^{1-\sigma}\right]^{\frac{1}{1-\sigma}}.
\end{equation}
Expenditure equals wage income plus tariff revenue rebated lump sum, $X_d = w_d \ell_d /\iota_d$, where $\iota_d \equiv \sum_o \pi_{od}/\tilde{\tau}_{od}\leq 1$. Labor-market clearing gives $w_o \ell_o = \sum_d X_d \pi_{od}/\tilde{\tau}_{od}$. Welfare is real income, $\omega_d = X_d/P_d = w_d\ell_d/(\iota_d P_d)$; in hat-algebra form $\hat\omega_d = \hat w_d\,(\iota_d/\iota'_d)/\hat P_d$, where the $\iota/\iota'$ wedge captures the tariff-revenue adjustment.

\textbf{Equilibrium in changes.} We solve counterfactuals by exact hat algebra \citep{Dekle2008-gc}: given a trade-cost shock $\hat{d}_{od}$, we iterate the hat-algebra system (Appendix~\ref{app:hat_algebra}) to a fixed point in $\{\hat w_d, \hat P_d, \hat\pi_{od}, \hat X_d\}$. One country's wage is fixed as numeraire.

\textbf{Calibration and trade-cost decomposition.} We implement the model on OECD Inter-Country Input-Output data for 78 countries, 1995--2021. Log trade costs are parameterized as $\ln d_{od,t} = \text{Geo}_{od,t} + \ln \tilde{\tau}_{od,t} + \boldsymbol{Z}_{od}\boldsymbol{\beta} + \varepsilon_{od,t}$, where $\text{Geo}_{od,t}$ denotes the geopolitical trade-cost component implied by alignment changes, $\boldsymbol{Z}_{od}$ collects time-invariant bilateral characteristics, and $\varepsilon_{od,t}$ is the residual barrier. Relative to base year 1995, log changes decompose into $\ln \hat{d}_{od,t} = \Delta \text{Geo}_{od,t} + \ln \hat{\tilde{\tau}}_{od,t} + \ln \hat{\varepsilon}_{od,t}$. We map alignment changes into cumulative trade-cost effects using the permanent-shock impulse response,
\begin{equation}
(1-\sigma)\Delta \text{Geo}_{od,t} =
\sum_{s=1}^{t-t_0}\hat{\beta}^{\text{geo}}_{t-t_0-s}\left(S_{od,t_0+s}-S_{od,t_0+s-1}\right),
\end{equation}
where $\hat\beta^{\text{geo}}_h$ is the permanent-shock response of log trade, in log points per unit of $S$, at horizon $h=t-t_0-s$ from Section~\ref{ss:trans_perm}. We estimate $\sigma=4$ from the tariff impulse response in Appendix~\ref{app:add_tp}. Following \citet{head2001increasing}, we recover the residual component from bilateral trade shares under symmetry ($\hat{\varepsilon}_{od}=\hat{\varepsilon}_{do}$) and unchanged domestic costs ($\hat{d}_{oo}=1$):
\begin{equation}
\hat{\varepsilon}_{od,t} =
\left[\frac{\hat{\pi}_{od,t}\hat{\pi}_{do,t}}{\hat{\pi}_{oo,t}\hat{\pi}_{dd,t}}\right]^{\frac{1}{2(1-\sigma)}}
\left[\hat{\tilde{\tau}}_{od,t}\hat{\tilde{\tau}}_{do,t}\right]^{-1/2}
\exp(-\Delta \text{Geo}_{od,t}),
\label{eq:unobserved_costs}
\end{equation}
so that total trade-cost changes decompose as
\begin{equation}
\hat{d}_{od,t} =
\hat{\varepsilon}_{od,t}\cdot \hat{\tilde{\tau}}_{od,t}\cdot \exp(\Delta \text{Geo}_{od,t}).
\label{eq:decomposition}
\end{equation}

\textbf{Counterfactuals.} For each year $t\in\{1996,\ldots,2021\}$, with productivity and labor endowments held at their 1995 values, we solve three specifications of the trade-cost shock (subscripts suppressed for compactness): baseline $\hat{d}^{\text{baseline}} = \hat{\varepsilon}\hat{\tilde\tau}\exp(\Delta\text{Geo})$, no-geo (alignment fixed at 1995) $\hat{d}^{\text{no-geo}} = \hat{\varepsilon}\hat{\tilde\tau}$, and no-tariff (tariffs fixed at 1995) $\hat{d}^{\text{no-tariff}} = \hat{\varepsilon}\exp(\Delta\text{Geo})$.

\subsection{Aggregate Trade Effects}

\begin{figure}[ht]
    \centering
    \caption{Aggregate Trade Effects: Trajectory and 2021 Decomposition}
    \includegraphics[width=\textwidth]{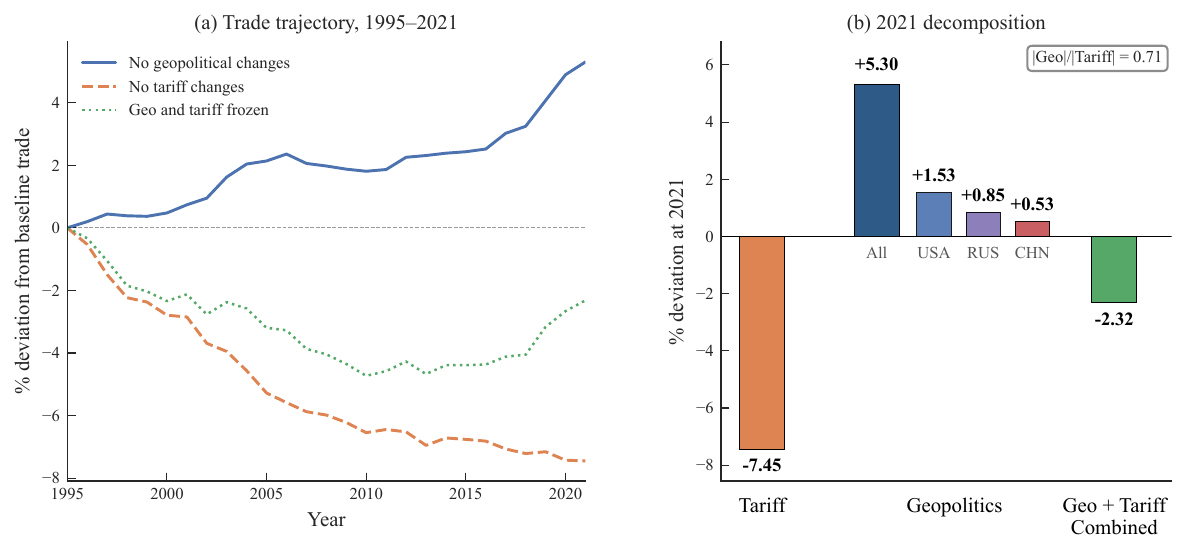}
    \label{fig:counterfactual_trade}
    \note{\emph{Notes:} Panel (a) plots year-by-year percent deviation from baseline global trade under three freezing counterfactuals: geopolitical alignment, tariffs, and both jointly (geo and tariff combined), each frozen at 1995 levels. Panel (b) reports the corresponding 2021 deviations, with the Geo bar decomposed into shares attributable to dyads involving the United States, Russia, and China; the remaining roughly 45 percent is attributable to other dyads. The Geo + Tariff Combined bar approximately equals the sum of the Tariff and Geo bars. Freezing-deviation signs are opposite to the contribution signs in the main text and Tables~\ref{tab:channel_attribution}--\ref{tab:robustness_r1}: a negative Tariff bar means tariff liberalization raised trade. Productivity and endowments are fixed at 1995 throughout.}
\end{figure}

Figure~\ref{fig:counterfactual_trade} reports the baseline decomposition. Under the calibrated model, tariff liberalization raised 2021 global trade by approximately 7.5 percent relative to a counterfactual with 1995 tariffs, while geopolitical deterioration reduced it by approximately 5.3 percent relative to a counterfactual with 1995 alignment. The two forces operate in opposite directions: geopolitical deterioration offsets roughly 70 percent of the trade expansion generated by tariff reduction. Panel (a) traces the buildup over time. Tariff liberalization accumulates steadily through the late 1990s and early 2000s as most of the major liberalization events (the Uruguay Round implementation, China's WTO accession, and regional free-trade agreements) take effect. Geopolitical deterioration builds up more slowly through the 2000s and accelerates after 2010, consistent with the fragmentation patterns documented in Section~\ref{s:geo_trade_patterns}.

No single major-power realignment dominates the aggregate geopolitical effect, though dyads involving the three largest powers together account for just over half. Panel (b) of Figure~\ref{fig:counterfactual_trade} decomposes the 5.3 percent aggregate by partner identity. Changes in U.S.\ alignment with its trading partners account for approximately 29 percent of the aggregate (about 1.5 percentage points); Russian alignment contributes about 16 percent (roughly 0.85 percentage points); Chinese alignment contributes about 10 percent (roughly 0.5 percentage points). Dyads involving the United States, Russia, or China together account for approximately 55 percent of the 2021 geopolitical trade effect, with no single power exceeding 29 percent. The remaining 45 percent is attributable to dyads that do not involve any of the three---a dispersed pattern of realignment across the other 75 economies in the sample. These magnitudes shed light on the slowdown in globalization documented by \citet{antras2021} and \citet{Goldberg2023-pb}: tariff liberalization continued to support trade integration, but geopolitical deterioration offset a substantial share of those gains.

\subsection{Decomposition and Robustness}
\label{ss:channel_attribution}

We first decompose the geopolitical contribution by separating direct trade-policy events from the rest of the alignment score, then probe the baseline's sensitivity to modeling choices and statistical uncertainty in the estimated elasticity.

\paragraph{Direct trade-policy events.}
Our baseline uses the full event-based alignment score because trade-policy actions, including tariffs, sanctions, export controls, FTAs, and commercial disputes, often carry geopolitical content beyond their direct policy effects. A tariff increase or sanction changes bilateral policy wedges, but also signals deterioration in bilateral trust, heightened retaliation risk, and future policy uncertainty, all dimensions not captured by the policy wedge alone. To decompose the geopolitical contribution by direct trade-policy events versus the rest, we recompute the Armington counterfactual using a non-trade alignment score that excludes these events while retaining investment, financial, diplomatic, and non-economic content. This partition is narrower than the economic/non-economic event split in Section~\ref{ss:robust_decomposition}, which removes all directly economic events. Because the tariff wedge in equation~\eqref{eq:decomposition} is calibrated from observed tariffs rather than from the event score, this exercise varies only the geopolitical component while holding the observed tariff path fixed.

\begin{table}[ht]
\centering
\caption{Decomposing the Geopolitical Contribution}
\label{tab:channel_attribution}
\begin{tabular*}{\textwidth}{@{\extracolsep{\fill}}lccc@{}}
\toprule
Alignment score used for geo wedge     & Tariff (\%) & Geo (\%) & $|\text{Geo}|/|\text{Tariff}|$ \\
\midrule
Full event score (baseline)            & $+7.45$ & $-5.30$ & $0.71$ \\
Excluding direct trade-policy events   & $+7.45$ & $-4.26$ & $0.57$ \\
\bottomrule
\end{tabular*}
\begin{minipage}{\textwidth}
\vspace{4pt}
{\footnotesize \textit{Notes:} Entries are contributions to 2021 global trade under the Armington counterfactual with $\sigma=4$; positive values raised trade, negative values reduced it. The full event score is the baseline alignment measure. The non-trade score excludes tariff, sanction, export-control, FTA, and commercial-dispute events from the alignment measure. Using unrounded values, the absolute difference between the two geo contributions is $1.05$ percentage points, equal to $19.8$ percent of the baseline geopolitical contribution.}
\end{minipage}
\end{table}

Excluding direct trade-policy events leaves a geopolitical contribution of $-4.26$ percent, about 80 percent of the baseline $-5.30$ percent (Table~\ref{tab:channel_attribution}). The implied offset relative to tariff liberalization is 57 percent rather than 71 percent. The direct-trade-policy component, at $19.8$ percent of the baseline, is of similar magnitude to the roughly 21 percent implied by the LP-based tariff and sanctions calculations in Section~\ref{ss:mechanisms}.

Table~\ref{tab:robustness_r1} reports robustness across alternative specifications and conventions.

\begin{table}[t]
\centering
\caption{Robustness of the Trade Decomposition}
\label{tab:robustness_r1}
\begin{tabular*}{\textwidth}{@{\extracolsep{\fill}}lccc@{}}
\toprule
Specification & Tariff (\%) & Geo (\%) & $|\text{Geo}|/|\text{Tariff}|$ \\
\midrule
Baseline ($\sigma=4$, 32-country IRF, full history)                & $+7.45$  & $-5.30$ & $0.71$ \\
Baseline IRF + 78-minus-32 IRF                                     & $+7.45$  & $-5.04$ & $0.68$ \\
\addlinespace[3pt]
\multicolumn{4}{@{}l}{Elasticity of substitution} \\
\quad \textit{$\sigma = 3$}                                         & $+4.70$  & $-5.36$ & $1.14$ \\
\quad \textit{$\sigma = 6$}                                         & $+12.34$ & $-5.25$ & $0.43$ \\
\addlinespace[3pt]
\multicolumn{4}{@{}l}{IRF statistical uncertainty (95\% CI, $\sigma=4$)} \\
\quad \textit{Lower 95\% CI on IRF}                                 & $+7.45$  & $-3.41$ & $0.46$ \\
\quad \textit{Upper 95\% CI on IRF}                                 & $+7.45$  & $-7.36$ & $0.99$ \\
\addlinespace[3pt]
Evolving fundamentals                                               & $+11.37$ & $-5.12$ & $0.45$ \\
\addlinespace[3pt]
Partial equilibrium (LP convolution, no GE feedback)                & \,---\,  & $-8.27$ & \,---\, \\
\bottomrule
\end{tabular*}
\begin{minipage}{\textwidth}
\vspace{4pt}
{\footnotesize \textit{Notes:} Entries are contributions to 2021 global trade; positive values raised trade, negative values reduced it. ``Baseline IRF + 78-minus-32 IRF'' applies the 32-country LP elasticity to major--major dyads and a separate LP elasticity to other dyads. IRF uncertainty uses the 95\% bootstrap band of the permanent IRF. ``Evolving fundamentals'' rebases the counterfactual to year-$t$ data; see text. Partial equilibrium holds wages and prices fixed; the Tariff column is not computed.}
\end{minipage}
\end{table}

\paragraph{Trade elasticity.} The geopolitical contribution is stable across alternative trade elasticities, implying a trade reduction of $5.25$ to $5.36$ percent across $\sigma \in \{3, 6\}$.\footnote{Changing $\sigma$ rescales the implied iceberg-cost change through the $1/(\sigma-1)$ mapping, while the LP coefficient disciplines the implied trade response.} The tariff contribution, by contrast, rises with $\sigma - 1$, so the ratio $|\text{Geo}|/|\text{Tariff}|$ ranges from $0.43$ to $1.14$. Applying a separate elasticity to dyads involving at least one non-major economy yields a trade reduction of $5.04$ percent, close to the baseline.\footnote{This reflects trade-weighted aggregation: major-major dyads are a small share of dyad counts but carry the bulk of world trade.}

\paragraph{Statistical uncertainty.} 
We assess statistical uncertainty by recomputing the Armington counterfactual using the lower and upper bounds of the 95 percent bootstrap band for the permanent IRF. The implied 2021 geopolitical trade reduction ranges from 3.41 to 7.36 percent, relative to the baseline estimate of 5.30 percent. The uncertainty affects the magnitude of the contribution, but not its sign.

\paragraph{Evolving fundamentals.} The baseline counterfactual holds productivity and labor endowments at their 1995 values, so trade-cost shocks are evaluated on 1995 bilateral weights. As an additional robustness check, we rebase the counterfactual year by year, using observed year-$t$ trade shares, value added, and tariffs as the base, so that each shock is evaluated on year-$t$ weights. The geopolitical contribution remains similar, at $-5.12$ percent. The tariff contribution rises to $+11.37$ percent, reflecting that tariff reductions receive greater weight when evaluated on the contemporaneous trade network. The implied $|\text{Geo}|/|\text{Tariff}|$ ratio is $0.45$, smaller than the baseline $0.71$ but still consistent with a substantial geopolitical offset to tariff liberalization.

\paragraph{Partial and General Equilibrium.}
A partial-equilibrium calculation that holds wages and prices fixed implies an 8.27 percent reduction in global trade, larger than the 5.30 percent reduction in the baseline general-equilibrium counterfactual.\footnote{The difference reflects multilateral substitution: when geopolitical alignment deteriorates between two countries, import demand reallocates toward alternative suppliers, partially offsetting the direct bilateral effect.} Across the decomposition and robustness exercises, geopolitical deterioration remains a quantitatively large offset to the trade gains from tariff liberalization.

\subsection{Country-Level Welfare Effects}

The aggregate offset translates into country-level welfare effects that differ sharply across the two forces. For country $d$, welfare changes are given by $\hat{\omega}_d=\hat{X}_d/\hat{P}_d$, where $\hat{X}_d$ incorporates both wage changes and tariff-revenue adjustments.

\begin{figure}[ht]
    \centering
    \caption{Country-Level Welfare Effects, 2021}
    \includegraphics[width=\textwidth]{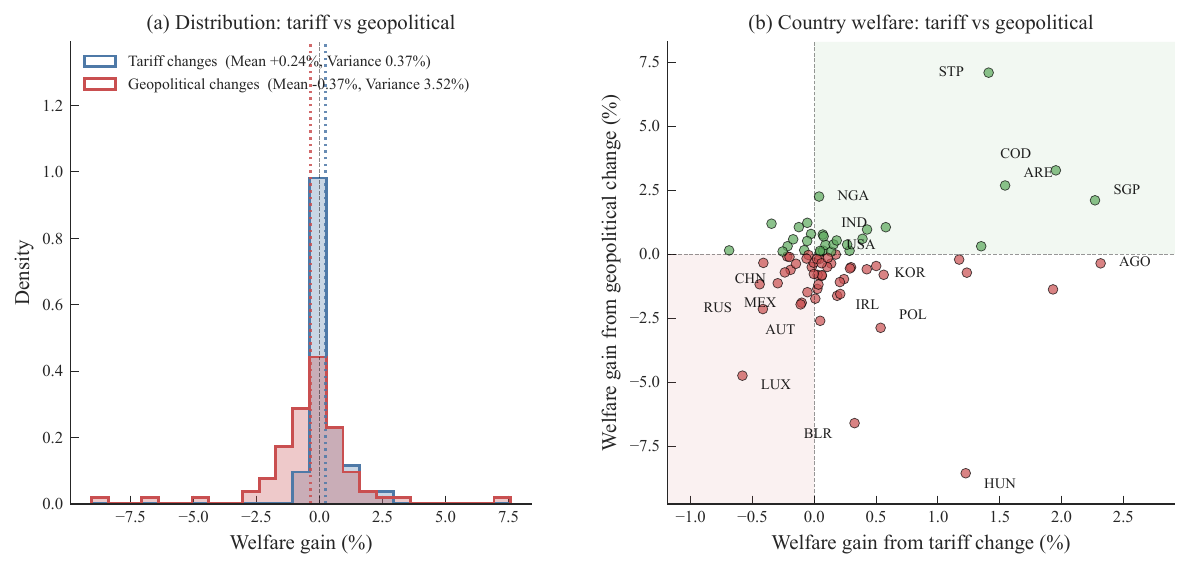}
    \label{fig:welfare_scatter}
    \note{\emph{Notes:} Panel (a) shows cross-country distributions of welfare gains from tariff and geopolitical changes as density-normalized step histograms, with dotted lines marking each mean; the legend reports cross-country mean (percentage points) and variance (squared percentage points; standard deviations cited in the text are the square roots). Panel (b) plots country-level welfare gains from tariff changes (horizontal axis) against gains from geopolitical changes (vertical axis), with selected countries labeled; dashed lines at zero separate the four quadrants, and the shaded green and red regions highlight the ``gain--gain'' and ``lose--lose'' quadrants, respectively.}
\end{figure}

Figure~\ref{fig:welfare_scatter} reports country-level welfare effects in 2021. The welfare effect of factor $k$ is defined as
\[
\left(\frac{\hat{\omega}_d^{\text{Baseline}}}{\hat{\omega}_d^{\text{No-}k}}-1\right)\times 100,
\]
so that a positive value indicates that the factor raised welfare.

Tariff changes generate modest average welfare gains (mean +0.24 percent, standard deviation 0.61 percent) with 50 of 78 countries gaining. The gains are smaller than wage-based measures would imply because lower tariffs also reduce tariff revenue; the largest gains accrue to economies with high initial tariff exposure, such as Angola, Singapore, and the Democratic Republic of the Congo.

Geopolitical realignment generates a much wider cross-country distribution of welfare effects. The mean is $-0.37$ percent (cross-country standard deviation 1.88 percent), and the range extends from $-8.56$ percent to $+7.09$ percent; 48 of 78 countries experience welfare losses.

Panel (a) overlays the tariff and geopolitical welfare distributions; Panel (b) plots country-level geopolitical welfare gains against tariff welfare gains, with shaded regions highlighting the ``gain--gain'' and ``lose--lose'' quadrants. Hungary records the largest welfare loss from geopolitical change, reflecting the widening divergence of its bilateral alignment from the rest of the European Union; Belarus loses 6.60 percent, reflecting the consequences of its alignment with Russia. At the other extreme, S\~ao Tom\'e and Pr\'incipe gains 7.09 percent and the United Arab Emirates gains 2.68 percent. Approximately 38 percent of countries gain from tariff liberalization but lose from geopolitical change, while another 26 percent gain from both forces.

Overall, tariff liberalization generates small and broadly shared gains, partly offset by lost tariff revenue, whereas geopolitical change produces more dispersed welfare effects. The cross-sectional standard deviation of welfare effects is roughly three times larger for geopolitics than for tariffs. In this Armington accounting, geopolitical change therefore produces more heterogeneous welfare consequences than tariff policy, even though its contribution to the cross-sectional variance of bilateral trade-cost changes is modest.\footnote{The single-sector Armington framework is intentionally parsimonious. It abstracts from intermediate inputs, sectoral heterogeneity, and dynamic adjustment. These omissions could affect magnitudes: intermediate inputs would likely amplify trade-cost effects \citep{Caliendo2015-xv}, sectoral structure could redistribute effects across industries, and dynamic adjustment could alter the timing of responses. We therefore interpret the exercise as a transparent aggregation of the estimated trade elasticity, rather than as a full structural quantification of all channels.}

\section{Conclusion} \label{s:conclusion}

Drawing on an event-based measure covering 833,485 political events across 193 countries from 1950 to 2024, this paper shows that bilateral geopolitical alignment is a quantitatively important determinant of trade. A one-standard-deviation permanent improvement in alignment raises bilateral trade by roughly 22 percent in the long run. In an Armington framework, geopolitical deterioration has offset roughly 70 percent of the trade expansion generated by tariff liberalization since 1995, with non-trade events accounting for about 80 percent of the geopolitical contribution. The associated welfare effects are more dispersed than those of tariffs and leave most countries worse off.
 
These findings are subject to important caveats. Our identification strategy exploits within-dyad variation and therefore speaks to the effects of changes in geopolitical alignment, not the level effects of sustained cooperation or isolation. The event-based measure, while more comprehensive than alternatives based on UN voting or binary indicators, relies on LLM-compiled data that may introduce measurement error. The absence of pre-trends, robustness to observable bilateral controls, event-window concordance around leadership transitions, and the reverse-direction diagnostic support the identifying interpretation; residual bilateral confounders that the two-way fixed-effects design absorbs imperfectly remain the main threat.
 
The central implication is that the post-2007 slowdown in globalization cannot be understood from tariff policy alone: bilateral political realignment has become a first-order trade friction in the accounting exercise. An important avenue for future research is to study the microeconomic channels through which geopolitical fragmentation affects trade, including how firms adjust supply chains and governments deploy policy instruments in response to geopolitical shocks.


\bibliography{\bib}

\newpage
\appendix

\newpage
\section{Additional Empirical Results} \label{appendix_A}

\subsection{Trade Decomposition: Major Countries} \label{app:major_countries}

\begin{figure}[htbp]
    \centering
    \caption{Trade Flow Decomposition by Country Groups, 1962--2024}
    \includegraphics[width=\linewidth]{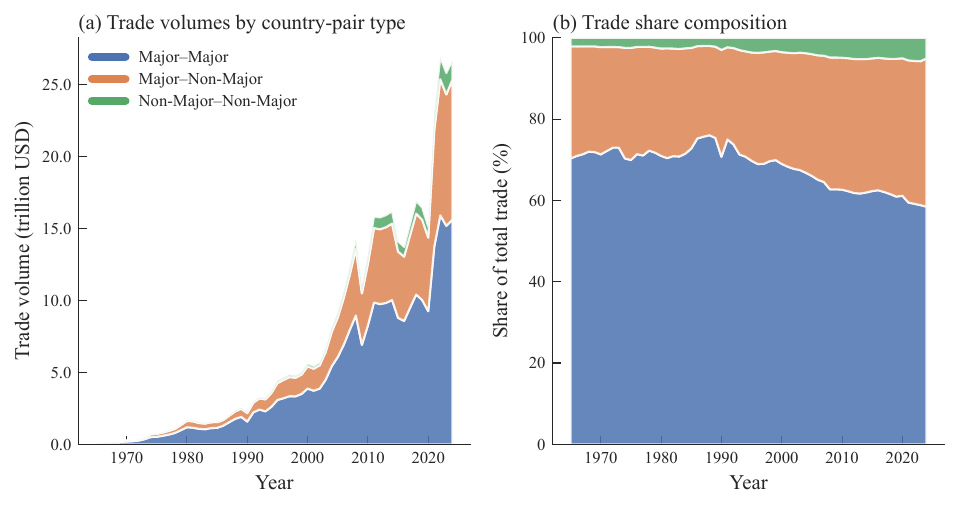}
    \label{fig:trade_major_decomp}
    \note{\emph{Notes:} Panel A shows absolute bilateral trade volumes by country-pair type (major--major, major--non-major, and non-major--non-major). Panel B displays the corresponding shares of total world trade.}
\end{figure}

This subsection documents the trade share of the 32 major economies (defined in Section~\ref{ss:measuring_alignment}). The restriction mitigates extensive-margin concerns in trade data \citep{Helpman2008-xj,head2014gravity} while capturing the bulk of global commerce. Figure~\ref{fig:trade_major_decomp} demonstrates that these economies dominate global trade patterns.

Figure~\ref{fig:trade_major_decomp} documents two patterns. First, Panel A shows that trade among major economies accounts for the majority of world trade throughout our sample period, despite these dyads representing fewer than 3\% of potential country pairs. The concentration reflects both size effects and deeper economic integration among the largest economies. Second, Panel B reveals that the major-economy share has declined over time---from approximately 71\% in the 1980s to 59\% in the 2020s---reflecting the integration of emerging markets into global value chains and the geographic dispersion of production networks.


\subsection{UNGA Voting Similarity and Trade Globalization}

\begin{figure}[htbp]
    \centering
    \caption{UNGA Voting Similarity and Trade Globalization, 1995--2021}
    \includegraphics[width=0.5\linewidth]{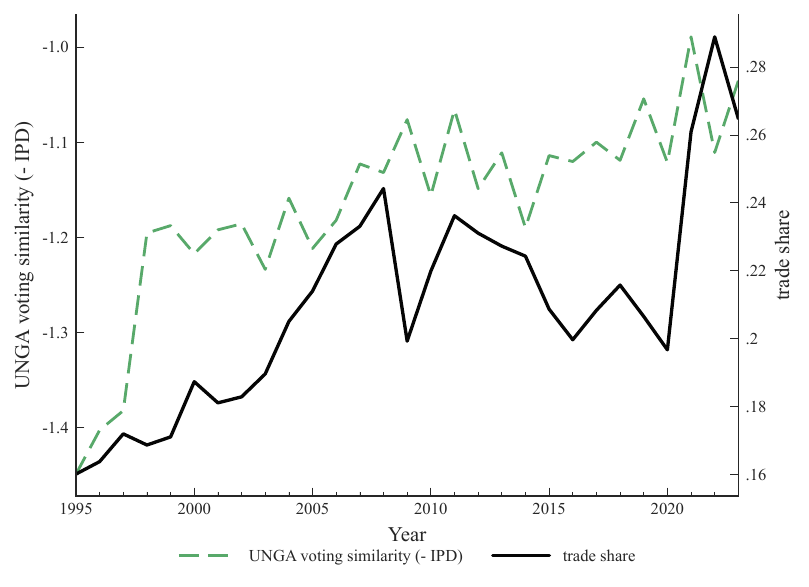}
    \label{fig:ipd_trade}
    \note{\emph{Notes:} The green dashed line plots the trade-weighted average UNGA voting similarity (negative Ideal Point Distance) across country pairs. The black solid line plots trade as a share of GDP.}
\end{figure}

Figure~\ref{fig:ipd_trade} illustrates a divergence between multilateral voting patterns and trade flows. From 1995 to 2007, UNGA voting similarity and global trade share moved broadly in tandem, both rising with deepening post–Cold War integration. After 2007, however, the two series diverge: UNGA voting similarity continues to increase through 2022, while trade as a share of GDP plateaus and subsequently declines. This pattern is consistent with the evidence in Section~\ref{ss:bilateral_multilateral} that UNGA voting primarily reflects multilateral positioning rather than the bilateral relationships most relevant for trade. The continued convergence in voting may instead capture the growing influence of developing-country coalitions in multilateral forums, a dynamic that is largely orthogonal to the bilateral geopolitical frictions that increasingly constrain trade flows.

\subsection{Additional Case Studies: U.S. Trading Partners}
\label{app:add_cases_usa}

We examine four distinct U.S. bilateral relationships—China, Brazil, Saudi Arabia, and Colombia—that illustrate how geopolitical alignment and trade co-move across different political contexts: strategic rivalry, ideological divergence, energy interdependence, and security cooperation. These cases complement the main text by demonstrating that the relationship between residualized geopolitical alignment and residualized bilateral trade persists across a wide range of U.S. partnerships, despite substantial heterogeneity in the underlying political drivers.

\begin{figure}[H]
    \centering
    \caption{Bilateral Dyads: Geopolitical Alignment and Trade}
    \begin{subfigure}[b]{0.49\textwidth}
        \centering
        \caption{U.S.\ Trading Partners}
        \includegraphics[width=\textwidth]{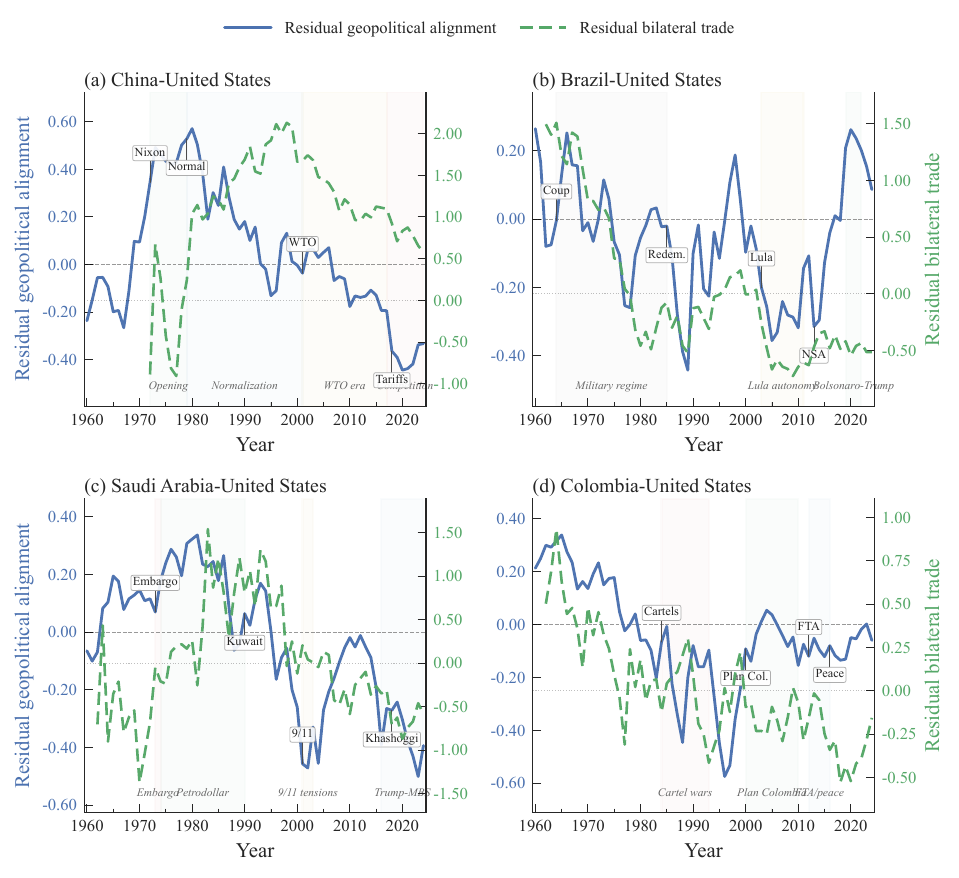}
        \label{fig:usa_dyads}
    \end{subfigure}
    \hfill
    \begin{subfigure}[b]{0.49\textwidth}
        \centering
        \caption{Asian Trading Networks}
        \includegraphics[width=\textwidth]{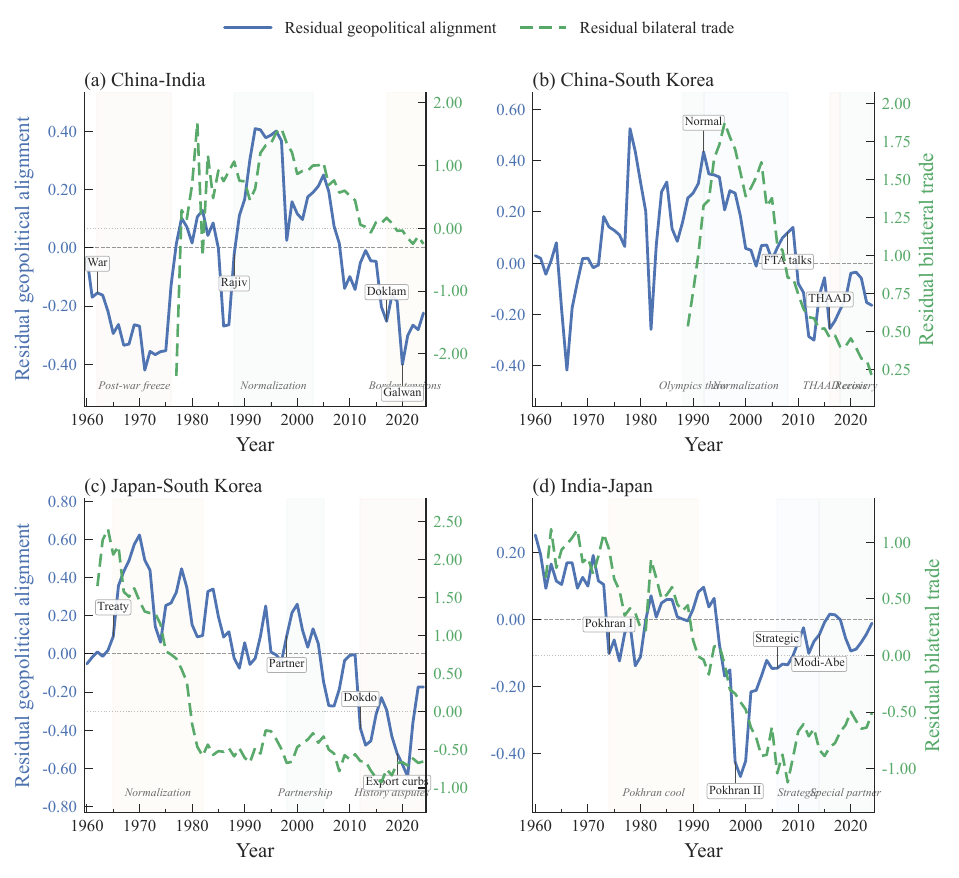}
        \label{fig:asia_dyads}
    \end{subfigure}
    \label{fig:case_study_dyads}
    \note{\emph{Notes:} Each panel plots residualized geopolitical alignment against residualized bilateral trade flows. All variables are residualized following the methodology in Figure~\ref{fig:geo_trade_cases}. In Panel A, the China–U.S. trade series begins in 1970, following the normalization of relations under the Nixon administration. In Panel B, the China–South Korea trade series begins in 1985, when indirect trade through Hong Kong became quantitatively significant, despite formal normalization occurring in 1992. Shaded regions indicate major diplomatic periods with distinct geopolitical characteristics.}
\end{figure}

\textbf{China-U.S.: Strategic rivalry and weakening economic ties.}
Figure~\ref{fig:usa_dyads}a highlights the evolution from engagement to strategic rivalry. The period following the Nixon opening and around WTO accession is characterized by strong trade growth alongside relatively favorable geopolitical alignment. By contrast, since the late 2000s, intensifying strategic competition is mirrored in a joint decline in both alignment and trade. Consistent with the notion of strategic rivalry, the figure suggests that deep commercial interdependence provides only limited insulation against deteriorating geopolitical relations.

\textbf{Brazil-U.S.: Ideological distance and attenuated trade integration.}
Figure~\ref{fig:usa_dyads}b illustrates how shifts in ideological distance shape economic linkages within the Western Hemisphere. In particular, the late 1970s and parts of the first Lula administration represent periods of heightened ideological divergence between Brazil and the United States. In both episodes, greater ideological distance is associated with weaker geopolitical alignment and reduced trade integration. By contrast, periods characterized by closer ideological alignment coincide with stronger bilateral trade flows.

\textbf{Saudi Arabia-U.S.: Energy interdependence and political volatility.}
Figure~\ref{fig:usa_dyads}c shows a relationship marked by substantial political volatility alongside enduring economic linkages. The 1973 embargo, post-9/11 tensions, and the Khashoggi episode coincide with declines in geopolitical alignment and are accompanied by noticeable reductions in bilateral trade, even if trade does not collapse entirely. Relative to other U.S.\ dyads, the Saudi case suggests that while strong complementarities in strategic commodities sustain the relationship, trade remains meaningfully affected by geopolitical strain rather than fully insulated from it.

\textbf{Colombia-U.S.: Security cooperation and economic integration.}
Figure~\ref{fig:usa_dyads}d presents a case where sustained security cooperation aligns with deepening economic integration. The strengthening of bilateral ties around Plan Colombia and the subsequent free trade agreement is associated with improvements in both geopolitical alignment and trade. In line with the notion of security cooperation, this case suggests that strategic partnerships can reinforce broader economic linkages.

Taken together, these cases reinforce that the relationship between geopolitics and trade is not confined to a single type of bilateral interaction. While the underlying mechanisms differ—ranging from rivalry and ideology to resource dependence and security alliances—each case exhibits a consistent pattern: stronger geopolitical alignment is associated with stronger bilateral trade.

\subsection{Additional Case Studies: Asian Trading Networks}
\label{app:add_cases_asia}

Intra-Asian trade accounts for a substantial share of global commerce, yet it unfolds within a political environment that differs markedly from many U.S.-centered relationships. We examine four dyads—China–India, China–South Korea, Japan–South Korea, and India–Japan—that illustrate three recurring features of Asian trade networks: persistent historical tensions, security considerations that often outweigh economic complementarities, and pronounced volatility in bilateral relations. Despite these regional characteristics, residualized geopolitical alignment and residualized bilateral trade display a systematic positive association across these Asian dyads.

\textbf{China-India: Territorial disputes and constrained integration.}
Figure~\ref{fig:asia_dyads}a illustrates how persistent geopolitical tensions shape economic interactions between the world’s two most populous economies. The 1962 war established a fragile political baseline, and subsequent episodes—such as the Doklam standoff in 2017 and the Galwan Valley clash in 2020—have reinforced this pattern. Consistent with these geopolitical frictions, residualized alignment remains persistently low, and residualized trade also weakens in the latter part of the sample. Overall, the figure underscores how enduring strategic rivalry continues to constrain the economic relationship between these neighboring markets.

\textbf{China-South Korea: Security tensions and economic exposure.}
Figure~\ref{fig:asia_dyads}b highlights how geopolitical tensions shape a highly integrated trade relationship. Following normalization in 1992, both geopolitical alignment and trade strengthened substantially. However, the relationship begins to deteriorate in the 2000s, when rising security tensions coincide with weaker bilateral trade residuals. The figure underscores how shifts in the geopolitical environment can translate into a deterioration of economic ties, even in deeply integrated markets.

\textbf{Japan-South Korea: Historical memory and repeated bilateral strain.}
Figure~\ref{fig:asia_dyads}c shows that even between two advanced U.S.\ allies, historical grievances continue to shape bilateral economic relations. Residualized trade declines alongside geopolitical alignment after 1970 and remains persistently low thereafter, with no clear recovery trend. Overall, the pattern suggests that unresolved historical disputes have had a lasting constraining effect on bilateral economic integration.

\textbf{India-Japan: Strategic convergence.}
Figure~\ref{fig:asia_dyads}d presents the case of Japan and India, where trade co-moves closely with geopolitical alignment. Following India’s nuclear tests, bilateral relations deteriorated, accompanied by a decline in trade. Since the mid-2000s, however, the gradual strengthening of ties—reflected in deeper security cooperation, more frequent high-level engagement, and agreements such as the Strategic and Global Partnership—coincides with improved trade performance. Overall, the pattern suggests that closer alignment in foreign policy priorities and security interests can support economic integration.

Taken together, these Asian cases show that geopolitical alignment and bilateral trade comove across a range of political contexts, from territorial disputes to security partnerships.

\subsection{Variance Decomposition of Gravity Determinants} \label{app:variance_decomp}

We decompose the variance of bilateral trade flows to quantify the relative importance of geopolitical alignment vis-à-vis traditional gravity determinants. Specifically, by excluding each determinant from equation \ref{eq:gravity}, we assess its marginal contribution to explaining cross-dyadic variation in trade. Table~\ref{tab:variance_decomp_app} reports the estimated coefficients alongside the reduction in $R^2$ associated with the exclusion of each variable. Two main findings emerge.

\begin{table}[H]
\centering
\caption{Gravity Equation Coefficients and Variance Decomposition}
\label{tab:variance_decomp_app}
\resizebox{\textwidth}{!}{
\begin{tabular}{lcccccccc}
\toprule
& \multicolumn{4}{c}{Regression Coefficients (S.E.)} & \multicolumn{4}{c}{$R^2$ Loss (\%)} \\
\cmidrule(lr){2-5} \cmidrule(lr){6-9}
Sample & Geo Alignment & Distance (log) & Border & Linguistic Dist. & Geo Alignment & Distance & Border & Ling. Dist. \\
\midrule
\multicolumn{9}{l}{\textit{Panel A: All Countries}} \\
Full Period (1962--2024) & 1.10 (0.027) & -1.45 (0.016) & 0.84 (0.079) & -1.70 (0.075) & 3.4 & 42.5 & 1.1 & 2.6 \\
Pre-1990 & 1.08 (0.043) & -1.29 (0.023) & 0.54 (0.104) & -0.88 (0.112) & 4.8 & 48.1 & 0.4 & 0.9 \\
Post-1990 & 1.08 (0.032) & -1.50 (0.017) & 1.02 (0.082) & -2.10 (0.076) & 2.5 & 40.0 & 1.1 & 3.2 \\
\addlinespace
\multicolumn{9}{l}{\textit{Panel B: Major Economies}} \\
Full Period & 0.46 (0.106) & -0.81 (0.052) & 0.23 (0.171) & -1.10 (0.330) & 2.2 & 49.7 & 0.6 & 2.8 \\
Pre-1990 & 0.78 (0.141) & -0.68 (0.069) & 0.10 (0.213) & -0.91 (0.422) & 8.6 & 44.8 & 0.0 & 2.6 \\
Post-1990 & 0.20 (0.139) & -0.92 (0.052) & 0.33 (0.172) & -1.24 (0.332) & 0.4 & 50.6 & 0.8 & 3.1 \\
\midrule
Full $R^2$ (Full Period) & \multicolumn{4}{c}{All: 0.268 \quad Major: 0.179} & \multicolumn{4}{c}{Observations: 1,087,543 / 58,948} \\
\bottomrule
\end{tabular}
}
\note{\emph{Notes:} Left panel: coefficients from gravity equation with origin-year, destination-year, and standard gravity controls. Standard errors clustered by dyad in parentheses. Right panel: percentage point reduction in $R^2$ when variable excluded.}
\end{table}

First, geopolitical alignment is quantitatively comparable to standard trade frictions such as language. In the full sample, it accounts for 3.4\% of the variation in bilateral trade, exceeding the contribution of linguistic distance (2.6\%) and more than tripling that of shared borders (1.1\%). Among major economies, geopolitical alignment explains 2.2\% of trade variation, a magnitude broadly similar to that of linguistic distance (2.8\%).

Second, the decomposition reveals substantial heterogeneity across samples, pointing to important structural shifts over time. Among major economies, the sensitivity of trade to geopolitical alignment declines markedly: the estimated coefficient falls from 0.78 in the pre-1990 period to 0.20 thereafter, while its explanatory power decreases from 8.6\% to 0.4\%. This attenuation is consistent with the view that global integration after the Cold War insulated economies from geopolitical fluctuations. However, recent episodes—such as sanctions on Russia and the decoupling between the United States and China—suggest that such insulation is incomplete, and that geopolitical constraints reassert themselves when political tensions intensify.


\subsection{Additional Cross-Sectional Estimates} \label{app:additional_static}

This section provides robustness checks for the cross-sectional gravity results in Section~\ref{ss:cross_country_gravity}.

\paragraph{Country-Pair Fixed Effects.}
Table~\ref{tab:static_odfe} adds country-pair fixed effects, so that identification relies on within-pair temporal variation rather than cross-sectional differences. The coefficient on geopolitical alignment remains positive and precisely estimated. For major economies, the estimate (0.595) is larger than the baseline without pair fixed effects, suggesting that time-invariant confounders, if anything, attenuate the cross-sectional relationship. The UNGA voting similarity coefficient switches sign and becomes positive in this specification. In the post-1995 sample, where tariff data are available, the results are reported in Columns 3 and 4 for major country pairs. The coefficients on $\ln(1+\tau)$ imply an elasticity of substitution of around 3, and the effect of geopolitical alignment remains largely unchanged after controlling for tariffs.

\begin{table}[H]
\small
\begin{center}
\caption{The Effect of Geopolitical Alignment on Trade: with Country Pair FEs}
\label{tab:static_odfe}
\scalebox{0.9}{
\begin{tabular}{lcccccccc}
\hline \hline
& (1) & (2) & (3) & (4) & (5) & (6) & (7) & (8)\\
Dependent Variable: & \multicolumn{8}{c}{log Trade Value} \\
\cmidrule(lr){2-5}\cmidrule(lr){6-9}
 & \multicolumn{4}{c}{Major Countries} & \multicolumn{4}{c}{All Countries} \\
\hline
Geopolitical Alignment & 0.595  & & 0.256 & 0.250  & 0.357 &  & 0.164 & 0.165\\
 & (0.077)  & & (0.085) & (0.084) & (0.017) & & (0.019) & (0.019) \\
UNGA Voting Similarity: & & 0.323 & & & & 0.170 & & \\
$\quad$ Negative Ideal Point Distance & & (0.045) & & & & (0.011) & & \\
$\ln$(1+Tariff Rate) & & & & -1.827 & & & & -1.899\\
& & & & (0.555) & & & & (0.079)\\
Mean Dep. Var. & 12.52 & 12.72 & 14.00 & 14.00 & 7.23  & 7.27 & 7.46 & 7.46\\
Observations   & 58,948 & 50,946 & 24,751 & 24,751 & 1,087,543 &  991,266 & 614,173 & 614,173\\
\hline
Origin $\times$ Year FE & Yes & Yes & Yes & Yes &Yes & Yes & Yes & Yes \\
Destination $\times$ Year FE & Yes & Yes & Yes & Yes & Yes & Yes & Yes & Yes\\
Origin $\times$ Destination FE & Yes & Yes & Yes & Yes & Yes & Yes & Yes & Yes\\
\hline \hline
\end{tabular}}
\end{center}
\note{\emph{Notes:} The unit of observation is an origin-destination country pair in a year. Columns 1--4 report results for country pairs among 32 major countries, while Columns 5--8 include all country pairs. Standard errors are clustered at the country pair level.}
\end{table}

\paragraph{PPML and Inverse Hyperbolic Sine.}
Table~\ref{tab:static_ppml} addresses concerns about zero trade flows. PPML estimates (columns 1--4) confirm a positive effect of geopolitical alignment on trade, though the coefficient for major economies is imprecisely estimated. The inverse hyperbolic sine specification (columns 5--8) yields qualitatively similar results, with larger coefficients in the full sample. In both specifications, UNGA voting similarity enters with the wrong sign, consistent with the main text finding that it captures multilateral positioning rather than bilateral relations.

\begin{table}[H]
\centering
\caption{The Effect of Geopolitical Alignment on Trade: PPML and Inverse Hyperbolic Sine}
\label{tab:static_ppml}
\resizebox{\textwidth}{!}{
\begin{tabular}{lcccccccc}
\hline \hline
& \multicolumn{4}{c}{PPML (Dep.\ Var.: Trade Value)} & \multicolumn{4}{c}{IHS (Dep.\ Var.: $\text{asinh}$(Trade Value))} \\
\cmidrule(lr){2-5}\cmidrule(lr){6-9}
& \multicolumn{2}{c}{Major Countries} & \multicolumn{2}{c}{All Countries} & \multicolumn{2}{c}{Major Countries} & \multicolumn{2}{c}{All Countries} \\
\cmidrule(lr){2-3}\cmidrule(lr){4-5}\cmidrule(lr){6-7}\cmidrule(lr){8-9}
& (1) & (2) & (3) & (4) & (5) & (6) & (7) & (8) \\
\hline
Geopolitical Alignment & 0.161 & & 0.236 & & 0.499 & & 1.490 & \\
 & (0.092) & & (0.076) & & (0.144) & & (0.030) & \\
UNGA Voting Similarity: & & -0.020 & & -0.029 & & -0.423 & & -0.163 \\
$\quad$ Negative Ideal Point Distance & & (0.041) & & (0.028) & & (0.056) & & (0.015) \\
Mean Dep. Var. & 43.1$^{a}$ & 46.7$^{a}$ & 1.82$^{a}$ & 2.41$^{a}$ & 12.5 & 12.8 & 3.71 & 4.87 \\
Observations & 62,496 & 53,554 & 2,334,528 & 1,599,544 & 62,496 & 53,554 & 2,334,528 & 1,599,544 \\
\hline
Origin $\times$ Year FE & Yes & Yes & Yes & Yes & Yes & Yes & Yes & Yes \\
Destination $\times$ Year FE & Yes & Yes & Yes & Yes & Yes & Yes & Yes & Yes \\
\hline \hline
\end{tabular}}
\note{\emph{Notes:} The unit of observation is an origin-destination country pair in a year. Odd columns use our geopolitical alignment measure; even columns use UNGA voting similarity. Columns 1--2 and 5--6 report results for country pairs among 32 major countries; columns 3--4 and 7--8 include all country pairs. Columns 1--4 report PPML estimates; columns 5--8 use the inverse hyperbolic sine transformation. All specifications control for geographic distance, contiguity, and linguistic distance (coefficients not reported). $^{a}$Mean dependent variable $\times 10^5$. Standard errors are clustered at the country pair level.}
\end{table}

\paragraph{Alternative Trade Data Sources.}
Table~\ref{tab:static_baci_imf} reports the estimates using BACI (1995--2023) and IMF Direction of Trade Statistics (1948--2024) as alternative data sources. The positive effect of geopolitical alignment is robust across both datasets, with coefficients of similar magnitude to the baseline UN Comtrade estimates. The IMF data, which extend back to 1948, show that the relationship holds over a longer time horizon.

\begin{table}[H]
\centering
\caption{The Effect of Geopolitical Alignment on Trade: Alternative Trade Measures}
\label{tab:static_baci_imf}
\resizebox{\textwidth}{!}{
\begin{tabular}{lcccccccccccc}
\hline \hline
\multirow{3}{*}{Dependent Variable: log Trade Value} & \multicolumn{6}{c}{BACI (1995--2023)} & \multicolumn{6}{c}{IMF (1948--2024)} \\
\cmidrule(lr){2-7}\cmidrule(lr){8-13}
& \multicolumn{3}{c}{Major Countries} & \multicolumn{3}{c}{All Countries} & \multicolumn{3}{c}{Major Countries} & \multicolumn{3}{c}{All Countries} \\
\cmidrule(lr){2-4}\cmidrule(lr){5-7}\cmidrule(lr){8-10}\cmidrule(lr){11-13}
& (1) & (2) & (3) & (4) & (5) & (6) & (7) & (8) & (9) & (10) & (11) & (12) \\
\hline
Geopolitical Alignment & 0.276 & 0.177 & & 2.806 & 1.114 & & 0.660 & 0.446 & & 2.483 & 1.022 & \\
 & (0.172) & (0.149) & & (0.050) & (0.034) & & (0.106) & (0.088) & & (0.042) & (0.027) & \\
UNGA Voting Similarity: & & & -0.210 & & & -0.170 & & & -0.219 & & & -0.018 \\
$\quad$ Negative Ideal Point Distance& & & (0.053) & & & (0.014) & & & (0.041) & & & (0.012) \\
Mean Dep. Var. & 13.94 & 13.94 & 14.03 & 7.57 & 7.57 & 7.60 & 12.26 & 12.26 & 12.49 & 7.45 & 7.45 & 7.48 \\
Observations & 27,451 & 27,451 & 26,601 & 668,849 & 668,849 & 632,568 & 64,691 & 64,691 & 54,701 & 1,078,076 & 1,078,076 & 947,751 \\
\hline
Origin $\times$ Year FE & Yes & Yes & Yes & Yes & Yes & Yes & Yes & Yes & Yes & Yes & Yes & Yes \\
Destination $\times$ Year FE & Yes & Yes & Yes & Yes & Yes & Yes & Yes & Yes & Yes & Yes & Yes & Yes \\
\hline \hline
\end{tabular}}
\note{\emph{Notes:} The unit of observation is an origin-destination country pair in a year. Columns 1–3 and 7--9 report results for country pairs among 32 major countries, while Columns 4–6 and 10--12 include all country pairs. All specifications control for geographic distance, contiguity, and linguistic distance (coefficients not reported). Standard errors are clustered at the country pair level.}
\end{table}

\paragraph{Alternative Geopolitical Measures.}
Table~\ref{tab:static_other_geo} shows that the results are robust to alternative constructions of the geopolitical alignment measure. The coefficient on geopolitical alignment is positive and significant across all specifications: raw scores without smoothing, alternative depreciation rates ($\lambda = 0.1$ and $\lambda = 0.5$), four-period moving averages, and cumulative sum scores. The coefficients attenuate when gravity controls are added (even columns) but remain precisely estimated throughout.

\begin{table}[H]
\centering
\caption{The Effect of Geopolitical Alignment on Trade: Alternative Alignment Measures}
\label{tab:static_other_geo}
\resizebox{\textwidth}{!}{
\begin{tabular}{lcccccccccc}
\hline \hline
\multirow{3}{*}{Dependent Variable:} & \multicolumn{10}{c}{log Trade Value, Major Countries} \\
\cmidrule(lr){2-11}
& \multicolumn{2}{c}{Raw Score} & \multicolumn{2}{c}{$\lambda=0.1$} & \multicolumn{2}{c}{$\lambda=0.5$} & \multicolumn{2}{c}{MA(4)} & \multicolumn{2}{c}{Sum Score} \\
\cmidrule(lr){2-3}\cmidrule(lr){4-5}\cmidrule(lr){6-7}\cmidrule(lr){8-9}\cmidrule(lr){10-11}
& (1) & (2) & (3) & (4) & (5) & (6) & (7) & (8) & (9) & (10) \\
\hline
Geopolitical Alignment & 0.388 & 0.223 & 0.902 & 0.555 & 0.599 & 0.368 & 0.823 & 0.488 & 0.115 & 0.063 \\
 & (0.058) & (0.049) & (0.178) & (0.158) & (0.095) & (0.083) & (0.115) & (0.100) & (0.014) & (0.012) \\
Mean Dep. Var. & 12.52 & 12.52 & 12.52 & 12.52 & 12.52 & 12.52 & 12.52 & 12.52 &12.52 & 12.52 \\
Observations & 58948 & 58948 & 58948 & 58948 & 58948 & 58948 & 58948 & 58948 & 58948 & 58948 \\
\hline
Origin $\times$ Year FE & Yes & Yes & Yes & Yes & Yes & Yes & Yes & Yes & Yes & Yes \\
Destination $\times$ Year FE & Yes & Yes & Yes & Yes & Yes & Yes & Yes & Yes & Yes & Yes \\
\hline \hline
\end{tabular}}
\note{\emph{Notes:} The unit of observation is an origin-destination country pair in a year. All columns report results for country pairs among 32 major countries. Columns 1--2 use the raw geopolitical alignment score. Columns 3--4 use a smoothed geopolitical alignment score with $\lambda = 0.1$, while columns 5--6 use a smoothed score with $\lambda = 0.5$. Columns 7--8 report results based on a four-period moving average of the raw score. Columns 9--10 use the smoothed sum score with $\lambda = 0.3$. Even columns control for geographic distance, contiguity, and linguistic distance (coefficients not reported). Standard errors are clustered at the country pair level.}
\end{table}

\begin{figure}[H]
    \centering
    \caption{The Effect of Geopolitical Alignment on Trade: Gravity by Year}
    \begin{subfigure}[b]{0.49\textwidth}
        \centering
        \caption{Major-Major Country Pairs}
        \includegraphics[width=\textwidth]{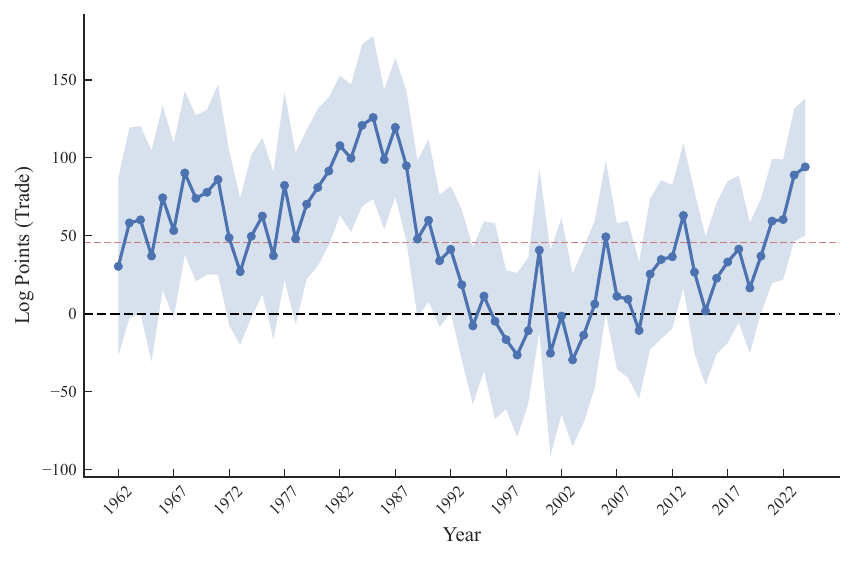}
    \end{subfigure}
    \hfill
    \begin{subfigure}[b]{0.49\textwidth}
        \centering
        \caption{All Country Pairs}
        \includegraphics[width=\textwidth]{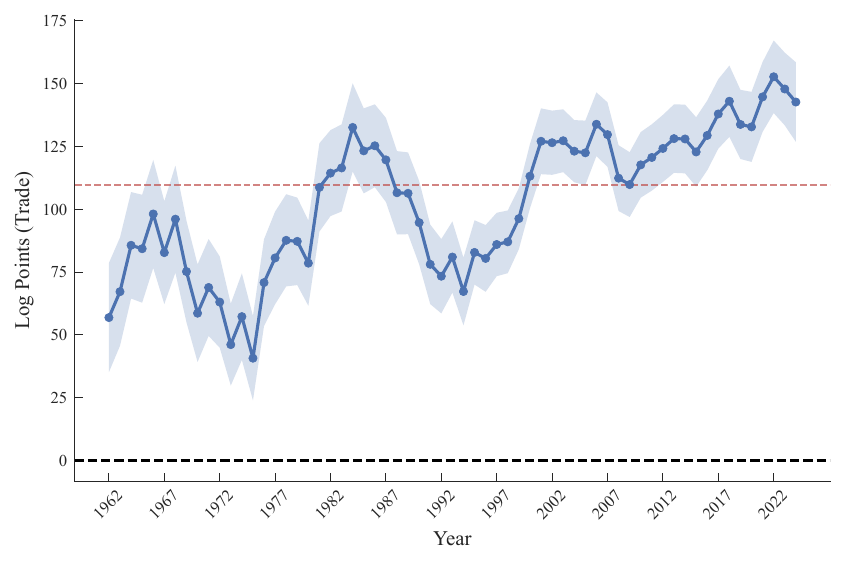}
    \end{subfigure}
    \label{fig:static_by_year_combined}
    \note{\emph{Notes:} This figure plots the estimated coefficients of geopolitical alignment on log trade values by year. Specifically, we estimate $\ln X_{odt} = \delta_{ot} +\delta_{dt}+ \sum_{\tau=1962}^{2024}\beta_\tau S_{odt}\times \mathbf{1}[t=\tau]+\text{Controls}+\epsilon_{odt}$, where controls include geographic distance, contiguity, and linguistic distance. The red dashed lines depict the corresponding estimates obtained from pooling all years. Panel A shows results for major-major country pairs (corresponding to column 2 in Table~\ref{tab:static}), while Panel B shows results for all countries (corresponding to column 6 in Table~\ref{tab:static}). Standard errors are clustered at the country-pair level.}
\end{figure}

\paragraph{Gravity Estimates by Year.}
Figure~\ref{fig:static_by_year_combined} plots the estimated coefficient of geopolitical alignment on trade by year. The coefficients are consistently positive and display a U-shaped pattern: larger during the Cold War, smaller during the globalization period, and rising again after 2018. This temporal variation suggests that while the geopolitics-trade relationship persists across international regimes, its magnitude responds to the prevailing level of geopolitical tension.

\subsection{Robustness of Dynamic Results} \label{app:add_dynamic}

This section provides robustness tests supporting the main dynamic results from Section~\ref{ss:dynamic_trade}.

\begin{figure}[H]
    \centering
    \caption{Dynamic Effect of Geopolitical Alignment on Trade: Alternative Lags}
    \includegraphics[width=0.48\textwidth]{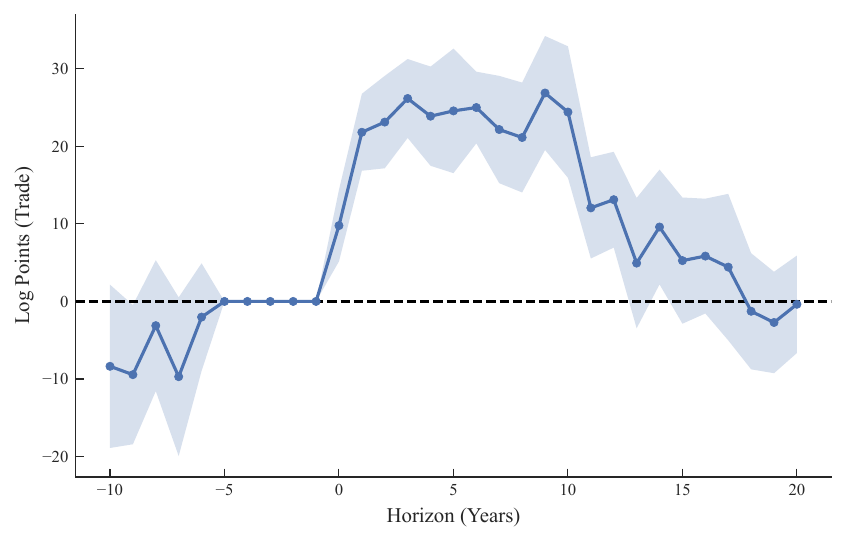}
    \label{fig:dynamic_lag}
    \note{\emph{Notes:} This figure reports estimates $\{\beta_h\}$ from $\ln X_{od,t+h} = \beta_h S_{od,t} + \sum_{\ell=1}^{5} \gamma_{h,\ell} \ln X_{od,t-\ell} + \sum_{\ell=1}^{5} \beta_{h,\ell} S_{od,t-\ell} + \delta_{od} + \delta_{ot} + \delta_{dt} + \varepsilon_{od,t+h}$. The sample includes country pairs among 32 major economies. Estimated coefficients are shown with 95\% confidence intervals based on Driscoll--Kraay standard errors.}
\end{figure}

\paragraph{Alternative Lag Structure.}
Figure~\ref{fig:dynamic_lag} assesses our baseline three-lag specification by extending to five lags. The trade impulse responses remain virtually unchanged, with identical peak timing and comparable magnitudes throughout the horizon. This stability indicates that three lags capture the relevant dynamics without overfitting.

\begin{figure}[H]
    \centering
    \caption{Dynamic Effect of Geopolitical Alignment on Trade: Alternative Inference}
    \begin{subfigure}[b]{0.48\textwidth}
        \caption{Clustered s.e.}
        \includegraphics[width=\textwidth]{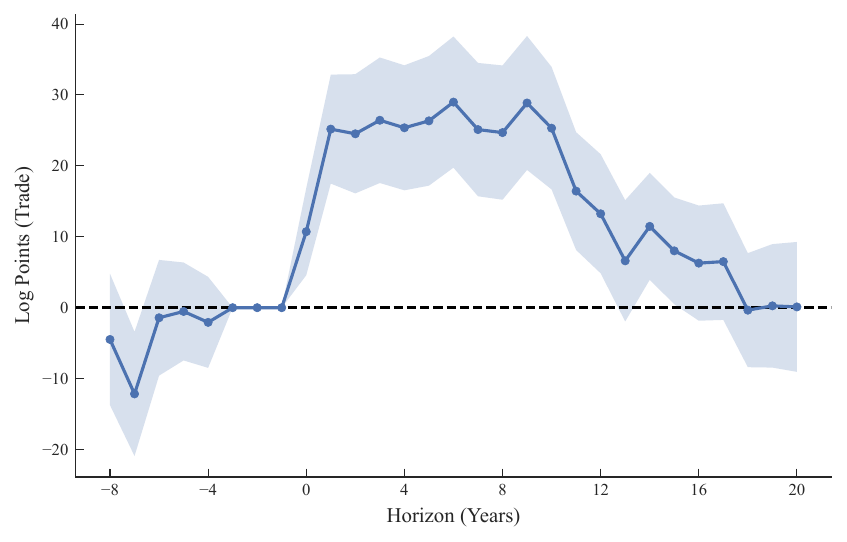}
    \end{subfigure}
    \hfill
    \begin{subfigure}[b]{0.48\textwidth}
        \caption{Bootstrap}
        \includegraphics[width=\textwidth]{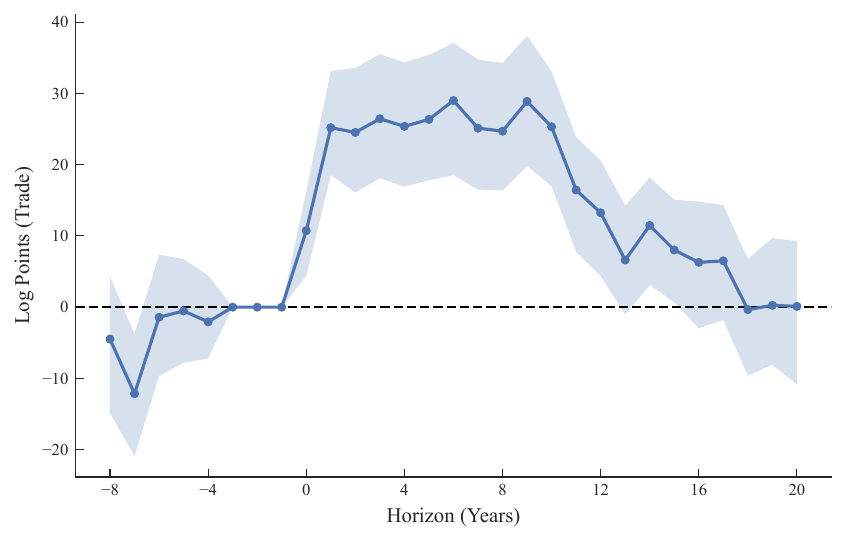}
    \end{subfigure}
    \label{fig:dynamic_inference}
    \note{\emph{Notes:} This figure reports estimates $\{\beta_h\}$ from $\ln X_{od,t+h} = \beta_h S_{od,t} + \sum_{\ell=1}^{3} \gamma_{h,\ell} \ln X_{od,t-\ell} + \sum_{\ell=1}^{3} \beta_{h,\ell} S_{od,t-\ell} + \delta_{od} + \delta_{ot} + \delta_{dt} + \varepsilon_{od,t+h}$. The sample includes country pairs among 32 major economies. Panel A reports estimated coefficients and 95\% confidence intervals with standard errors clustered at the country-pair level. Panel B reports estimated coefficients and 95\% confidence intervals from 200 bootstrap iterations with country-pair block resampling.}
\end{figure}

\paragraph{Alternative Inference.}
Figure~\ref{fig:dynamic_inference} examines robustness to alternative standard error constructions. Panel A uses standard errors clustered at the country-pair level. Panel B implements block bootstrap resampling (200 iterations). Both methods yield confidence intervals closely matching our baseline Driscoll--Kraay approach, with bootstrap intervals widening marginally at longer horizons.

\begin{figure}[H]
    \centering
    \caption{Dynamic Effect of Geopolitical Alignment on Trade: IHS}
    \begin{subfigure}[b]{0.48\textwidth}
        \caption{Major Countries}
        \includegraphics[width=\textwidth]{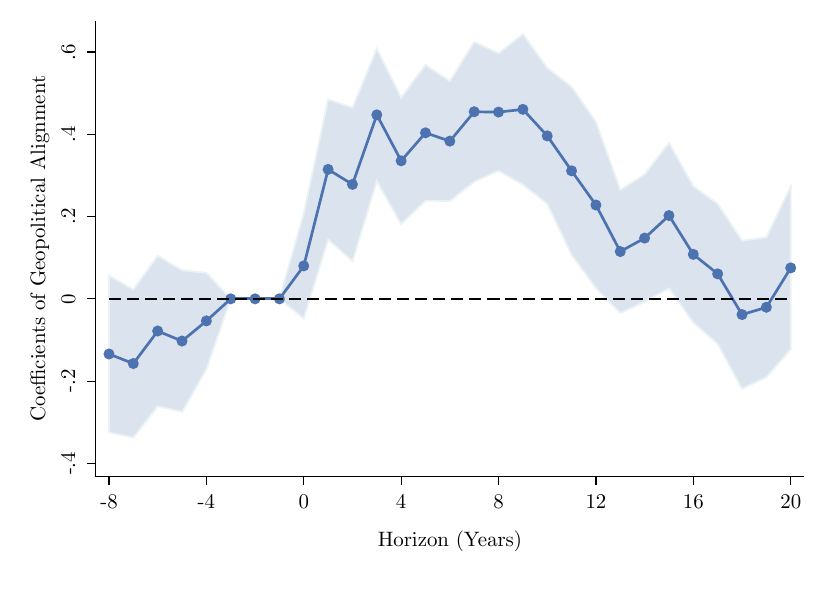}
    \end{subfigure}
    \hfill
    \begin{subfigure}[b]{0.48\textwidth}
        \caption{All Countries}
        \includegraphics[width=\textwidth]{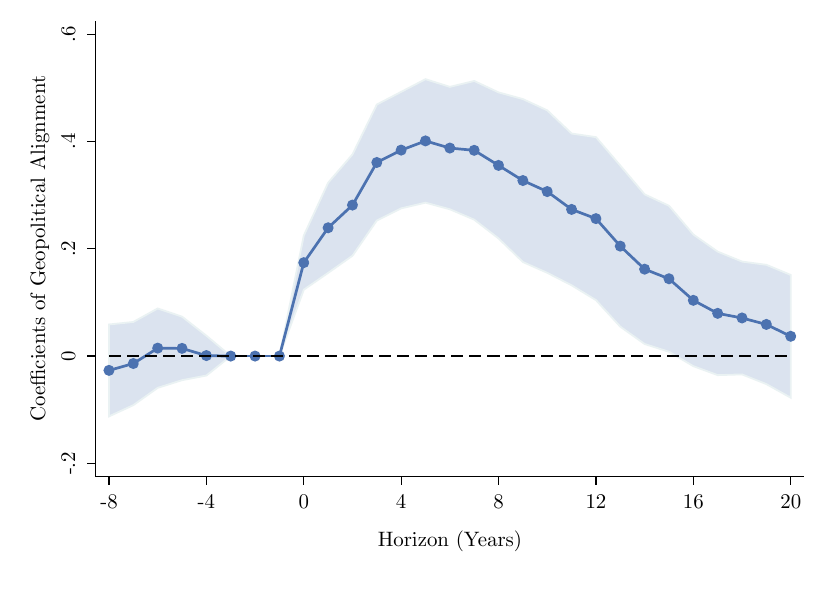}
    \end{subfigure}
    \label{fig:dynamic_ihs}
    \note{\emph{Notes:} This figure reports estimates $\{\beta_h\}$ from $\text{sinh}^{-1}( X_{od,t+h}) = \beta_h S_{od,t} + \sum_{\ell=1}^{3} \gamma_{h,\ell} \text{sinh}^{-1}( X_{od,t-\ell}) + \sum_{\ell=1}^{3} \beta_{h,\ell} S_{od,t-\ell} + \delta_{od} + \delta_{ot} + \delta_{dt} + \varepsilon_{od,t+h}$. The sample includes country pairs among 32 major economies in panel A and all country pairs in panel B. Shaded bands are 95\% confidence intervals based on Driscoll--Kraay standard errors.}
\end{figure}

\paragraph{Inverse Hyperbolic Sine Specification.}
Figure~\ref{fig:dynamic_ihs} examines robustness using the inverse hyperbolic sine of trade as the dependent variable, following \cite{boehm2023long}. Panel A reports results for 32 major country pairs, while Panel B includes the full sample of country pairs. The estimates are similar to the baseline. (Recall that for large trade values, the inverse hyperbolic sine transformation approximates twice the natural logarithm.)

\begin{figure}[H]
    \centering
    \caption{Dynamic Effect of Geopolitical Alignment on Trade: Time-Varying Bilateral Controls}
    \begin{subfigure}[b]{0.48\textwidth}
        \caption{Aid, sanctions, migration}
        \includegraphics[width=\textwidth]{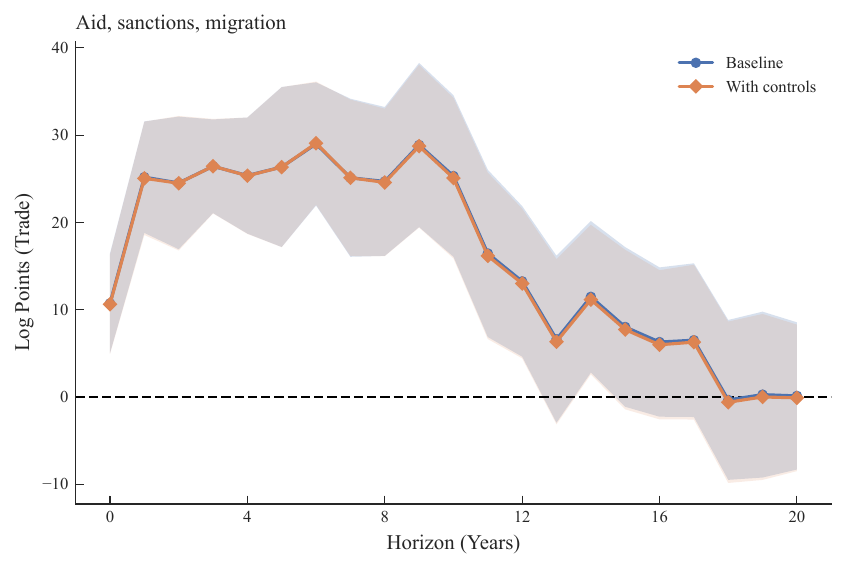}
    \end{subfigure}
    \hfill
    \begin{subfigure}[b]{0.48\textwidth}
        \caption{+ FDI, risk, tariffs}
        \includegraphics[width=\textwidth]{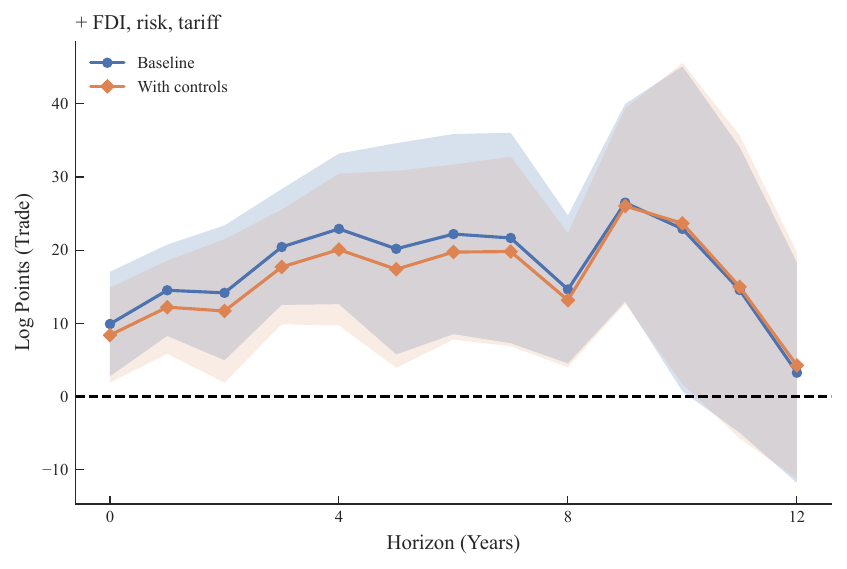}
    \end{subfigure}
    \label{fig:dynamic_controls}
    \note{\emph{Notes:} This figure reports estimates $\{\beta_h\}$ from augmented versions of the baseline local projection, $\ln X_{od,t+h} = \beta_h S_{od,t} + \sum_{\ell=1}^{3} \gamma_{h,\ell} \ln X_{od,t-\ell} + \sum_{\ell=1}^{3} \beta_{h,\ell} S_{od,t-\ell} + \sum_{\ell=1}^{3} \boldsymbol{\gamma}_{h,\ell}'\mathbf{W}_{od,t-\ell} + \delta_{od} + \delta_{ot} + \delta_{dt} + \varepsilon_{od,t+h}$, where $\mathbf{W}_{od,t}$ is a vector of time-varying bilateral controls. Panel A includes gross bilateral aid, trade and financial sanctions, and migration (horizon $h=-8,\ldots,20$). Panel B additionally includes bilateral FDI, country risk, and applied tariffs (horizon $h=-5,\ldots,12$, reflecting the shorter availability of these controls). In each panel, the blue series reports the baseline specification without controls, and the red series reports the augmented specification; both are estimated on the same sample. The sample includes country pairs among 32 major economies. Shaded bands are 95\% confidence intervals based on Driscoll--Kraay standard errors.}
\end{figure}

\paragraph{Time-Varying Bilateral Controls.}
A residual concern with the two-way fixed-effects design is that time-varying bilateral shocks correlated with alignment could drive the trade response. Figure~\ref{fig:dynamic_controls} addresses this concern by augmenting the baseline specification with three lags of observable bilateral variables. Panel A adds gross bilateral aid, sanctions, and migration, preserving the full horizon. Panel B additionally adds bilateral FDI, country risk \citep{Hassan2024-cr}, and applied tariffs; the horizon shortens to $h = -5$ through $h = 12$ because these controls are available only after 2001. In both panels, the impulse response with controls tracks the baseline closely, indicating that observable bilateral time-varying variables account for only a small share of the estimated trade response.

\begin{figure}[H]
    \centering
    \caption{Dynamic Effect of Geopolitical Alignment on Trade: Raw Alignment Scores}
    \begin{subfigure}[b]{0.48\textwidth}
        \caption{Autocorrelation of Geo}
        \includegraphics[width=\textwidth]{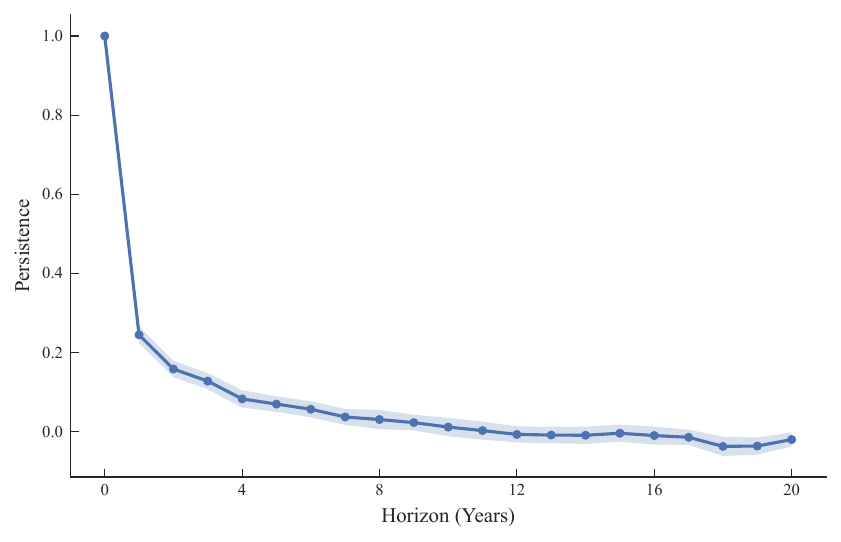}
    \end{subfigure}
    \hfill
    \begin{subfigure}[b]{0.48\textwidth}
        \caption{Local Projection of Trade on Geo}
        \includegraphics[width=\textwidth]{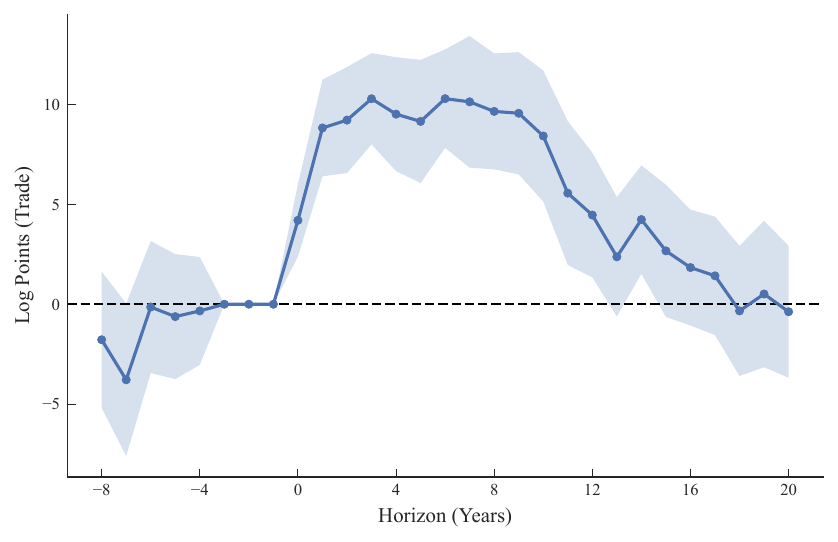}
    \end{subfigure}
    \begin{subfigure}[b]{0.48\textwidth}
        \caption{IRF of Transitory Shock}
        \includegraphics[width=\textwidth]{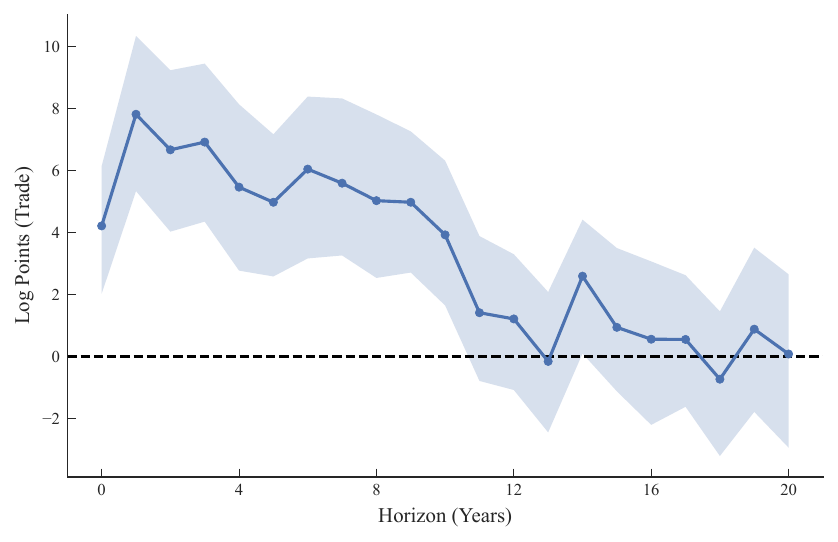}
    \end{subfigure}
    \hfill
    \begin{subfigure}[b]{0.48\textwidth}
        \caption{IRF of Permanent Shock}
        \includegraphics[width=\textwidth]{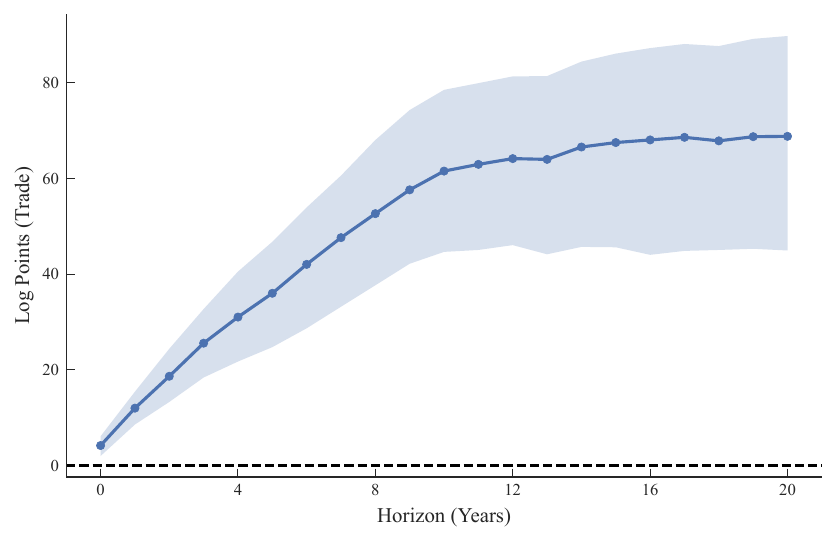}
    \end{subfigure}
    \label{fig:dynamic_nonsmooth}
    \note{\emph{Notes:} This figure uses the raw geopolitical alignment score. Panel A reports estimates $\{\phi_h\}$ from: $S_{od,t+h} = \phi_h S_{od,t}+ \sum_{\ell=1}^{3} \phi_{h,\ell} S_{od,t-\ell} +\delta_{ot} +\delta_{dt}+\delta_{od}+ \varepsilon_{od,t+h}$. Panel B shows estimates $\{\beta_h\}$ from $\ln X_{od,t+h} = \beta_h S_{od,t} + \sum_{\ell=1}^{3} \gamma_{h,\ell} \ln X_{od,t-\ell} + \sum_{\ell=1}^{3} \beta_{h,\ell} S_{od,t-\ell} + \delta_{od} + \delta_{ot} + \delta_{dt} + \varepsilon_{od,t+h}$. The sample includes country pairs among 32 major economies. Both panels report estimated coefficients with 95\% confidence intervals based on Driscoll--Kraay standard errors. Panel C presents the impulse response of log trade to a purely transitory unit shock in the raw geopolitical alignment score. Panel D shows the cumulative response to a permanent unit shock. 95\% confidence intervals are from 200 bootstrap iterations with country-pair block resampling.}
\end{figure}

\paragraph{Alternative Geopolitical Measures.}
Figure~\ref{fig:dynamic_nonsmooth} examines robustness using unsmoothed average event scores. These unsmoothed measures exhibit low persistence, with autocorrelation falling below 0.2 after two years (Panel A), yet they still generate economically meaningful trade responses, with peak elasticities of around 0.12 at horizons of 4–6 years (Panel B). Although the local projection produces somewhat smaller estimates, a one-unit permanent geopolitical shock implies a cumulative increase in trade of about 70 log points after 20 years (Panel D), close to the baseline estimate of 78 log points. This similarity indicates that permanent geopolitical realignments yield comparable long-run trade effects regardless of the measurement approach.

\begin{figure}[H]
    \centering
    \caption{Dynamic Effect of Geopolitical Alignment on Trade: Alternative Alignment Scores}
    \begin{subfigure}[b]{0.48\textwidth}
        \caption{Autocorrelation of Geo: $\lambda=0.1$}
        \includegraphics[width=\textwidth]{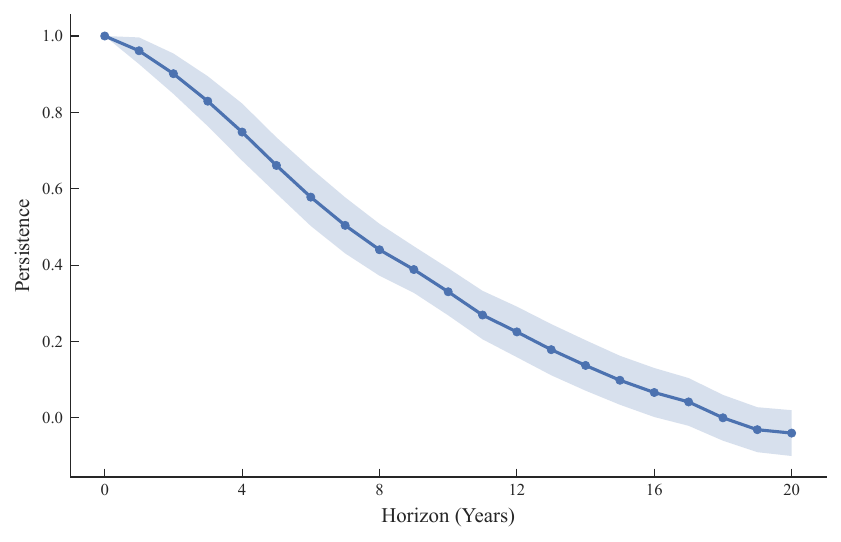}
    \end{subfigure}
    \hfill
    \begin{subfigure}[b]{0.48\textwidth}
        \caption{Local Projection of Trade on Geo: $\lambda=0.1$}
        \includegraphics[width=\textwidth]{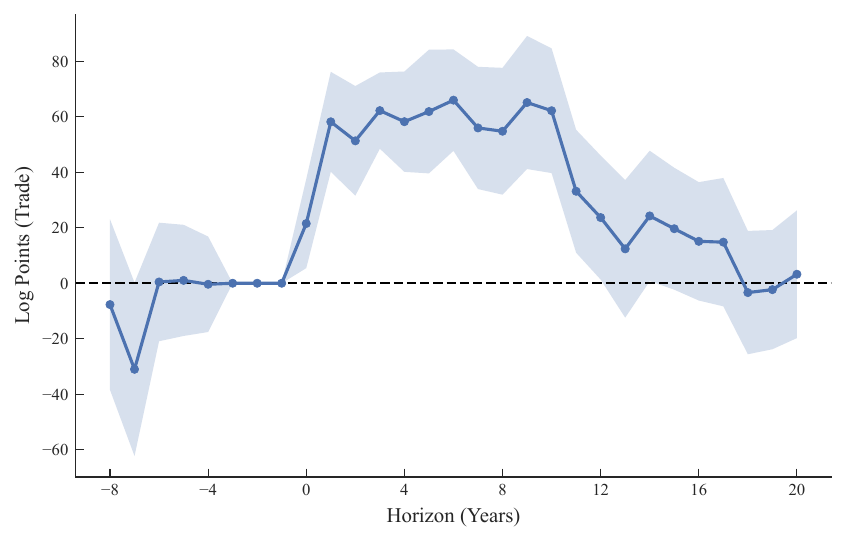}
    \end{subfigure}\\
    \begin{subfigure}[b]{0.48\textwidth}
        \caption{Autocorrelation of Geo: $\lambda=0.5$}
        \includegraphics[width=\textwidth]{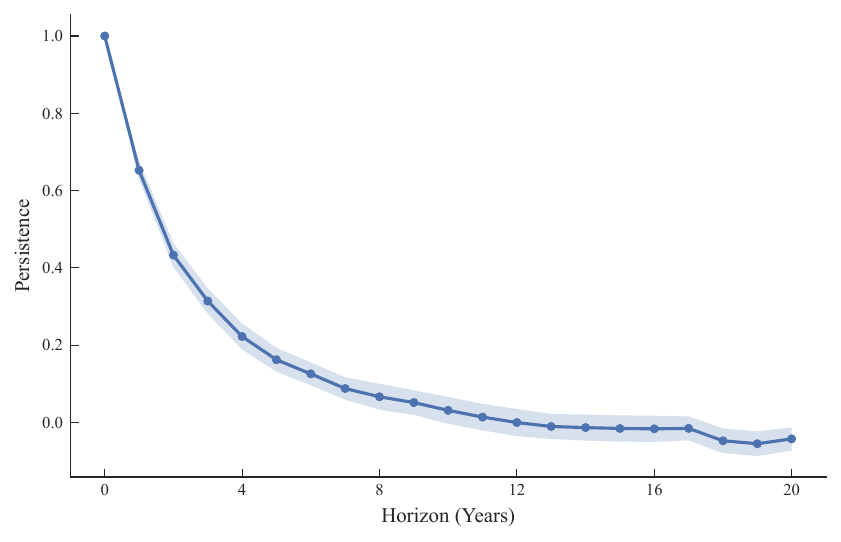}
    \end{subfigure}
    \hfill
    \begin{subfigure}[b]{0.48\textwidth}
        \caption{Local Projection of Trade on Geo: $\lambda=0.5$}
        \includegraphics[width=\textwidth]{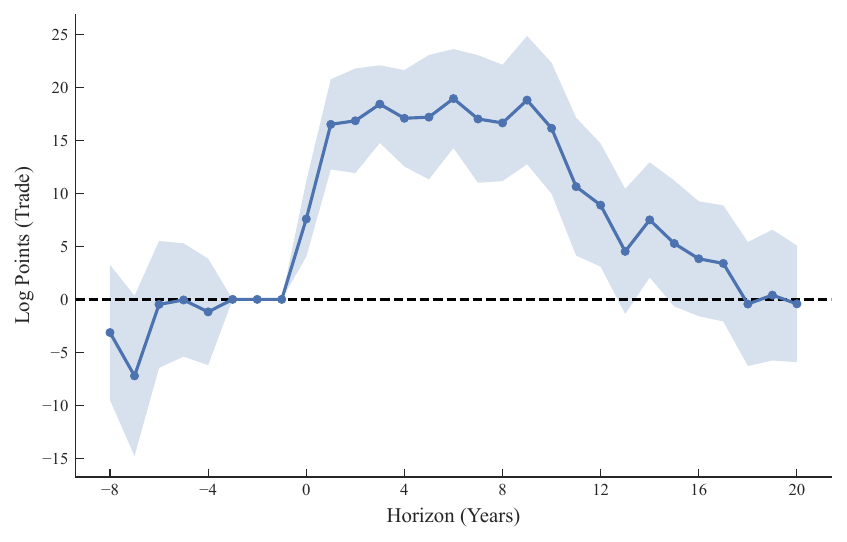}
    \end{subfigure}\\
    \begin{subfigure}[b]{0.48\textwidth}
        \caption{Autocorrelation of Geo: MA(4)}
        \includegraphics[width=\textwidth]{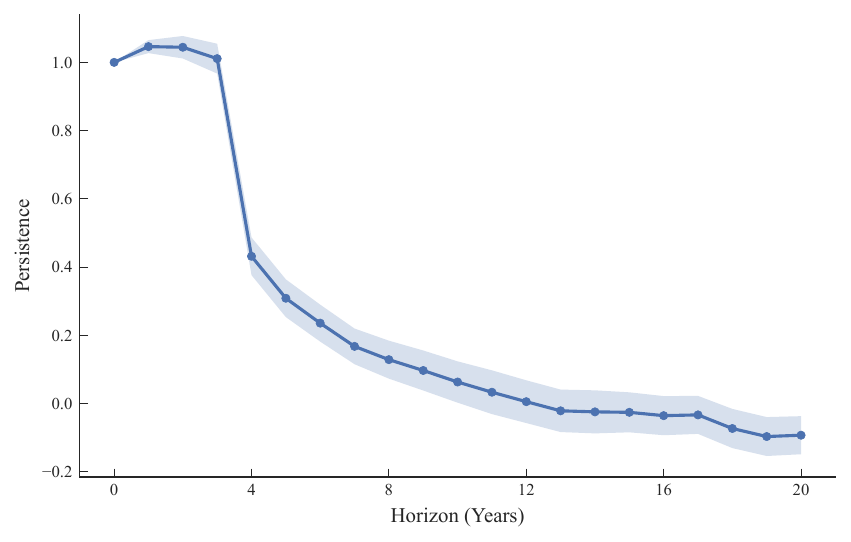}
    \end{subfigure}
    \hfill
    \begin{subfigure}[b]{0.48\textwidth}
        \caption{Local Projection of Trade on Geo: MA(4)}
        \includegraphics[width=\textwidth]{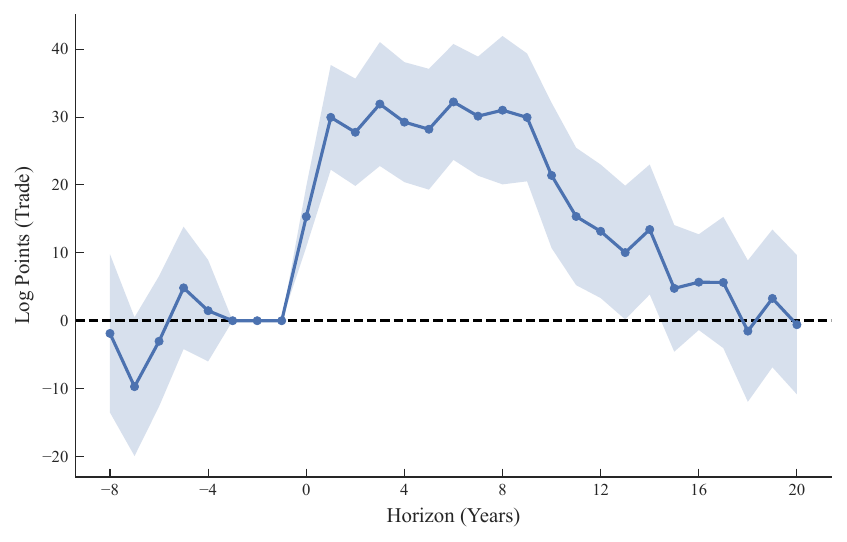}
    \end{subfigure}
    \label{fig:dynamic_robust_specification}
    \note{\emph{Notes:} Panel A, C, E report estimates $\{\phi_h\}$ from: $S_{od,t+h} = \phi_h S_{od,t}+ \sum_{\ell=1}^{3} \phi_{h,\ell} S_{od,t-\ell} +\delta_{ot} +\delta_{dt}+\delta_{od}+ \varepsilon_{od,t+h}$. Panel B, D, F show estimates $\{\beta_h\}$ from $\ln X_{od,t+h} = \beta_h S_{od,t} + \sum_{\ell=1}^{3} \gamma_{h,\ell} \ln X_{od,t-\ell} + \sum_{\ell=1}^{3} \beta_{h,\ell} S_{od,t-\ell} + \delta_{od} + \delta_{ot} + \delta_{dt} + \varepsilon_{od,t+h}$. The sample includes country pairs among 32 major economies. Panel A and B use a smoothed geopolitical alignment score with $\lambda=0.1$, while Panel C and D use a smoothed score with $\lambda=0.5$. Panel E and F report results based on a four-period moving average of the raw score. All panels report estimated coefficients with 95\% confidence intervals based on Driscoll--Kraay standard errors.}
\end{figure}

Figure~\ref{fig:dynamic_robust_specification} examines robustness to alternative depreciation rates ($\lambda = 0.1$, $\lambda = 0.5$) and a four-period moving average. Results remain qualitatively similar across specifications, showing that our findings are not sensitive to the particular smoothing parameter.

\subsection{Decomposition of Transitory and Permanent Shocks} \label{app:irf_transitory_persistent}

The local projection estimates in Section~\ref{ss:dynamic_trade} combine the initial shock's direct impact with effects from its subsequent persistence. We decompose these components by constructing counterfactual responses to purely transitory versus permanent geopolitical improvements.

\subsubsection{Methodology}

Following \citet{Sims1986-mt} and \citet{bilal2024macroeconomic}, we implement a two-step decomposition. First, we estimate the autocorrelation function of geopolitical alignment:
\begin{equation}
S_{od,t+h} = \phi_h S_{od,t} + \sum_{\ell=1}^{L} \phi_{h,\ell} S_{od,t-\ell} + \delta_{od} + \delta_{ot} + \delta_{dt} + \mu_{od,t+h}
\label{eq:geo_autocorr}
\end{equation}
where $\{\phi_h\}_{h=0}^{H}$ captures shock persistence.

To generate a purely transitory shock (one at $h=0$, zero thereafter), we solve for the auxiliary shock sequence $\mathbf{p} = \{p_0, p_1, \ldots, p_H\}$:
\begin{equation}
\mathbf{p} = (\boldsymbol{\Phi})^{-1} \mathbf{e}_1
\end{equation}
where $\boldsymbol{\Phi}$ is the lower triangular matrix with elements $\Phi_{ij} = \phi_{i-j}$ for $i \geq j$ (zero otherwise), and $\mathbf{e}_1 = (1, 0, \ldots, 0)'$ represents the targeted transitory shock pattern.

The trade response to this transitory geopolitical shock becomes:
\begin{equation}
\tilde{\beta}_h = \sum_{s=0}^{h} p_s \beta_{h-s}
\end{equation}
where $\{\beta_h\}$ are baseline local projection estimates from equation~\eqref{eq:lp_trade}. For permanent shocks, we compute cumulative responses: $\sum_{s=0}^{h} \tilde{\beta}_s$.

\subsubsection{Implementation and Inference}

We implement this decomposition using local projection estimates from our baseline specification with $L=3$. Statistical inference employs block bootstrap with 200 iterations: 1. Resample country pairs with replacement (block bootstrap); 2. Re-estimate both geopolitical autocorrelation \eqref{eq:geo_autocorr} and trade responses \eqref{eq:lp_trade}; 3. Compute decomposed impulse responses for each bootstrap sample; 4. Construct 95\% confidence intervals using the 2.5th and 97.5th percentiles. This procedure accounts for estimation uncertainty in both stages while preserving within-pair correlation structure.

\subsection{Temporal Variation in IPD-Based Estimates}

\begin{figure}[htbp]
    \centering
    \caption{Dynamic Effect of Negative IPD on Trade by Period}
    \begin{subfigure}[b]{0.48\textwidth}
        \caption{Local Projection of Trade on -IPD: 1965--1994}
        \includegraphics[width=\textwidth]{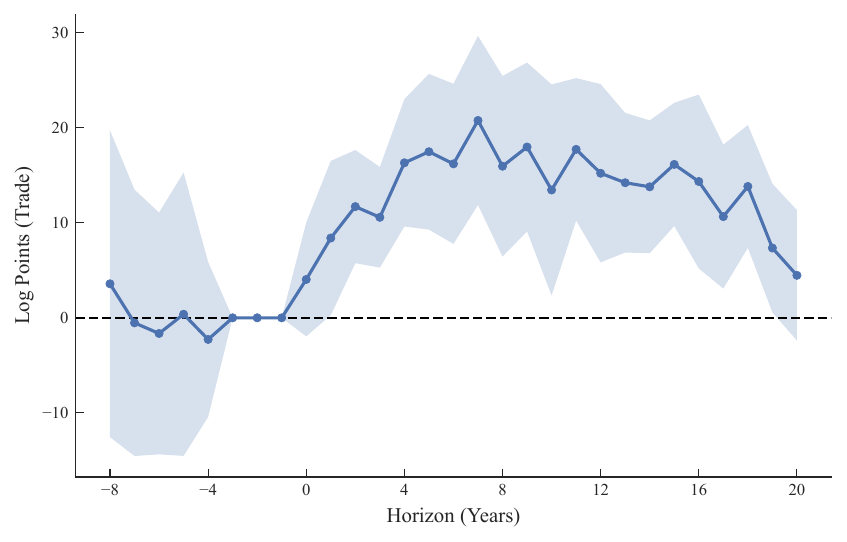}
    \end{subfigure}
    \hfill
    \begin{subfigure}[b]{0.48\textwidth}
        \caption{Local Projection of Trade on -IPD: 1995--2024}
        \includegraphics[width=\textwidth]{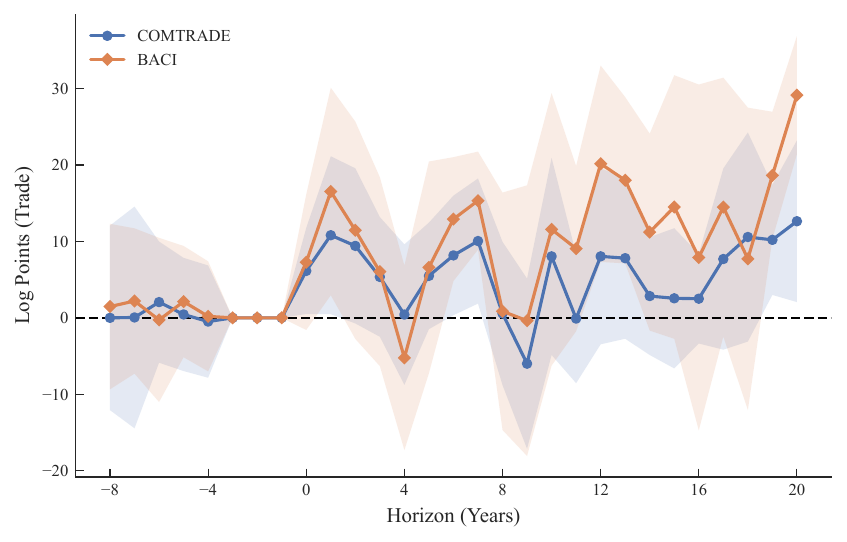}
    \end{subfigure}
    \label{fig:dynamic_time_ipd}
    \note{\emph{Notes:} This figure uses negative IPD as the measure of geopolitical alignment. Both panels report estimates $\{\beta_h\}$ from the regression $\ln X_{od,t+h} = \beta_h S_{od,t} + \sum_{\ell=1}^{3} \gamma_{h,\ell} \ln X_{od,t-\ell} + \sum_{\ell=1}^{3} \beta_{h,\ell} S_{od,t-\ell} + \delta_{od} + \delta_{ot} + \delta_{dt} + \varepsilon_{od,t+h}$. The sample includes country pairs among 32 major economies. Panel A focuses on 1965--1994; Panel B covers 1995--2024. Both panels report estimated coefficients with 95\% confidence intervals based on Driscoll--Kraay standard errors.}
\end{figure}

Figure~\ref{fig:dynamic_time_ipd} illustrates how the relationship between UN voting--based measures and trade varies across periods. We use negative IPD as a measure of geopolitical alignment, so that higher values correspond to closer alignment. In the earlier period, which largely coincides with the Cold War (Panel A), the local projection estimates display the expected relationship, with lower alignment (that is, more divergent UN voting) associated with lower trade. This pattern is consistent with the evidence in Section~\ref{ss:bilateral_multilateral} that UNGA voting primarily captures multilateral positioning, which during the Cold War was more closely linked to bilateral relations because bloc membership shaped both voting behavior and trade linkages. In the post--Cold War period (Panel B), the estimated trade responses are smaller and less precisely estimated, consistent with a weaker connection between multilateral voting patterns and bilateral economic relationships as trade networks became less dependent on geopolitical blocs.

The comparison with our event-based measure is useful in highlighting the different information contained in the two measures. While the association between IPD and trade varies across periods, our event-based measure yields positive trade elasticities in both subsamples (see figures in Section~\ref{ss:dynamic_trade}). This pattern is consistent with the broader finding in Section~\ref{ss:bilateral_multilateral} that bilateral and multilateral dimensions of geopolitical alignment are distinct and need not move together over time.

\FloatBarrier

\subsection{Heterogeneity: Alliances and Sectors} \label{app:het_sector_alliance}

\begin{figure}[H]
    \centering
    \caption{Dynamic Trade Responses: EU and NATO}
    \label{fig:dynamic_alliance}
    \begin{subfigure}[b]{0.48\textwidth}
        \caption{EU}
        \includegraphics[width=\textwidth]{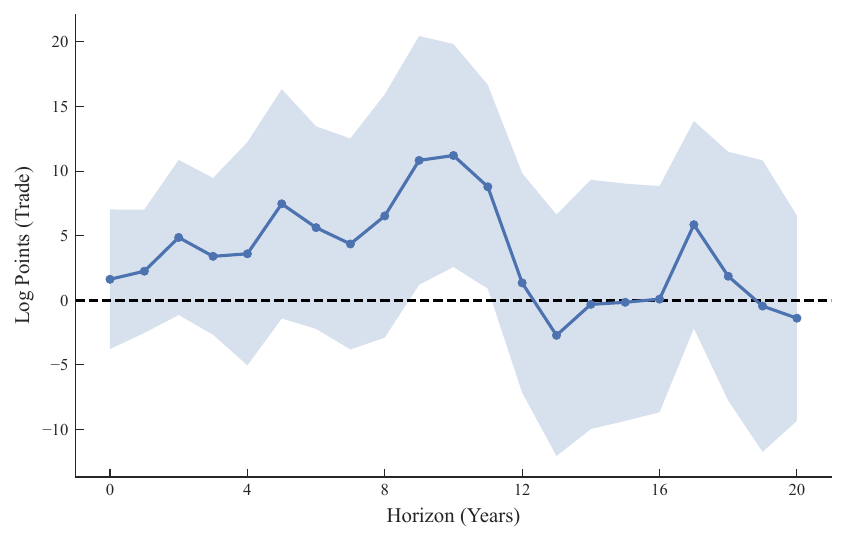}
    \end{subfigure}
    \hfill
    \begin{subfigure}[b]{0.48\textwidth}
        \caption{NATO}
        \includegraphics[width=\textwidth]{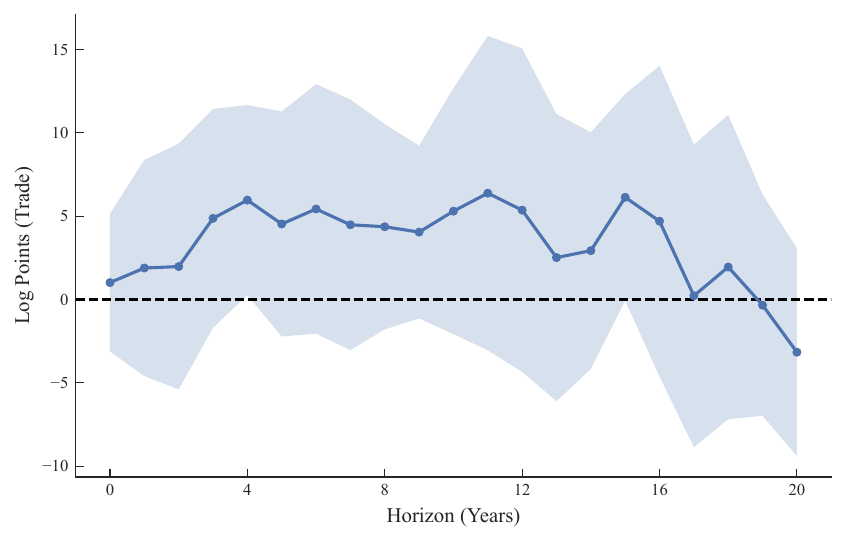}
    \end{subfigure}
    \note{\emph{Notes:} This figure reports estimates ${\beta_h}$ from the regression $\ln X_{od,t+h} = \beta_h S_{od,t} + \sum_{\ell=1}^{3} \gamma_{h,\ell} \ln X_{od,t-\ell} + \sum_{\ell=1}^{3} \beta_{h,\ell} S_{od,t-\ell} + \delta_{od} + \delta_{ot} + \delta_{dt} + \varepsilon_{od,t+h}$. The sample consists of country pairs among EU in panel A, and among NATO in panel B. All panels display coefficient estimates with 95\% confidence intervals based on Driscoll--Kraay standard errors.}
\end{figure}

Figure~\ref{fig:dynamic_alliance} examines heterogeneity by EU and NATO membership. Among country pairs in which both partners are EU members (Panel A) or both are NATO members (Panel B), the estimated responses are smaller in magnitude and statistically indistinguishable from zero, consistent with the view that formal institutional ties reduce the sensitivity of trade to bilateral geopolitical shocks.

\begin{figure}[H]
    \centering
    \caption{Dynamic Trade Responses: Sector Heterogeneity}
    \hfill
    \begin{subfigure}[b]{0.48\textwidth}
        \caption{NAICS 11}
        \includegraphics[width=\textwidth]{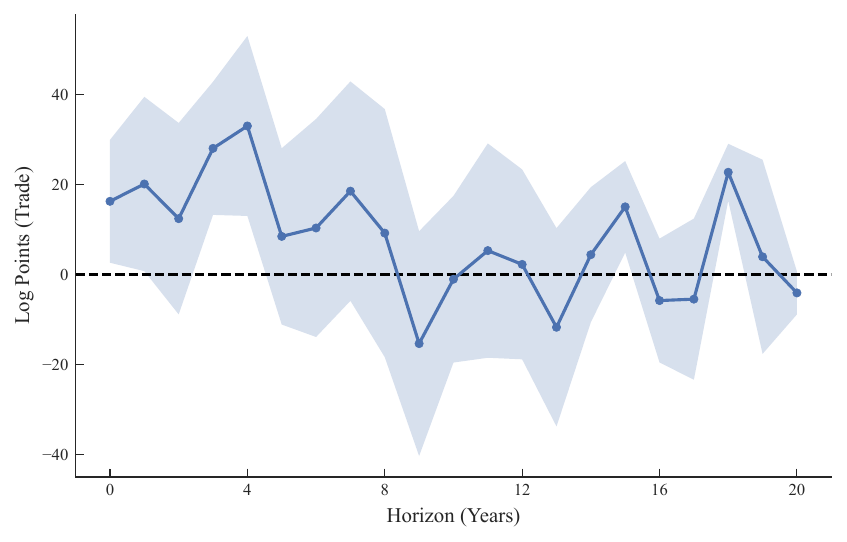}
    \end{subfigure}
    \hfill
    \begin{subfigure}[b]{0.49\textwidth}
        \caption{NAICS 21}
        \includegraphics[width=\textwidth]{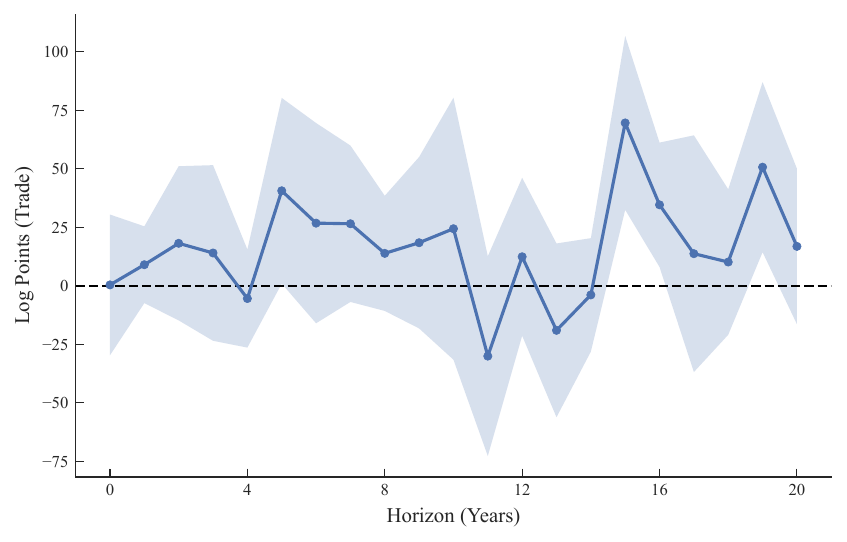}
    \end{subfigure}\\
    \begin{subfigure}[b]{0.49\textwidth}
        \caption{NAICS 31}
        \includegraphics[width=\textwidth]{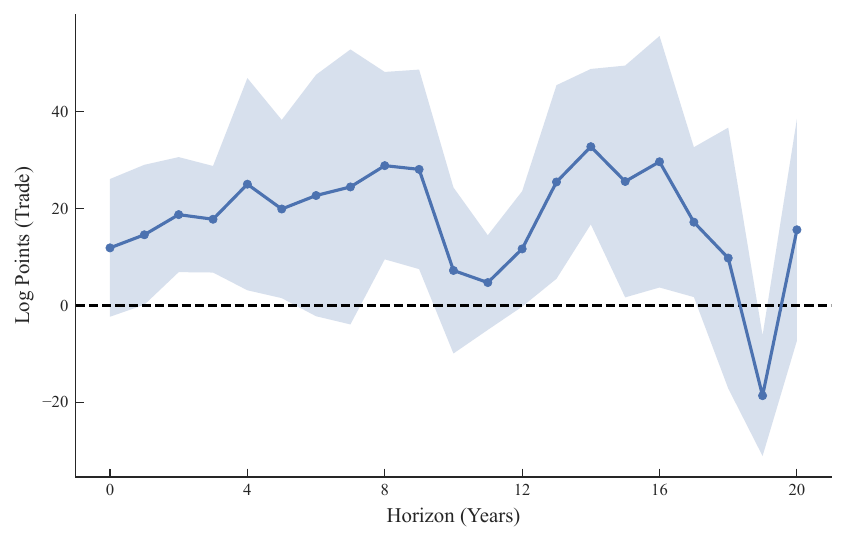}
    \end{subfigure}
    \hfill
    \begin{subfigure}[b]{0.49\textwidth}
        \caption{NAICS 32}
        \includegraphics[width=\textwidth]{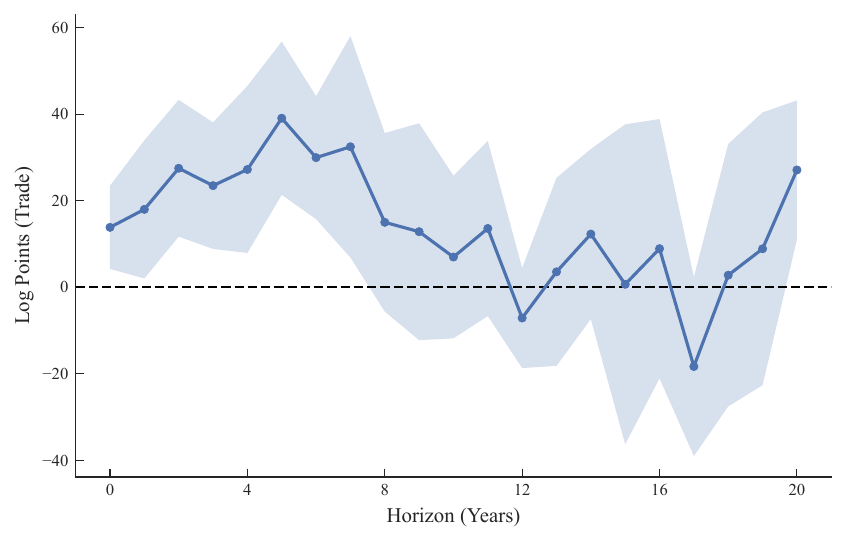}
    \end{subfigure}\\
    \begin{subfigure}[b]{0.49\textwidth}
        \caption{NAICS 33}
        \includegraphics[width=\textwidth]{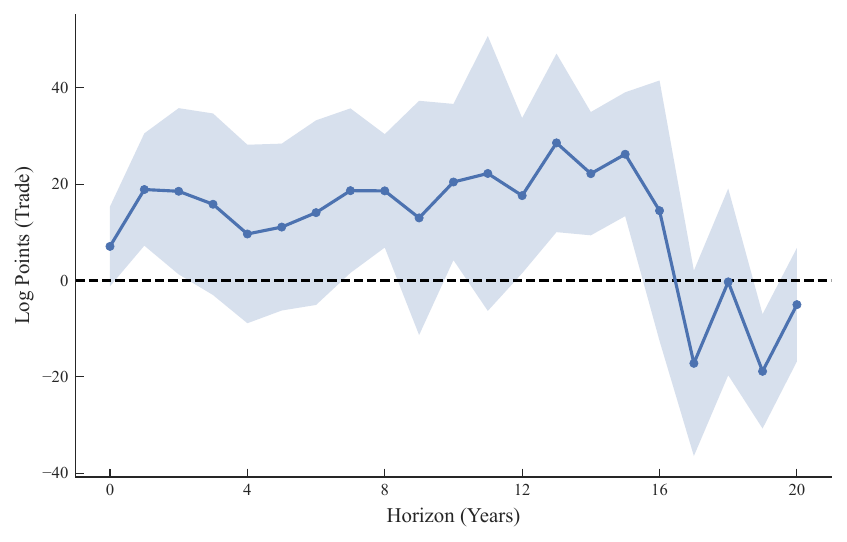}
    \end{subfigure}
    \hfill
    \begin{subfigure}[b]{0.48\textwidth}
        \caption{Critical vs Non-Critical}
        \includegraphics[width=\textwidth]{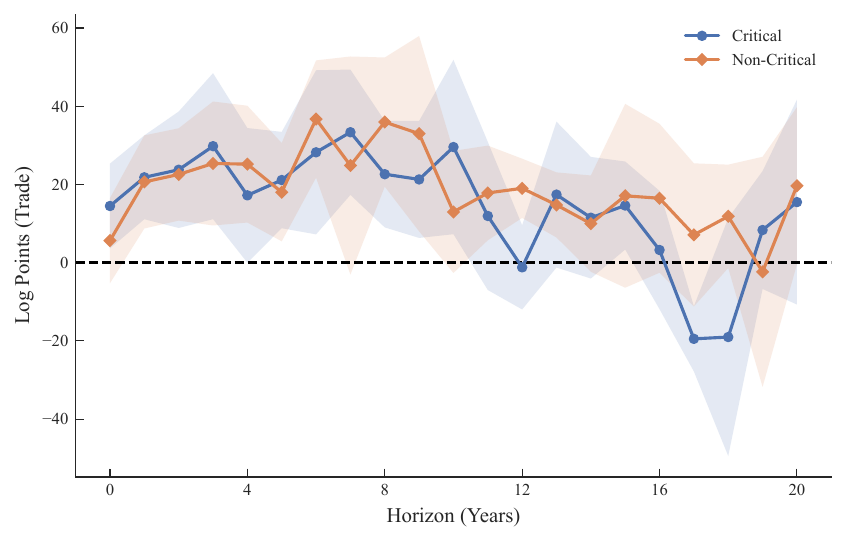}
    \end{subfigure}
    \label{fig:trade_sector}
    \note{\emph{Notes:} This figure reports estimates ${\beta_h}$ from the regression $\ln X_{od,t+h} = \beta_h S_{od,t} + \sum_{\ell=1}^{3} \gamma_{h,\ell} \ln X_{od,t-\ell} + \sum_{\ell=1}^{3} \beta_{h,\ell} S_{od,t-\ell} + \delta_{od} + \delta_{ot} + \delta_{dt} + \varepsilon_{od,t+h}$, using sector-specific trade flows. The sample consists of country pairs among 32 major economies. Panels A–E report estimates for NAICS sectors 11, 21, 31, 32, and 33, respectively. Panel F reports estimates for trade in critical and non-critical sectors. All panels display coefficient estimates with 95\% confidence intervals based on Driscoll--Kraay standard errors.}
\end{figure}

Figure~\ref{fig:trade_sector} reports dynamic trade responses by NAICS sector. The effects are positive across all sectors except energy and mining (NAICS 21). The largest and most persistent responses occur in manufacturing, particularly NAICS 33 (primary metals, machinery, electronics, and transportation equipment). Panel F shows that the effects are similar in critical and non-critical sectors, suggesting that the results are not driven by targeted government intervention.

\FloatBarrier

\subsection{The Trade Effects of Trade Policies} \label{app:add_tp}

\subsubsection{Econometric Specification}
We estimate dynamic trade responses to tariff and sanction shocks using local projections with country-pair, origin-year, and destination-year fixed effects. The trade response to tariffs is
\begin{equation*}
\ln X_{od,t+h} = \beta_h \ln (1+\tau_{od,t})+\sum_{\ell=1}^{3} \gamma_{h,\ell} \ln X_{od,t-\ell}+\sum_{\ell=1}^{3} \beta_{h,\ell} \ln (1+\tau_{od,t-\ell})+\delta_{ot} +\delta_{dt}+\delta_{od}+\varepsilon_{od,t+h},
\end{equation*}
and the trade response to sanctions is estimated analogously,
\begin{equation*}
\ln X_{od,t+h} = \beta_h \mathbf{1}[\text{sanction}_{od,t}]+\sum_{\ell=1}^{3} \gamma_{h,\ell} \ln X_{od,t-\ell}+\sum_{\ell=1}^{3} \beta_{h,\ell} \mathbf{1}[\text{sanction}_{od,t-\ell}]+\delta_{ot} +\delta_{dt}+\delta_{od}+ \varepsilon_{od,t+h}.
\end{equation*}

\subsubsection{Results}

\begin{figure}[htbp]
    \centering
    \caption{Dynamic Trade Responses to Tariffs and Sanctions}
    \label{fig:dynamic_tp_effects}
    \begin{subfigure}[b]{0.48\textwidth}
        \caption{Local Projection of Trade on Tariffs}
        \includegraphics[width=\textwidth]{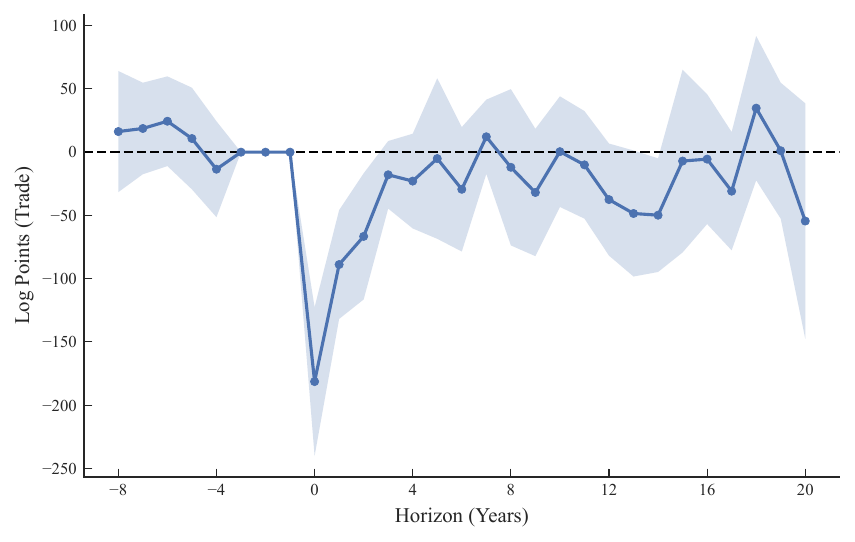}
    \end{subfigure}
    \hfill
    \begin{subfigure}[b]{0.48\textwidth}
        \caption{Local Projection of Trade on Sanctions}
        \includegraphics[width=\textwidth]{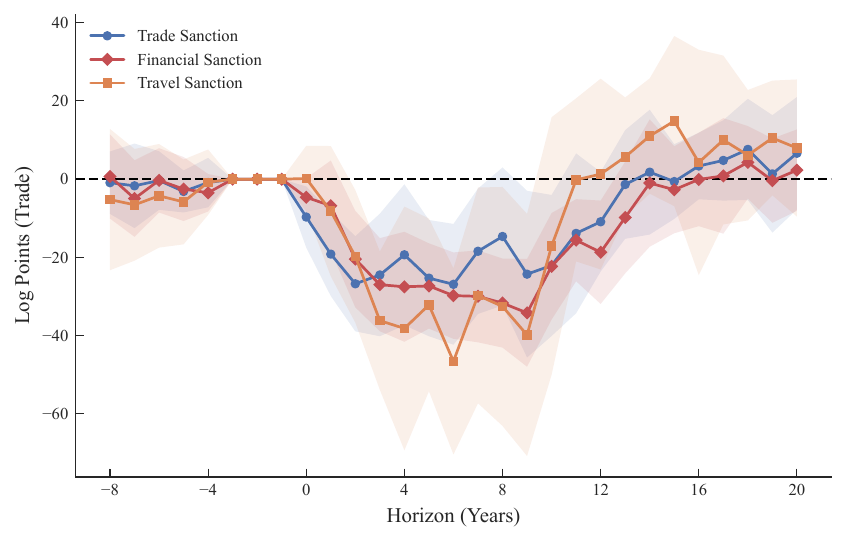}
    \end{subfigure}
    \note{\emph{Notes:} Panel A reports estimates $\{\beta_h\}$ from $\ln X_{od,t+h} = \beta_h \ln(1+\tau_{od,t}) + \sum_{\ell=1}^{3} \gamma_{h,\ell} \ln X_{od,t-\ell} + \sum_{\ell=1}^{3} \beta_{h,\ell} \ln(1+\tau_{od,t-\ell}) + \delta_{od} + \delta_{ot} + \delta_{dt} + \varepsilon_{od,t+h}$. Panel B reports estimates $\{\beta_h\}$ from $\ln X_{od,t+h} = \beta_h \mathbf{1}[\text{sanction}_{od,t}] + \sum_{\ell=1}^{3} \gamma_{h,\ell} \ln X_{od,t-\ell} + \sum_{\ell=1}^{3} \beta_{h,\ell} \mathbf{1}[\text{sanction}_{od,t-\ell}] + \delta_{od} + \delta_{ot} + \delta_{dt} + \varepsilon_{od,t+h}$, estimated separately for the trade, financial, and travel sanction indicators. The sample includes country pairs among 32 major economies. Both panels report coefficient estimates with 95\% confidence intervals based on Driscoll--Kraay standard errors.}
\end{figure}

Figure~\ref{fig:dynamic_tp_effects} reports the estimated dynamic responses; the two policy instruments display sharply different patterns. Panel~A shows that the elasticity of bilateral trade with respect to $\ln(1+\tau)$ is approximately $-2$ on impact, so a 1 percent increase in the tariff factor $1+\tau$ reduces trade by roughly 2 log points initially, with effects dissipating within 3--4 years. Panel~B shows that the onset of a trade, financial, or travel sanction reduces bilateral trade by approximately 20 log points on impact, with the response remaining large and persistent for over a decade.

These contrasting dynamics are consistent with institutional differences in how the two instruments operate. Tariffs are largely set within multilateral frameworks, and firms can adjust through supply-chain reorganization and product substitution, so effects dissipate within a few years. Sanctions, by contrast, signal broader diplomatic ruptures and trigger wider economic disengagement, generating persistent reductions in bilateral trade.

\begin{figure}[htbp]
    \centering
    \caption{IRF: Trade to Permanent Policy Shocks}
    \label{fig:irf_tp}
    \begin{subfigure}[b]{0.48\textwidth}
        \caption{Tariff}
        \includegraphics[width=\textwidth]{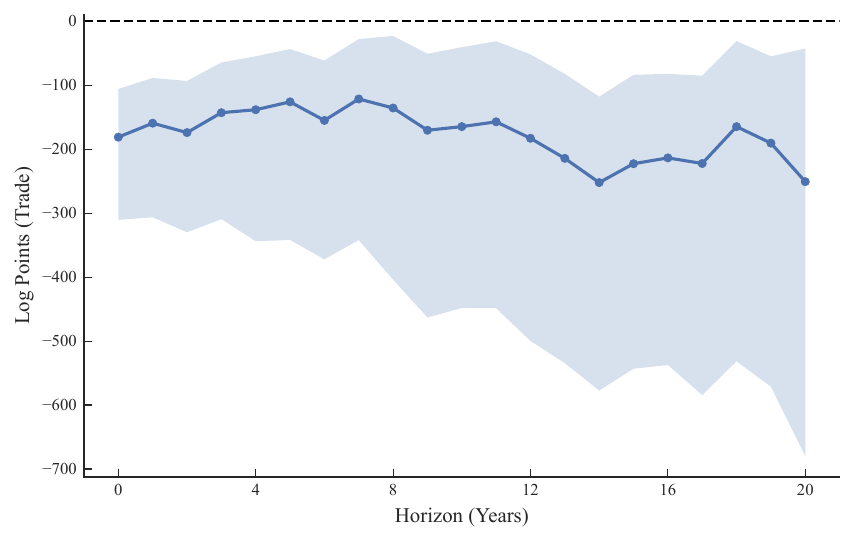}
    \end{subfigure}
    \hfill
    \begin{subfigure}[b]{0.48\textwidth}
        \caption{Sanctions}
        \includegraphics[width=\textwidth]{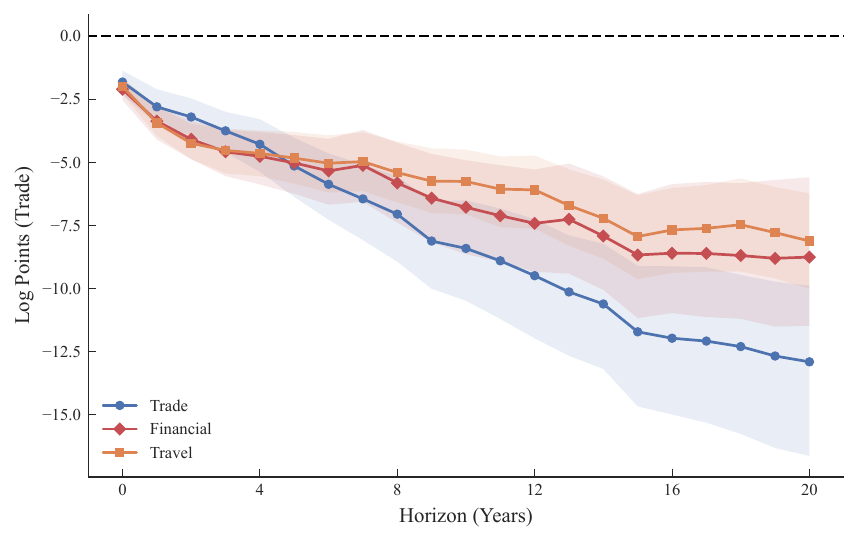}
    \end{subfigure}
    \note{\emph{Notes:} Panel A presents the cumulative impulse response of log trade to a permanent unit shock in $\ln(1+\tau)$. Panel B shows the cumulative response to a permanent unit shock in each sanction indicator (trade, financial, and travel), plotted separately. Both panels use the baseline specification with three lags and all two-way fixed effects. 95\% confidence intervals are from 200 bootstrap iterations with country-pair block resampling.}
\end{figure}

Figure~\ref{fig:irf_tp} reports the cumulative impulse responses to permanent policy shocks, the long-run elasticities used in the quantitative model. A permanent 1 percent increase in the tariff factor $1+\tau$ reduces trade by approximately 2 percent on impact and stabilizes at a 3 percent long-run decline, implying a trade elasticity of 3 and a substitution elasticity of $\sigma \approx 4$. Permanent sanctions produce substantially larger effects, reducing bilateral trade by 40--60 percent in the long run, consistent with the view that sanctions disrupt a broader set of commercial relationships than tariffs.

\FloatBarrier

\subsection{Non-Economic Geopolitical Alignment and Policies}

Table~\ref{tab:tp_nonecon} shows that the relationship between geopolitical alignment and policy instruments holds when using only the non-economic component of the geopolitical alignment score, which excludes trade agreements, sanctions, and commercial disputes.

\begin{table}[H]
\caption{Geopolitical Alignment and Policies: Non-Economic Alignment Score}\label{tab:tp_nonecon}
\begin{center}
\scalebox{0.9}{
\begin{tabular}{lcccccc}
\hline \hline
& (1)      & (2)      & (3)      &  (4)      & (5)   & (6)  \\
\multirow{2}{*}{Dependent Variable:}  & \multirow{2}{*}{log (1+Tariff Rate)} & \multicolumn{3}{c}{$\mathbf{1}$ [Sanction]} & \multicolumn{2}{c}{Restricting Trade Policy} \\
  & & Trade & Financial & Travel & $\geq 1$ Product & $\geq 20$ Products \\
\hline
Geopolitical Alignment   & -0.003 & -0.045 & -0.043 & -0.041 & -3.38 & -1.50 \\
        & (0.003)  & (0.008)  & (0.006)  &  (0.004) & (2.04)   & (0.79) \\
Mean Dep. Var. & 0.05 & 0.049 & 0.037 & 0.019 & 21.4 & 3.24\\
Observations   & 24760   & 74400  & 74400  & 74400 & 16864  & 16864 \\
\hline
Imposer $\times$ Year FE & Yes & Yes & Yes & Yes & Yes & Yes\\
Receiver $\times$ Year FE & Yes & Yes & Yes & Yes & Yes & Yes\\
Imposer $\times$ Receiver FE & Yes & Yes & Yes & Yes & Yes & Yes\\
\hline \hline
\end{tabular}}
\end{center}
\vspace{-5mm}
\note{\emph{Notes:} The unit of observation is an imposer--receiver country pair in a given year. The dependent variables are the log (1+tariff rate) (column 1); indicators for the presence of trade, financial, and travel sanctions (columns 2--4, respectively); and the number of restrictive trade policies (columns 5--6). The geopolitical alignment measure uses only non-economic events. The sample includes country pairs among 32 major countries. Standard errors are clustered at the country-pair level.}
\end{table}

\subsection{Hat Algebra} \label{app:hat_algebra}
Using exact hat algebra, we denote proportional changes as $\hat{x} = x'/x$ and define $\tilde{\tau}_{od} \equiv 1 + \tau_{od}$. 

\paragraph{Price index:}
\begin{equation*}
\hat{P}_d = \left[ \sum_{o=1}^{N} \pi_{od} \left( \hat{w}_o \hat{d}_{od} \right)^{1-\sigma} \right]^{\frac{1}{1-\sigma}}.
\end{equation*}

\paragraph{Trade shares:}
\begin{equation*}
\hat{\pi}_{od} = \left( \frac{\hat{w}_o \hat{d}_{od}}{\hat{P}_d} \right)^{1-\sigma}.
\end{equation*}

\paragraph{Budget constraint:}
\begin{equation*}
X_d' = \frac{\hat{w}_d \, w_d \, \ell_d}{\displaystyle\sum_{o=1}^{N} \frac{\hat{\pi}_{od} \, \pi_{od}}{\hat{\tilde{\tau}}_{od} \, \tilde{\tau}_{od}}}.
\end{equation*}

\paragraph{Labor market clearing:}
\begin{equation*}
\hat{w}_o \, w_o \, \ell_o = \sum_{d=1}^{N} \frac{\hat{\pi}_{od} \, \pi_{od}}{\hat{\tilde{\tau}}_{od} \, \tilde{\tau}_{od}} \, X_d'.
\end{equation*}

\newpage
\section{Construction of the Geopolitical Alignment Measure} \label{app:geo_measure}

\setcounter{theorem}{0}
\setcounter{proposition}{0} 
\setcounter{lemma}{0}
\setcounter{corollary}{0}
\setcounter{definition}{0}
\setcounter{assumption}{0}
\setcounter{remark}{0}
\setcounter{table}{0}
\setcounter{figure}{0}
\setcounter{equation}{0} 
%
\renewcommand{\thetheorem}{B\arabic{theorem}}
\renewcommand{\theproposition}{B\arabic{proposition}}
\renewcommand{\thelemma}{B\arabic{lemma}}
\renewcommand{\thecorollary}{B\arabic{corollary}}
\renewcommand{\thedefinition}{B\arabic{definition}}
\renewcommand{\theassumption}{B\arabic{assumption}}
\renewcommand{\theremark}{B\arabic{remark}}
\renewcommand{\thetable}{B\arabic{table}}
\renewcommand{\thefigure}{B\arabic{figure}}
\renewcommand{\theequation}{B\arabic{equation}}


This appendix details the construction and assessment of our bilateral geopolitical alignment measure. Using the event-based measurement framework of \citet{fan2026measuringgeopoliticalalignmenteconomic}, we compile an event database across all country pairs from 1950--2024 using large language models. Section~\ref{app:event_categories} presents the full taxonomy of event categories. Section~\ref{app:event_collection} describes event compilation and classification. Section~\ref{app:model_robustness} examines robustness to the choice of LLM. Section~\ref{app:validation} reports case-study assessments. 
Section~\ref{app:evo_geo_alignment} documents the evolution of geopolitical event scores globally. Section~\ref{app:llm_prompt} documents the LLM prompt design.

\subsection{Event Categories} \label{app:event_categories}

Table~\ref{tab:event_categories} presents the full taxonomy of bilateral geopolitical events used in our database. Events span six major categories and 22 subcategories, covering the principal dimensions of interstate interaction.

\subsection{Event Compilation Using Large Language Models} \label{app:event_collection}

Major bilateral geopolitical events are typically documented in news archives, government publications, international organizations, and scholarly sources. We employ Gemini 2.5 Pro with web search capabilities to systematically compile and analyze these events through structured prompt engineering. Figure~\ref{fig:geopolitical_analysis} illustrates our analysis procedure, with the complete prompt specification documented in Section~\ref{app:llm_prompt}.

\begin{figure}[H]
\centering
\caption{LLM Geopolitical Event Analysis Procedure}
\begin{tikzpicture}[
  startstop/.style={rectangle, rounded corners, minimum height=0.6cm, text width=1.8cm, text centered, draw=black, fill=red!10, font=\scriptsize},
  process/.style={rectangle, minimum height=0.6cm, text width=1.8cm, text centered, draw=black, fill=blue!10, font=\scriptsize},
  verify/.style={rectangle, minimum height=0.6cm, text width=1.8cm, text centered, draw=black, fill=red!10, font=\scriptsize},
  io/.style={trapezium, trapezium left angle=70, trapezium right angle=110, minimum height=0.6cm, text width=1.8cm, text centered, draw=black, fill=green!10, font=\scriptsize},
  arrow/.style={thick, -Stealth},
  every node/.style={inner sep=1pt}
]

\node[startstop] (start) at (0,-1.2) {Input \{countries, year\}};
\node[verify, right=0.4cm of start] (verify) {Verify Political Entities};
\node[process, right=0.4cm of verify] (search) {Google Search Political Events};
\node[process, right=0.4cm of search, yshift=1.2cm] (cameo) {Assign CAMEO Codes};
\node[process, below=0.4cm of cameo] (goldstein) {Estimate Goldstein Score};
\node[process, below=0.4cm of goldstein] (economic) {Classify Economic Event};
\node[io, right=0.4cm of goldstein, xshift=0.8cm] (output) {JSON Output};

\draw[arrow] (start) -- (verify);
\draw[arrow] (verify) -- (search);
\draw[arrow] (search) |- (cameo);
\draw[arrow] (cameo.east) -| ([xshift=0.4cm]goldstein.east) -- (output);
\draw[arrow] (goldstein.east) -- (output);
\draw[arrow] (economic.east) -| ([xshift=0.4cm]goldstein.east);

\end{tikzpicture}
\label{fig:geopolitical_analysis}
\end{figure}

The LLM performs five sequential tasks: (i) verify historical political entities accounting for state succession (e.g., Soviet Union to Russian Federation); (ii) search knowledge base and internet sources for major bilateral events from authoritative sources; (iii) classify events using the Conflict and Mediation Event Observations (CAMEO) framework; (iv) assign Goldstein scores from $-10$ (maximum conflict) to $+10$ (maximum cooperation); and (v) categorize economic content when applicable.

\begin{table}[H]
\centering
\caption{Categories of Bilateral Geopolitical Events}
\label{tab:event_categories}
\footnotesize
\resizebox{\linewidth}{!}{
\begin{tabular}{p{2cm}p{7cm}p{2cm}p{7cm}}
\toprule
\textbf{Subcategory} & \textbf{Representative Events} & \textbf{Subcategory} & \textbf{Representative Events} \\
\midrule
\multicolumn{2}{l}{\textit{A. Economic Relations}} & \multicolumn{2}{l}{\textit{D. Legal, Territorial \& Movement}} \\
\addlinespace[0.1cm]
A1. Trade Policy
& Tariff imposition/removal; non-tariff barriers; FTA negotiations, signing, ratification, withdrawal
& D1. Legal Actions
& ICJ/WTO disputes; international arbitration; extradition; law enforcement cooperation \\
A2. Financial Relations
& Asset freezes; SWIFT exclusions; capital market denial; currency swaps; FDI restrictions; investment treaties
& D2. Territorial \& Maritime
& EEZ/continental shelf disputes; freedom of navigation operations; border demarcation; sovereignty claims \\
A3. Economic Coercion
& Trade embargoes; sectoral sanctions; Entity List designations; export controls; foreign aid; debt relief
& D3. Movement of People
& Visa regime changes; travel bans; guest worker programs; refugee/asylum policies \\
A4. Strategic Sectors
& Energy supply agreements; pipeline projects; nuclear cooperation; 5G bans; rare earth controls
& & \\
A5. Integration
& BRI/B3W infrastructure; port concessions; friend-shoring; supply chain arrangements; customs unions
& \multicolumn{2}{l}{\textit{E. Multilateral \& Global Governance}} \\
\addlinespace[0.1cm]
A6. Other Economic
& Regulatory harassment; boycott campaigns; government procurement restrictions; gray zone pressures
& E1. Int'l Organizations
& UNSC confrontations; GA coalition building; regional bloc membership changes \\
& & E2. Global Issues
& Climate cooperation; pandemic coordination; vaccine diplomacy; human rights \\
\addlinespace[0.15cm]
\multicolumn{2}{l}{\textit{B. Diplomatic \& Political Relations}} & \multicolumn{2}{l}{\textit{F. Other Significant Events}} \\
\addlinespace[0.1cm]
B1. Formal Diplomacy
& Embassy/consulate openings and closures; ambassador recalls; staff expulsions; d\'{e}marches
& F1. Historical \& Symbolic
& Apologies for historical wrongs; memorial visits; monument disputes \\
B2. High-Level Interactions
& Presidential visits; bilateral summits; summit boycotts; ministerial meetings; strategic dialogues
& F2. Humanitarian
& Disaster aid offers/rejections; joint rescue operations; evacuation cooperation \\
B3. Public Diplomacy
& Policy speeches; parliamentary resolutions; propaganda campaigns
& F3. Sports \& Events
& Olympic boycotts; joint hosting; World Expo participation \\
B4. Cultural \& Educational
& Cultural agreements/boycotts; scholarship programs; university partnerships; student visa policies
& F4. Technology \& Space
& Joint space missions; tech theft accusations; research terminations \\
& & F5. Environment
& Transboundary river/pollution disputes; conservation conflicts \\
\addlinespace[0.15cm]
\multicolumn{2}{l}{\textit{C. Security \& Defense}}
& F6. Communications
& Journalist expulsions; broadcasting restrictions; information warfare \\
\addlinespace[0.1cm]
C1. Military Cooperation
& Alliances; defense pacts; base arrangements; arms sales/embargoes; joint exercises; military aid
& F7. Other
& Emerging forms of bilateral interaction \\
C2. Security Incidents
& Border skirmishes; airspace/maritime violations; naval encounters; hybrid warfare
& & \\
C3. Intelligence \& Cyber
& Espionage revelations; intel officer expulsions; intel-sharing pacts; state-sponsored cyber attacks
& & \\
\bottomrule
\end{tabular}
}
\end{table}

Table~\ref{tab:us_china_2001} illustrates a year of sharp swings in U.S.--China relations rather than uniformly rising cooperation or conflict. The Hainan Island incident ($-8.5$) and the U.S. arms sale to Taiwan ($-5.0$) generated severe security tensions in the first half of the year. Yet these shocks were followed by a marked improvement in bilateral relations through post-9/11 counterterrorism cooperation ($+6.0$), the Jiang-Bush APEC summit ($+5.0$), and China's WTO accession ($+8.0$). The 2001 pattern is informative for our measure because it shows that bilateral geopolitical alignment can incorporate both acute political crises and major trade-relevant breakthroughs within the same year. This combination of security conflict and economic integration is precisely the type of nuance that aggregate voting-based measures struggle to capture.

\begin{table}[H]
\centering
\caption{Major U.S.--China Bilateral Events in 2001: Recorded Events and Scores}
\label{tab:us_china_2001}
\footnotesize
\resizebox{\linewidth}{!}{
\begin{tabular}{@{}p{3.0cm}p{6.0cm}p{2.7cm}p{1.5cm}p{1.5cm}@{}}
\toprule
\textbf{Event Name} & \textbf{Event Description} & \textbf{CAMEO Class.} & \textbf{Econ. Type} & \textbf{Goldstein Score} \\
\midrule

Hainan Island Incident & 
A U.S. EP-3 surveillance aircraft collided with a Chinese fighter near Hainan on Apr. 1, triggering a major diplomatic crisis, emergency landing, and prolonged crew detention &
Material Conflict (17-173) & 
Not econ. & 
$-8.5$ \\
\addlinespace[0.3em]

U.S. Arms Sale to Taiwan & 
The Bush administration approved a major arms package for Taiwan on Apr. 24, including submarines, destroyers, and surveillance aircraft, provoking sharp objections from Beijing &
Material Conflict (17-170) & 
Not econ. & 
$-5.0$ \\
\addlinespace[0.3em]

Jiang-Bush APEC Summit & 
Presidents Jiang Zemin and George W. Bush met at the APEC summit in Shanghai on Oct. 19, helping stabilize relations after the Hainan crisis and opening space for renewed dialogue &
Verbal Coop. (04-043) & 
Not econ. & 
$+5.0$ \\
\addlinespace[0.3em]

China Joins the WTO & 
China formally entered the World Trade Organization on Dec. 11 after fifteen years of negotiations, with U.S. support playing a central role in the accession process &
Material Coop. (06-061) & 
Trade & 
$+8.0$ \\
\addlinespace[0.3em]

Post-9/11 Counterterrorism Cooperation & 
Following the Sept. 11 attacks, China expressed support for the U.S.-led campaign against terrorism, creating a temporary period of strategic cooperation despite earlier tensions &
Verbal Coop. (03-032) & 
Not econ. & 
$+6.0$ \\

\bottomrule
\end{tabular}
}
\end{table}

Comparing Table~\ref{tab:us_china_2001} with the 2024 U.S.--China example in the main text highlights the dynamic range of our measure within a single bilateral relationship. In 2001, severe security frictions---the Hainan incident ($-8.5$) and the Taiwan arms sale ($-5.0$)---coexisted with substantial cooperation through post-9/11 coordination, the Jiang-Bush summit, and China's WTO accession ($+8.0$). By 2024, that mixed pattern had shifted toward managed rivalry: conflict remained pronounced through tariffs, sanctions, export controls, and military tensions over Taiwan, while the cooperative events were narrower and primarily aimed at crisis management rather than deeper integration. This contrast shows that our event-based framework captures not only overall relationship valence, but also changes in the substantive channels through which geopolitics affects economic exchange.

\paragraph{Comparison with Existing Event Databases} 
Our approach differs from GDELT \citep{leetaru2013gdelt} and ICEWS \citep{boschee2015icews} along two dimensions. First, we leverage LLMs' contextual understanding to focus on major bilateral political events that define geopolitical relationships, rather than seeking comprehensive coverage of all international interactions.\footnote{GDELT and ICEWS collect all global events across actors and issue areas, complicating aggregation into meaningful bilateral relationship measures.} This targeted approach yields more precise measurement of relationship intensity. Second, our compilation spans 1950–2024, aligning with economic data availability for panel analysis.\footnote{GDELT begins in 1979 and ICEWS covers only 1995–present, limiting historical economic analysis.}
\subsubsection{Statistics of Geopolitical Events} \label{app_a_geo_events_stats}

\begin{figure}[ht]
\centering
\caption{Geopolitical Events Summary (1950--2024)}
\includegraphics[width=\textwidth]{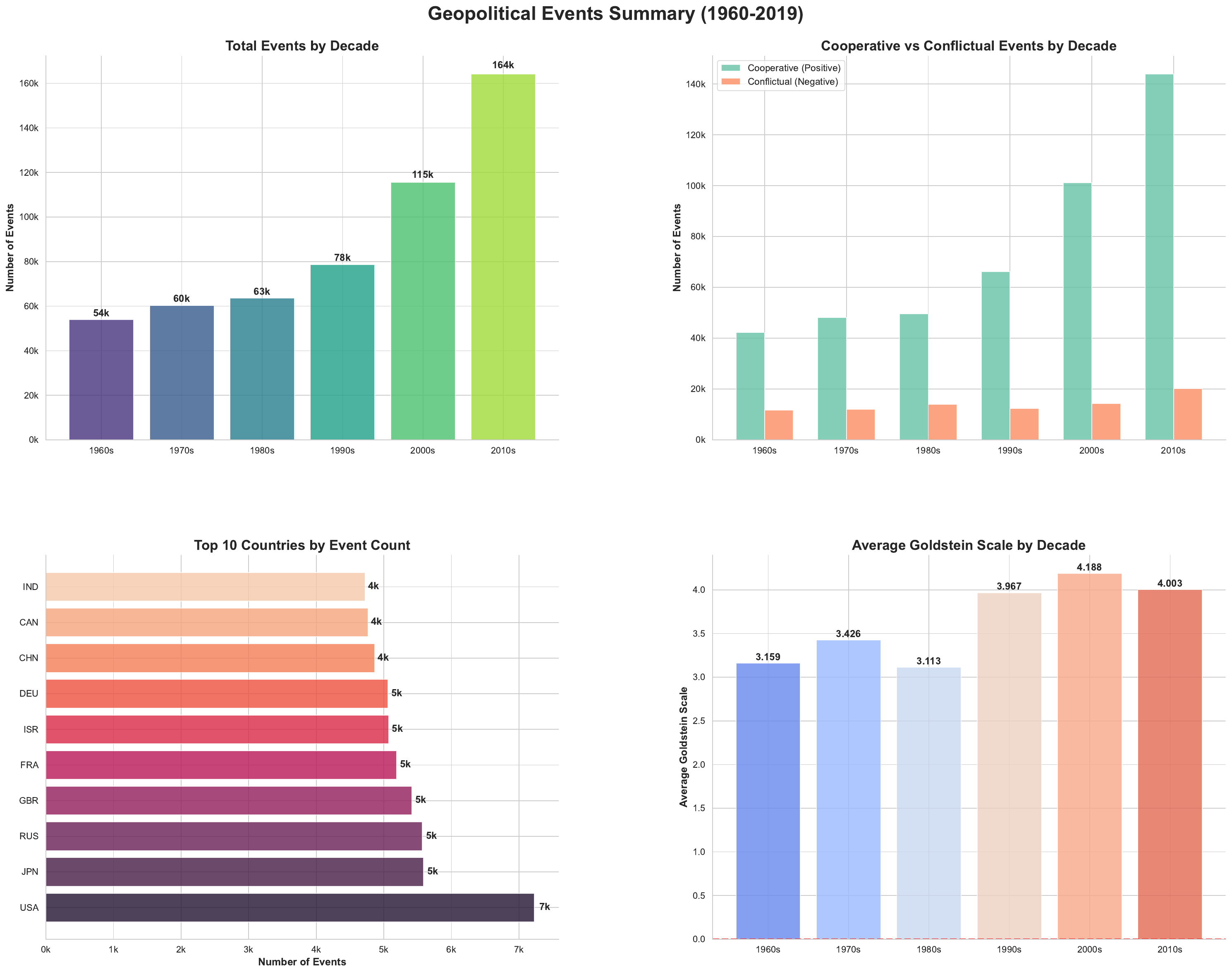}
\label{fig:geopolitical_events_summary}
\end{figure}

Our comprehensive compilation of bilateral geopolitical events spans seven and a half decades (1950--2024) and encompasses 833,485 individual events across all 193×192/2 country pairs. Table~\ref{tab:geopolitical_events_summary} and Figure~\ref{fig:geopolitical_events_summary} provide detailed statistics revealing both the scale and evolution of international political interactions over this extended period.

\begin{table}[ht]
\centering
\caption{Summary Statistics of Geopolitical Events by Decade, 1950--2024}
\label{tab:geopolitical_events_summary}
\resizebox{\textwidth}{!}{
\begin{tabular}{lccccccccc}
\toprule
 & 1950s & 1960s & 1970s & 1980s & 1990s & 2000s & 2010s & 2020s & Total \\
\midrule
\multicolumn{10}{l}{\textbf{CAMEO Event Classification}} \\
\addlinespace[0.1cm]
\quad Verbal Cooperation & 21,709 & 33,113 & 43,108 & 39,089 & 45,930 & 70,549 & 113,422 & 71,141 & 438,061 \\
\quad Material Cooperation & 13,423 & 18,314 & 22,087 & 22,664 & 28,743 & 44,521 & 58,045 & 32,899 & 240,696 \\
\quad Verbal Conflict & 11,484 & 13,953 & 12,810 & 14,064 & 8,241 & 11,863 & 16,754 & 10,104 & 99,273 \\
\quad Material Conflict & 5,598 & 6,924 & 6,505 & 7,782 & 6,705 & 6,606 & 9,556 & 5,779 & 55,455 \\
\addlinespace[0.2cm]
\multicolumn{10}{l}{\textbf{Goldstein Scale Statistics}} \\
\addlinespace[0.1cm]
\quad Mean & 2.02 & 2.41 & 3.06 & 2.61 & 3.83 & 4.00 & 3.76 & 3.62 & 3.38 \\
\quad Std. Dev. & 5.45 & 5.12 & 4.71 & 4.85 & 4.35 & 3.86 & 3.69 & 3.66 & 4.33 \\
\quad Median & 4.00 & 4.50 & 5.00 & 4.00 & 5.00 & 5.00 & 4.00 & 4.00 & 4.50 \\
\addlinespace[0.2cm]
\multicolumn{10}{l}{\textbf{Event Categories}} \\
\addlinespace[0.1cm]
\quad Economic Relations & 8,631 & 12,264 & 16,324 & 16,419 & 22,470 & 35,396 & 46,383 & 25,946 & 183,833 \\
\quad Diplomatic \& Political & 19,324 & 25,498 & 29,274 & 28,522 & 29,716 & 49,028 & 87,932 & 55,222 & 324,516 \\
\quad Security \& Defense & 9,267 & 9,978 & 9,498 & 9,817 & 7,857 & 10,398 & 14,940 & 8,483 & 80,238 \\
\quad Legal \& Territorial & 3,061 & 2,811 & 3,310 & 2,813 & 3,821 & 6,743 & 10,031 & 5,434 & 38,024 \\
\quad Multilateral Governance & 10,562 & 19,951 & 23,771 & 21,353 & 22,201 & 26,460 & 29,560 & 18,178 & 172,036 \\
\quad Other Events & 1,369 & 1,802 & 2,333 & 4,675 & 3,554 & 5,514 & 8,931 & 6,660 & 34,838 \\
\addlinespace[0.2cm]
\multicolumn{10}{l}{\textbf{Summary}} \\
\addlinespace[0.1cm]
\quad Total Events & 52,214 & 72,304 & 84,510 & 83,599 & 89,619 & 133,539 & 197,777 & 119,923 & 833,485 \\
\bottomrule
\end{tabular}
}
\note{\emph{Notes:} CAMEO classifications follow the Conflict and Mediation Event Observations framework. Goldstein Scale ranges from $-10$ (most conflictual) to $+10$ (most cooperative). Event Categories: Economic Relations, Diplomatic \& Political Relations, Security \& Defense, Legal \& Territorial, Multilateral Governance, and Other Events. The 2020s column covers 2020--2024. All figures represent event counts except Goldstein Scale statistics.
}
\end{table}

The data reveal a persistent cooperative skew in international relations, with cooperative events (both verbal and material) comprising 81.5\% of all recorded interactions (678,757 events) compared to 18.5\% for conflictual events (154,728 events). Verbal cooperation represents the single largest category with 438,061 events (52.6\%), followed by material cooperation with 240,696 events (28.9\%). Diplomatic communications and cooperative actions---such as trade agreements, aid provision, and joint initiatives---account for the majority of recorded interactions.

Recorded event frequency rises substantially over time, with total events nearly quadrupling from 52,214 in the 1950s to 197,777 in the 2010s. The 2020s show 119,923 events over five years (2020--2024). Part of this rise likely reflects greater documentation density and digital news coverage in recent decades rather than real-world event growth alone; our empirical specifications absorb such trends through dyad and time fixed effects. Notably, the growth concentrates in cooperative categories: verbal cooperation increases more than fivefold from the 1950s (21,709) to the 2010s (113,422), while material cooperation quadruples (from 13,423 to 58,045). Conflict events show more modest variation, with the ratio of cooperation to conflict remaining stable across decades.

The Goldstein Scale statistics document shifts in relationship intensity over time. The early Cold War (1950s) exhibits the lowest mean cooperation score (2.02) with the highest volatility (standard deviation of 5.45), reflecting the bipolar tensions and uncertainty of the period. Mean scores improve through the 1970s (3.06) before declining in the 1980s (2.61) amid renewed Cold War tensions. The post-Cold War transformation is pronounced: mean scores nearly double from 2.61 in the 1980s to 3.83 in the 1990s, peaking at 4.00 in the 2000s with reduced volatility (standard deviation of 3.86). The 2010s and early 2020s show modest declines (3.76 and 3.62 respectively), suggesting emerging fragmentation while maintaining cooperation levels well above Cold War norms.

Event categories reveal the multifaceted nature of contemporary international relations. Diplomatic and political relations predominate with 324,516 events (38.9\%), reflecting the primacy of state-to-state engagement. Economic relations account for 183,833 events (22.1\%), underscoring the centrality of economic interdependence---particularly relevant for our analysis of trade flows. The substantial growth in economic events from 8,631 in the 1950s to 46,383 in the 2010s parallels the expansion of global trade networks. Multilateral governance events (172,036 total, 20.6\%) highlight the rise of international institutions, while security and defense interactions (80,238 events, 9.6\%) remain relatively stable across decades, suggesting that military considerations, while important, no longer dominate bilateral relationships as they did during the Cold War.

\subsection{Model Robustness} \label{app:model_robustness}

A potential concern with LLM-based measurement is that results may depend on the specific model used or vary across runs. We address this through three complementary exercises using three frontier LLMs: Gemini 2.5 Pro (our baseline), GPT 5.4, and Claude (Opus 4.6 and Sonnet 4.6).

\begin{figure}[ht]
    \centering
    \caption{Model Robustness and Replicability of Geopolitical Alignment Scores}
    \label{fig:model_robustness}
    \includegraphics[width=\linewidth]{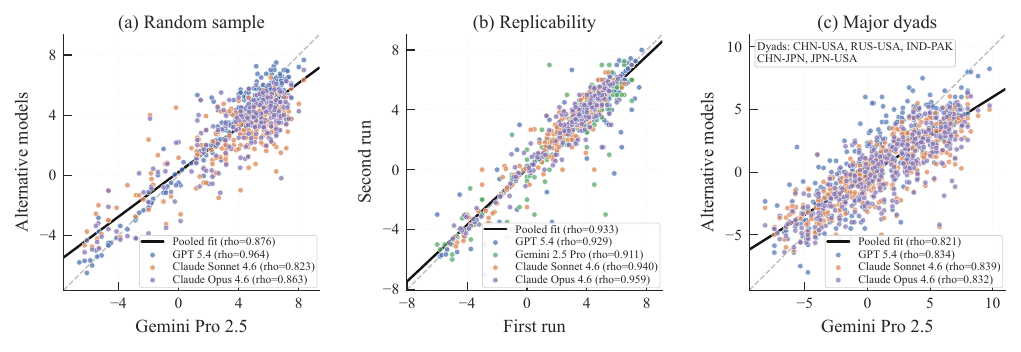}
    \note{\emph{Notes:} Each point represents the mean Goldstein score for a country-pair-year observation. Panel~(a) compares scores from alternative models (vertical axis) against Gemini 2.5 Pro (horizontal axis) for 300 randomly selected dyad-years. Panel~(b) tests within-model replicability by running each model twice on the same dyad-years. Panel~(c) compares scores across models for five major bilateral relationships (CHN-USA, RUS-USA, IND-PAK, CHN-JPN, JPN-USA) across all years.}
\end{figure}

Figure~\ref{fig:model_robustness} presents the results. Panel~(a) compares scores from alternative LLMs against our baseline Gemini 2.5 Pro for 300 randomly selected dyad-years. Scores cluster tightly along the 45-degree line, with correlation coefficients exceeding 0.82 in all pairwise comparisons (pooled $\rho = 0.88$). Panel~(b) tests within-model replicability by running each model twice on the same inputs.\footnote{All models are run with temperature $= 0.1$ to reduce stochastic variation while preserving some flexibility in event retrieval and summary.} Within-model correlations are higher than cross-model correlations (pooled $\rho = 0.93$), as expected: the only source of variation is stochastic decoding, while cross-model comparisons additionally reflect differences in training data and scoring tendencies. Newer frontier models achieve higher stability, with Opus 4.6 reaching $\rho = 0.96$ and Sonnet 4.6 reaching $\rho = 0.94$, compared to $\rho = 0.93$ for GPT 5.4 and $\rho = 0.91$ for Gemini 2.5 Pro. Panel~(c) provides a more demanding test by comparing scores for five major bilateral relationships across all years. The pooled correlation ($\rho = 0.82$) is lower than for random dyad-years, reflecting greater judgment involved in scoring volatile major-power relationships. Across all three exercises, correlation coefficients exceed 0.82, suggesting that major bilateral geopolitical events are sufficiently well-documented that different LLMs converge on similar assessments.

\subsection{Measure Assessment} \label{app:validation}

This section assesses our event-based measure through bilateral case studies, geographic visualization, and statistical tests of its economic relevance.

\subsubsection{Case Studies Across Regions}

Beyond the U.S.--China trajectory presented in Section~\ref{measure_alignment}, we examine our measure across 16 additional bilateral relationships spanning four regions. Each panel plots the dynamic alignment score alongside the yearly event score from 1950 to 2024, with key historical events annotated.

\begin{figure}[ht]
    \centering
    \caption{Geopolitical Alignment Trajectories: United States \& Others}
    \includegraphics[width=\linewidth]{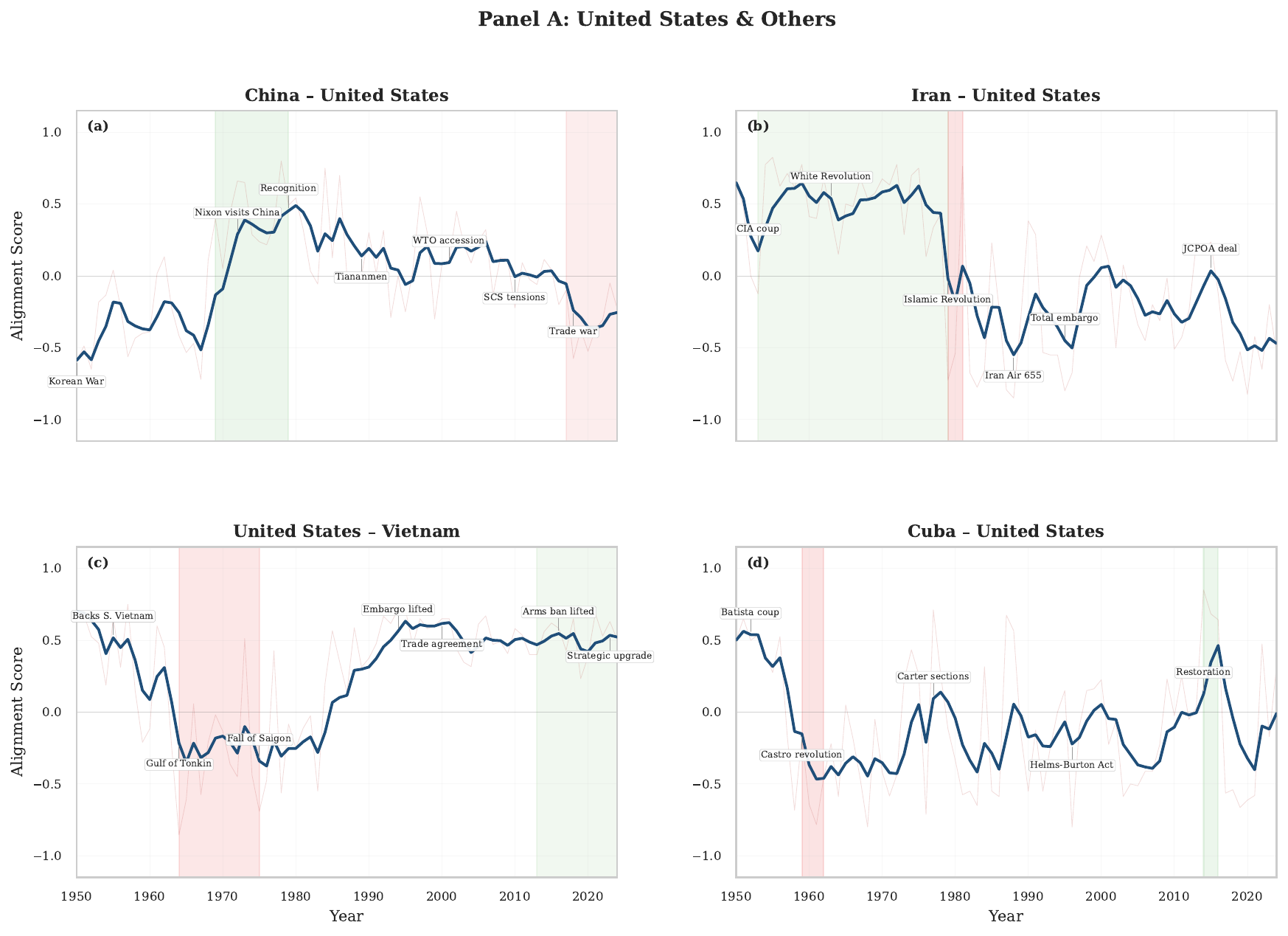}
    \label{fig:panel_us_others}
    \note{\emph{Notes:} Dynamic alignment scores (dark blue) and yearly event scores (light red) for four U.S. bilateral relationships, 1950--2024. Green shading denotes cooperative eras; red shading denotes conflictual eras. Key geopolitical events are annotated along the dynamic score series.}
\end{figure}

Figure~\ref{fig:panel_us_others} presents four U.S. bilateral relationships that illustrate the measure's ability to capture sharp reversals. The China--U.S. trajectory (panel a) traces the full arc from Korean War hostility through Nixon's rapprochement to the recent deterioration driven by trade war and strategic rivalry. The Iran--U.S. relationship (panel b) exhibits one of the sharpest discontinuities in our dataset: the 1979 Islamic Revolution transforms the score from sustained cooperation to deep hostility, with only the brief 2015 JCPOA episode registering a partial recovery. The U.S.--Vietnam trajectory (panel c) captures the mirror image---from Cold War conflict through post-embargo normalization to the 2023 strategic upgrade. Cuba--U.S. relations (panel d) show persistent hostility following the Castro revolution, punctuated by the short-lived Obama-era opening in 2014--2016 and its subsequent reversal.

\begin{figure}[ht]
    \centering
    \caption{Geopolitical Alignment Trajectories: Asia}
    \includegraphics[width=\linewidth]{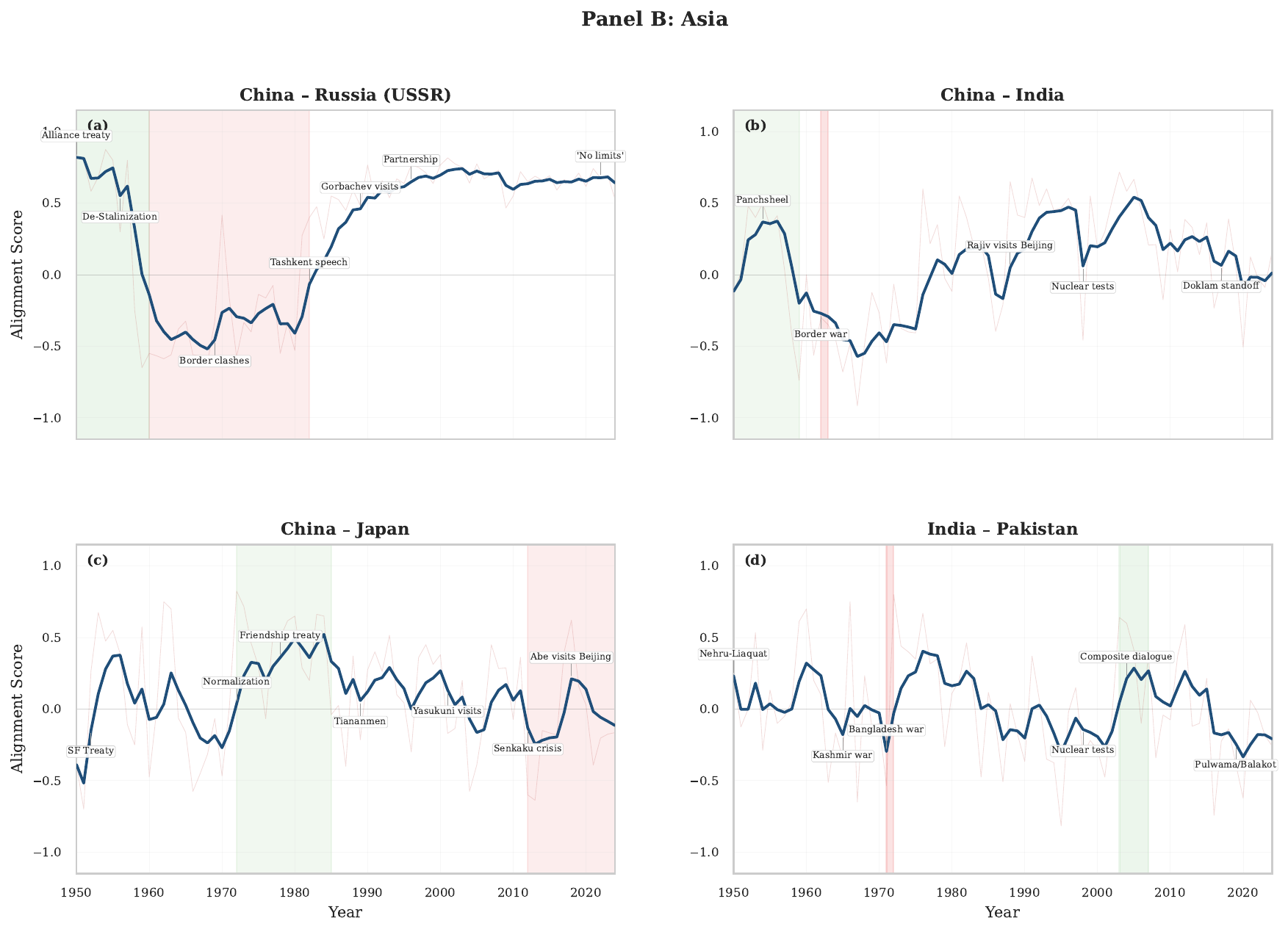}
    \label{fig:panel_asia}
    \note{\emph{Notes:} Dynamic alignment scores (dark blue) and yearly event scores (light red) for four Asian bilateral relationships, 1950--2024. Green shading denotes cooperative eras; red shading denotes conflictual eras.}
\end{figure}

Figure~\ref{fig:panel_asia} covers four Asian relationships characterized by ideological realignments and territorial disputes. The China--Russia trajectory (panel a) captures the Sino-Soviet alliance, the Sino-Soviet split and 1969 border clashes, gradual rapprochement, and the contemporary partnership. China--India relations (panel b) show an early cooperative phase interrupted by the 1962 border war, followed by decades of cautious engagement disrupted by standoffs at Doklam (2017) and Galwan Valley Clash (2020). The China--Japan score (panel c) tracks post-war hostility, the 1972 normalization, and the deterioration following the Senkaku/Diaoyu crisis of 2012. India--Pakistan (panel d) exhibits persistent tension punctuated by wars (1965, 1971), nuclear tests (1998), and rare cooperative windows.

\begin{figure}[ht]
    \centering
    \caption{Geopolitical Alignment Trajectories: Middle East}
    \includegraphics[width=\linewidth]{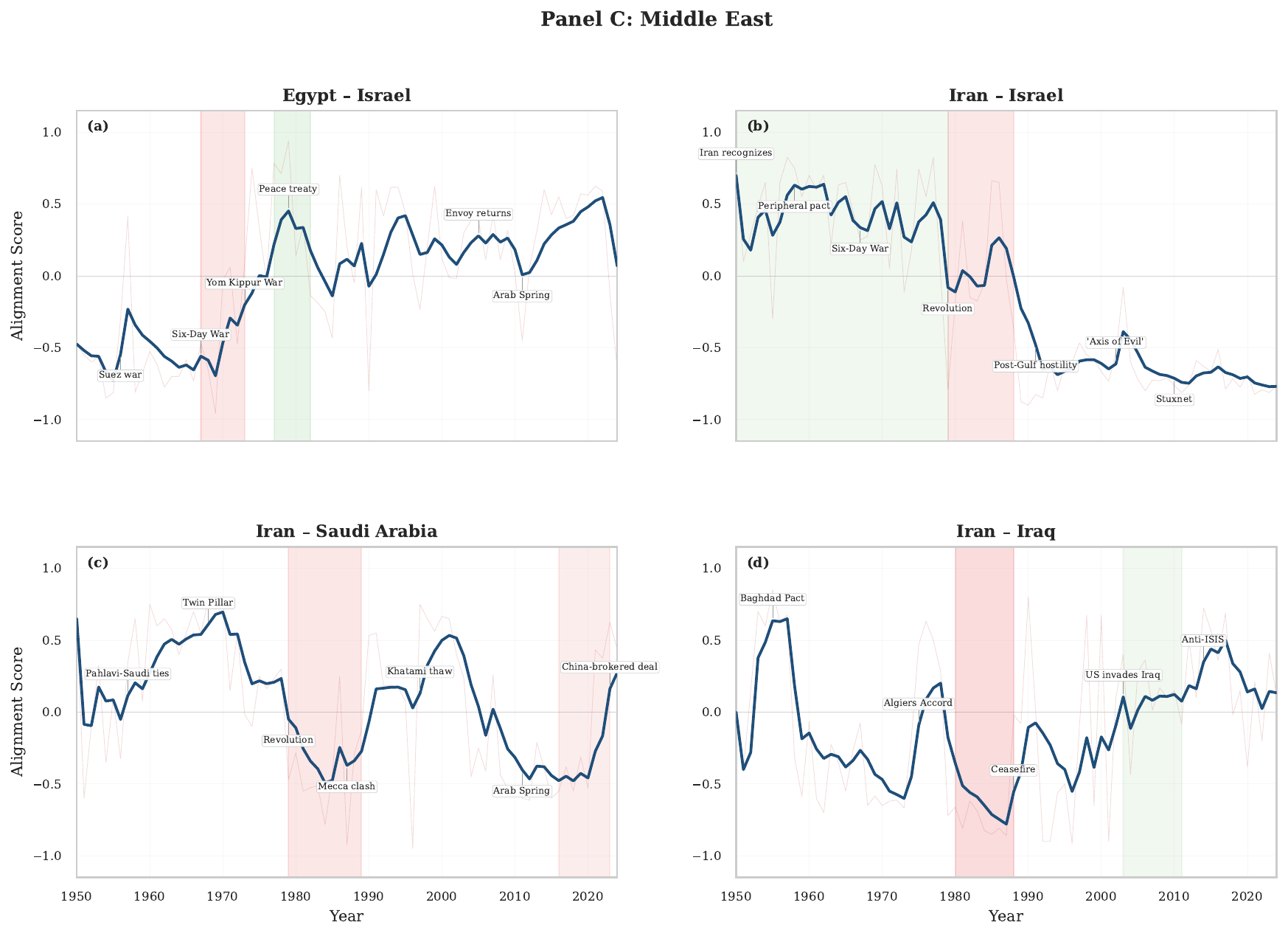}
    \label{fig:panel_middle_east}
    \note{\emph{Notes:} Dynamic alignment scores (dark blue) and yearly event scores (light red) for four Middle Eastern bilateral relationships, 1950--2024. Green shading denotes cooperative eras; red shading denotes conflictual eras.}
\end{figure}

Figure~\ref{fig:panel_middle_east} illustrates relationships shaped by revolutionary regime changes and prolonged conflicts. The Egypt--Israel trajectory (panel a) captures the transformation from repeated wars to the Camp David peace. Iran--Israel relations (panel b) exhibit another revolutionary discontinuity: the Pahlavi-era peripheral pact gives way to sustained enmity after 1979. Iran--Saudi Arabia (panel c) follows a similar pattern of pre-revolutionary cooperation, post-1979 hostility, and the 2023 China-brokered rapprochement. The Iran--Iraq score (panel d) captures the eight-year war (1980--1988) and the post-2003 reversal as regime change in Baghdad transformed the bilateral relationship.

\begin{figure}[ht]
    \centering
    \caption{Geopolitical Alignment Trajectories: Europe}
    \includegraphics[width=\linewidth]{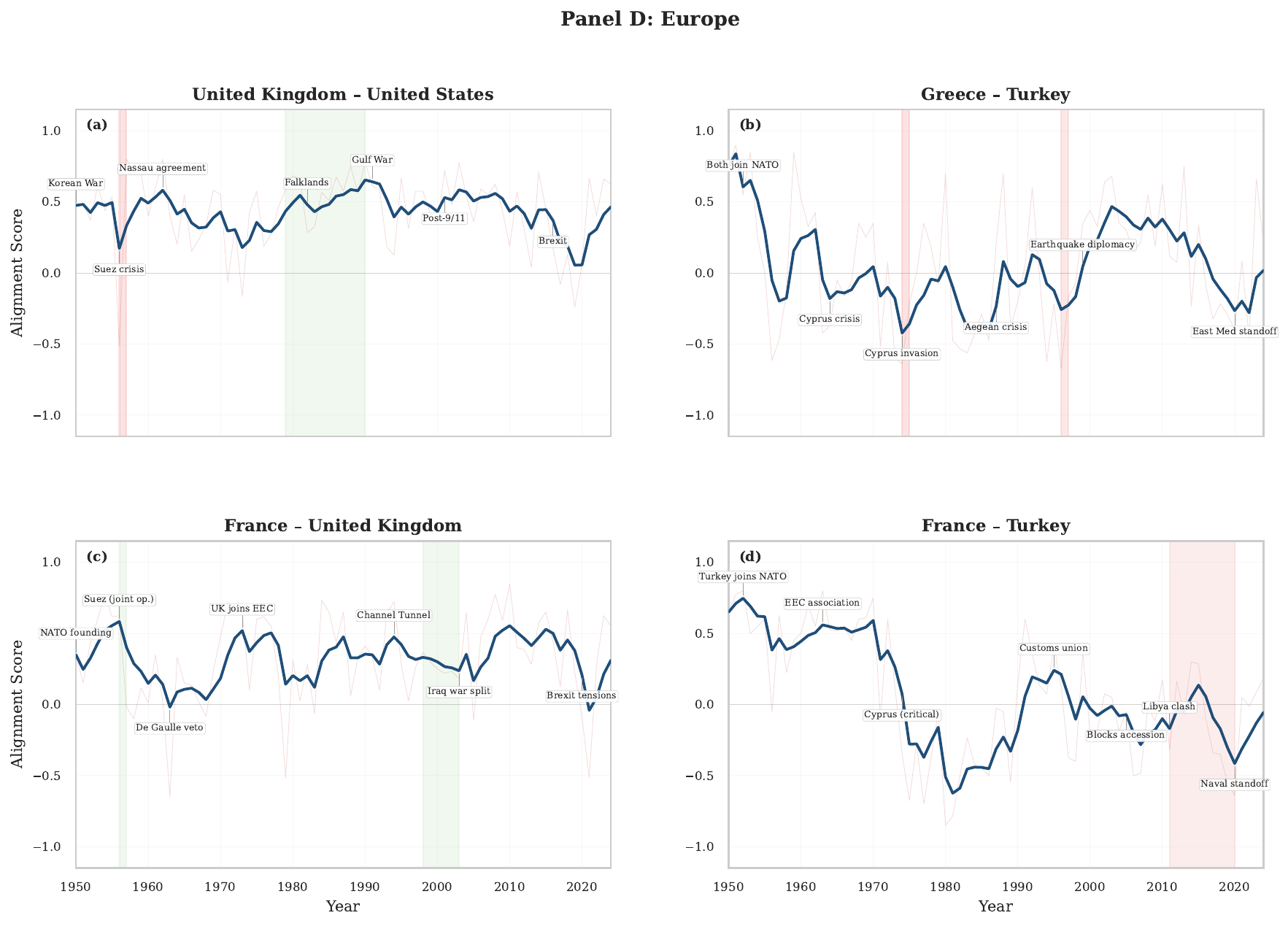}
    \label{fig:panel_europe}
    \note{\emph{Notes:} Dynamic alignment scores (dark blue) and yearly event scores (light red) for four European bilateral relationships, 1950--2024. Green shading denotes cooperative eras; red shading denotes conflictual eras.}
\end{figure}

Figure~\ref{fig:panel_europe} presents European relationships that exhibit meaningful variation driven by institutional integration and strategic disagreements. The UK--U.S. ``special relationship'' (panel a) maintains consistently positive scores, with the Suez crisis of 1956 representing the most notable disruption. Greece--Turkey relations (panel b) show the most volatile European trajectory, driven by the Cyprus crises, the Aegean disputes, and renewed tensions over Eastern Mediterranean resources. France--UK relations (panel c) fluctuate around a cooperative baseline, with De Gaulle's EEC veto and Brexit producing the largest dips. France--Turkey (panel d) exhibits a sustained decline from NATO-era cooperation through progressive disagreements over EU accession, Libya, and Eastern Mediterranean sovereignty.

Across all 16 dyads, the measure consistently tracks well-documented historical turning points, captures both gradual trends and sharp discontinuities, and distinguishes between sustained regime shifts (e.g., Iran--U.S. post-1979) and temporary disruptions (e.g., UK--U.S. during Suez).

\subsubsection{Geographic Assessment Through Maps}

We assess our measure's ability to capture geopolitical dynamics by examining three pivotal moments in great power competition: Cold War bipolarity (USA--USSR, 1980), emerging multipolarity (USA--China, 2019), and contemporary regional conflict (Ukraine--Russia, 2024).

\begin{figure}[htbp]
    \centering
    \caption{Cold War Bipolarity: USA vs. USSR, 1980}
    \begin{subfigure}[b]{0.49\textwidth}
        \caption{United States, 1980}
        \includegraphics[width=\textwidth]{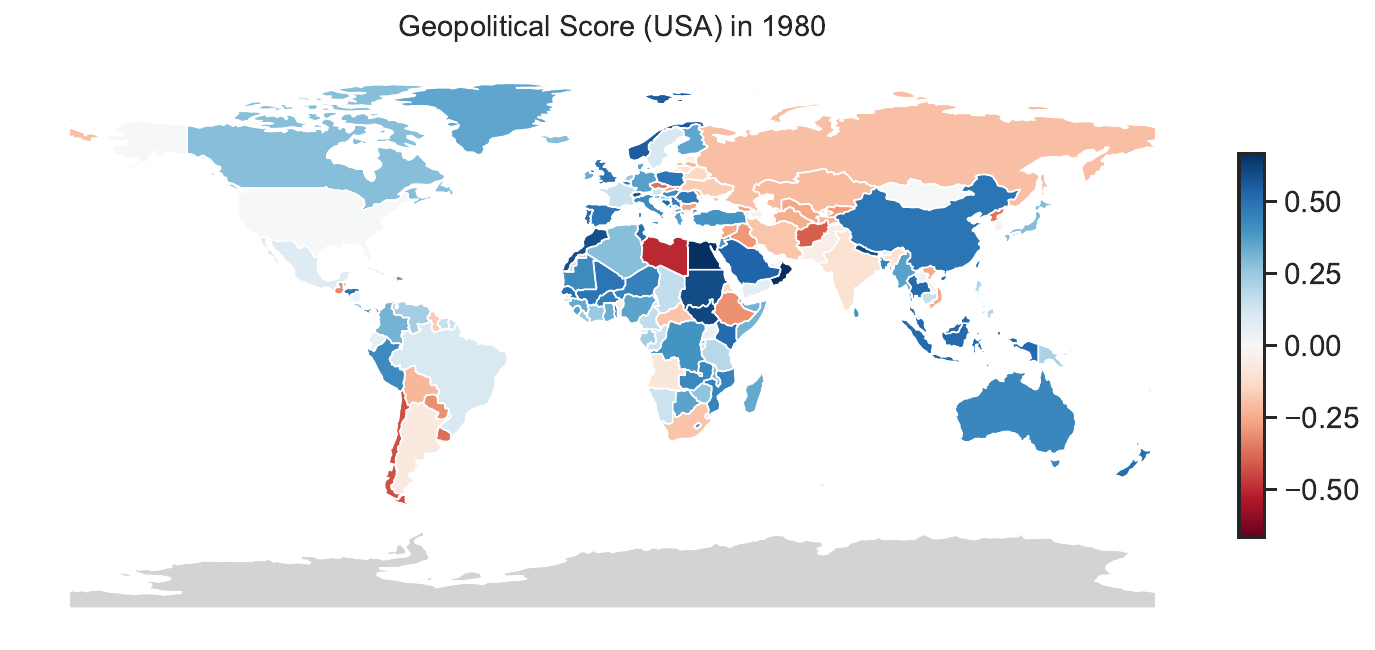}
    \end{subfigure}
    \hfill
    \begin{subfigure}[b]{0.49\textwidth}
        \caption{Soviet Union, 1980}
        \includegraphics[width=\textwidth]{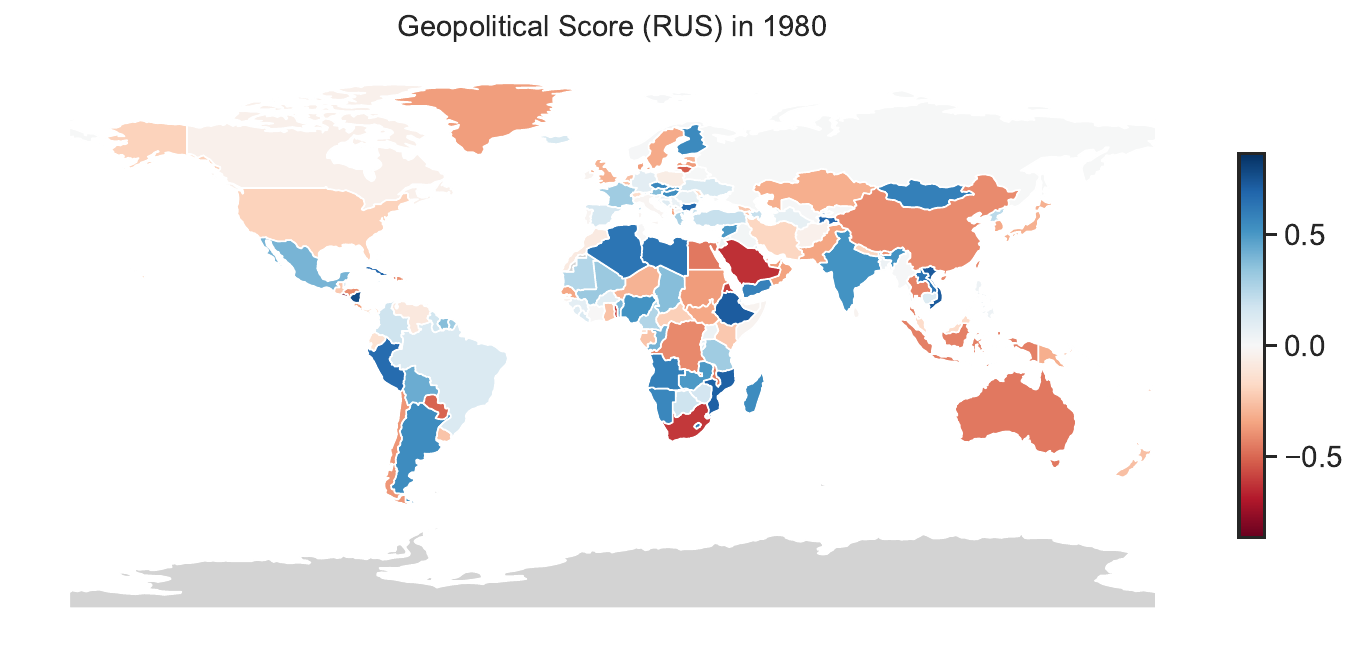}
    \end{subfigure}
    \label{fig:map_usa_rus_1980}
    \note{\emph{Notes:} Each map shades country $d$ by its bilateral alignment score with the focal country (subpanel caption) in the year indicated. Blue denotes positive alignment; red denotes antagonism; gray indicates missing data. The color scale is clipped at $\pm 0.5$ for visual contrast.}
\end{figure}

The 1980 maps (Figure~\ref{fig:map_usa_rus_1980}) exhibit canonical Cold War alignment patterns: the two superpowers' scores are strongly negatively correlated across partners, consistent with zero-sum competition. The United States records high alignment with NATO Europe, Pacific allies, and the Western Hemisphere, while the USSR records high alignment with Eastern Europe, Cuba, Vietnam, Ethiopia, and Afghanistan. Countries displaying positive relations with one superpower show negative relations with the other, consistent with our measure capturing mutually exclusive alignment. Notable exceptions such as India and Egypt are consistent with Non-Aligned Movement membership despite their practical tilts.

\begin{figure}[htbp]
    \centering
    \caption{Emerging Multipolarity: USA vs. China, 2019}
    \begin{subfigure}[b]{0.49\textwidth}
        \caption{United States, 2019}
        \includegraphics[width=\textwidth]{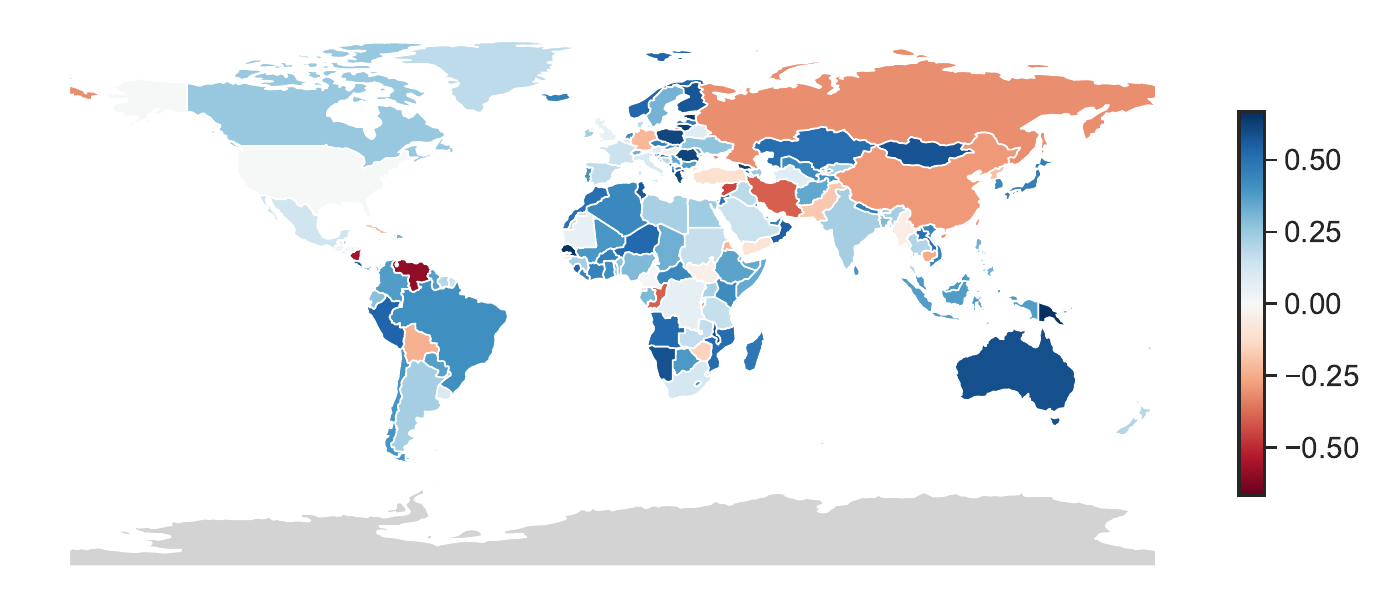}
    \end{subfigure}
    \hfill
    \begin{subfigure}[b]{0.49\textwidth}
        \caption{China, 2019}
        \includegraphics[width=\textwidth]{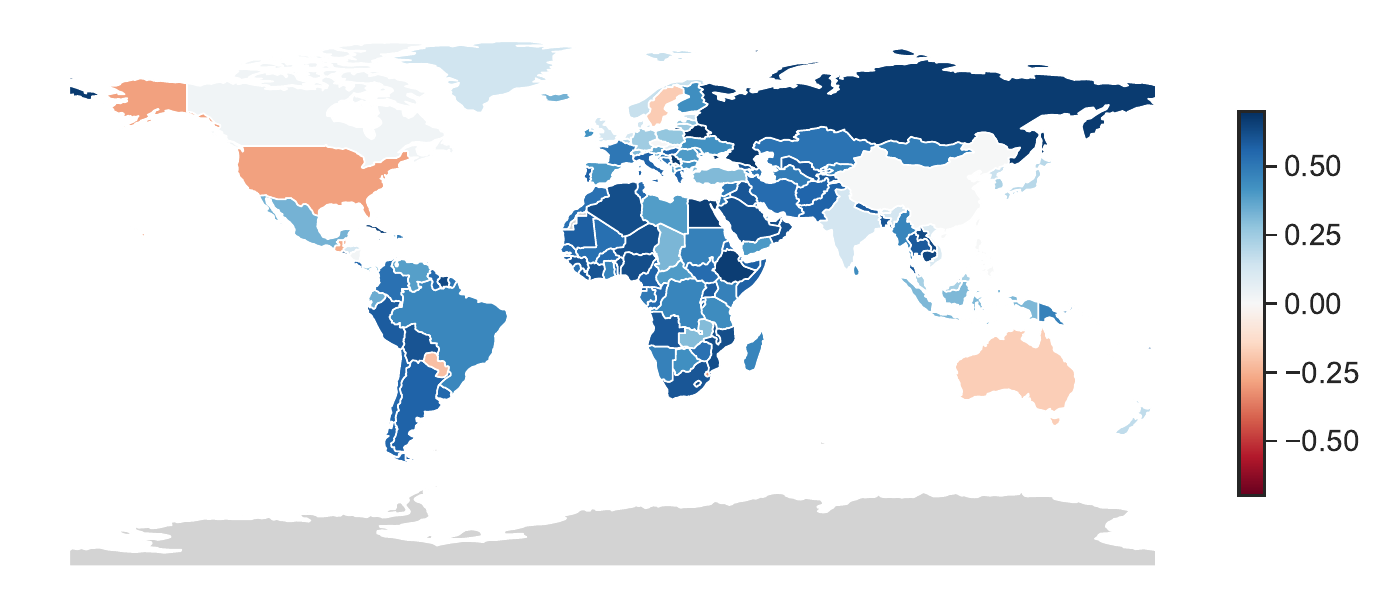}
    \end{subfigure}
    \label{fig:map_usa_chn_2019}
    \note{\emph{Notes:} See Figure~\ref{fig:map_usa_rus_1980} for color scale conventions.}
\end{figure}

The 2019 comparison (Figure~\ref{fig:map_usa_chn_2019}) reveals transformed competition patterns. While the United States maintains traditional alliances (NATO, Japan, Australia), sharp bipolar divisions have dissolved. China achieves positive relations across most of Africa and Southeast Asia through economic engagement rather than military alliances. Many countries maintain positive relations with both powers—the "connector states" that enable continued globalization despite strategic rivalry. China's most negative scores concentrate among U.S. treaty allies, while the United States shows deteriorated relations primarily with Russia, Iran, and Venezuela.

\begin{figure}[htbp]
    \centering
    \caption{Regional Conflict Polarization: Ukraine vs. Russia, 2024}
    \begin{subfigure}[b]{0.49\textwidth}
        \caption{Ukraine, 2024}
        \includegraphics[width=\textwidth]{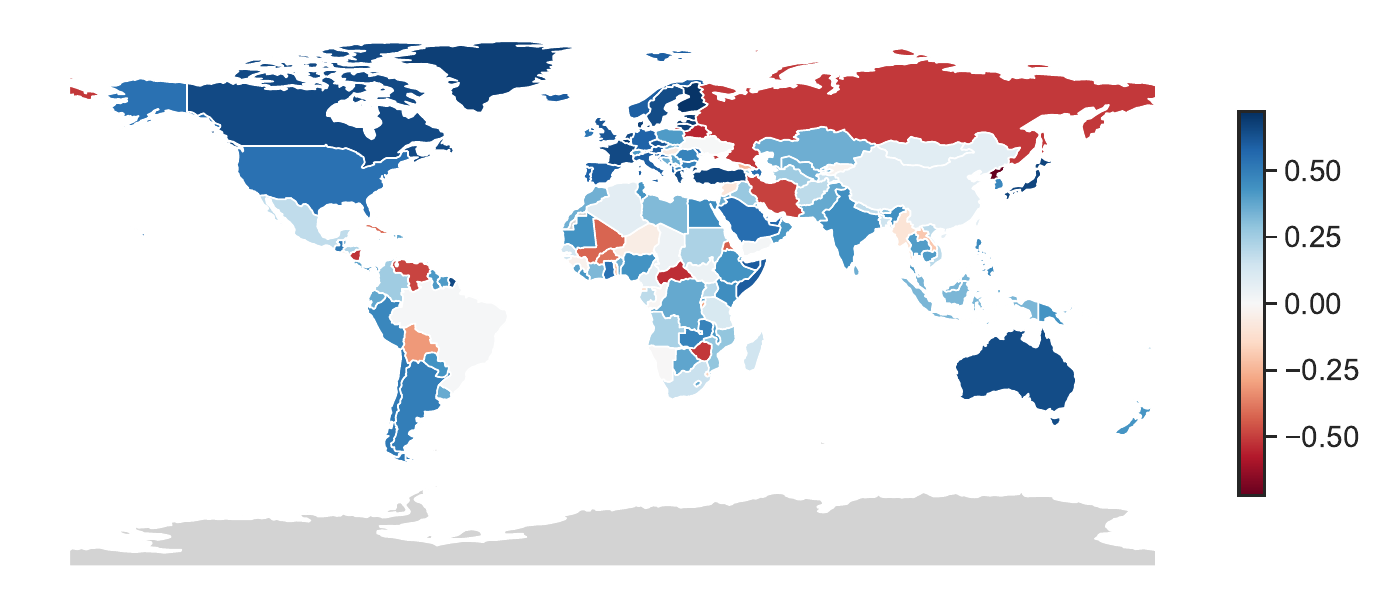}
    \end{subfigure}
    \hfill
    \begin{subfigure}[b]{0.49\textwidth}
        \caption{Russia, 2024}
        \includegraphics[width=\textwidth]{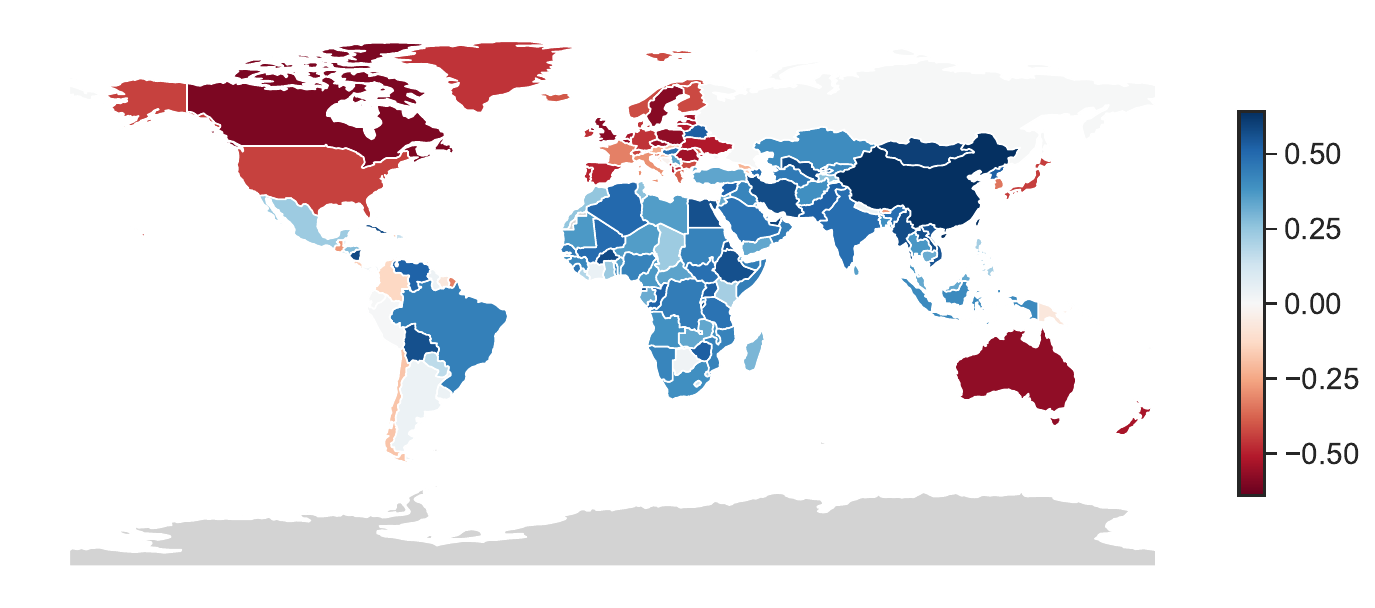}
    \end{subfigure}
    \label{fig:map_ukr_rus_2024}
    \note{\emph{Notes:} See Figure~\ref{fig:map_usa_rus_1980} for color scale conventions.}
\end{figure}

The 2024 Ukraine--Russia maps (Figure~\ref{fig:map_ukr_rus_2024}) demonstrate how regional conflicts generate global polarization. Ukraine achieves very high Western alignment—exceeding even U.S. levels across NATO members—reflecting wartime solidarity. Russia faces near-total isolation from developed democracies while maintaining positive relations with China, India, and parts of Africa resistant to Western pressure. The clear geographic clustering—democratic alignment with Ukraine versus authoritarian and non-aligned support for Russia—illustrates our measure's sensitivity to how regional crises restructure global alignments. Belarus and Central Asian states' positioning is consistent with their delicate balance between historical ties and sanctions pressure.

\paragraph{Comparison with Existing Measures}

Our event-based approach complements existing measures of bilateral geopolitical relations along three dimensions: coverage across countries and time, the timing and intensity of bilateral dynamics, and a continuous range from conflict to cooperation.

Existing literature predominantly relies on UN voting similarity to achieve broad coverage \citep{Signorino1999-yb, Bailey2017-po}. However, as shown in Section~\ref{ss:bilateral_multilateral}, UNGA voting captures countries' broad multilateral positioning but is largely orthogonal to dyad-specific variation in bilateral relations: the cross-sectional correlation with our raw bilateral measure averages only 0.06 after the 1960s, and rises to at most 0.38 when comparing against variance-weighted principal components.

Our measure also complements categorical approaches that classify relationships using discrete indicators such as \textit{Strategic Rivalry} \citep{thompson2001identifying}, \textit{Sanctions} \citep{Ahn2020-ng, Felbermayr2021-ws}, \textit{Formal Alliance} \citep{Gibler2008-xr}, and \textit{Treaties} \citep{broner2025hegemonic}. While these binary classifications capture important institutional milestones, they necessarily focus on specific relationship thresholds rather than continuous evolution. Our framework incorporates these landmark events—military rivalries, alliance formations, treaty signings—while situating them within a broader spectrum of bilateral interactions.

\FloatBarrier

\subsection{Evolution of Global Geopolitical Events} \label{app:evo_geo_alignment}

This section documents the evolution of geopolitical event scores globally and across UN geographical regions.

\begin{figure}[H]
    \centering
    \caption{Global Mean Goldstein Scores with Major Conflicts, 1950--2024}
    \includegraphics[width=\linewidth]{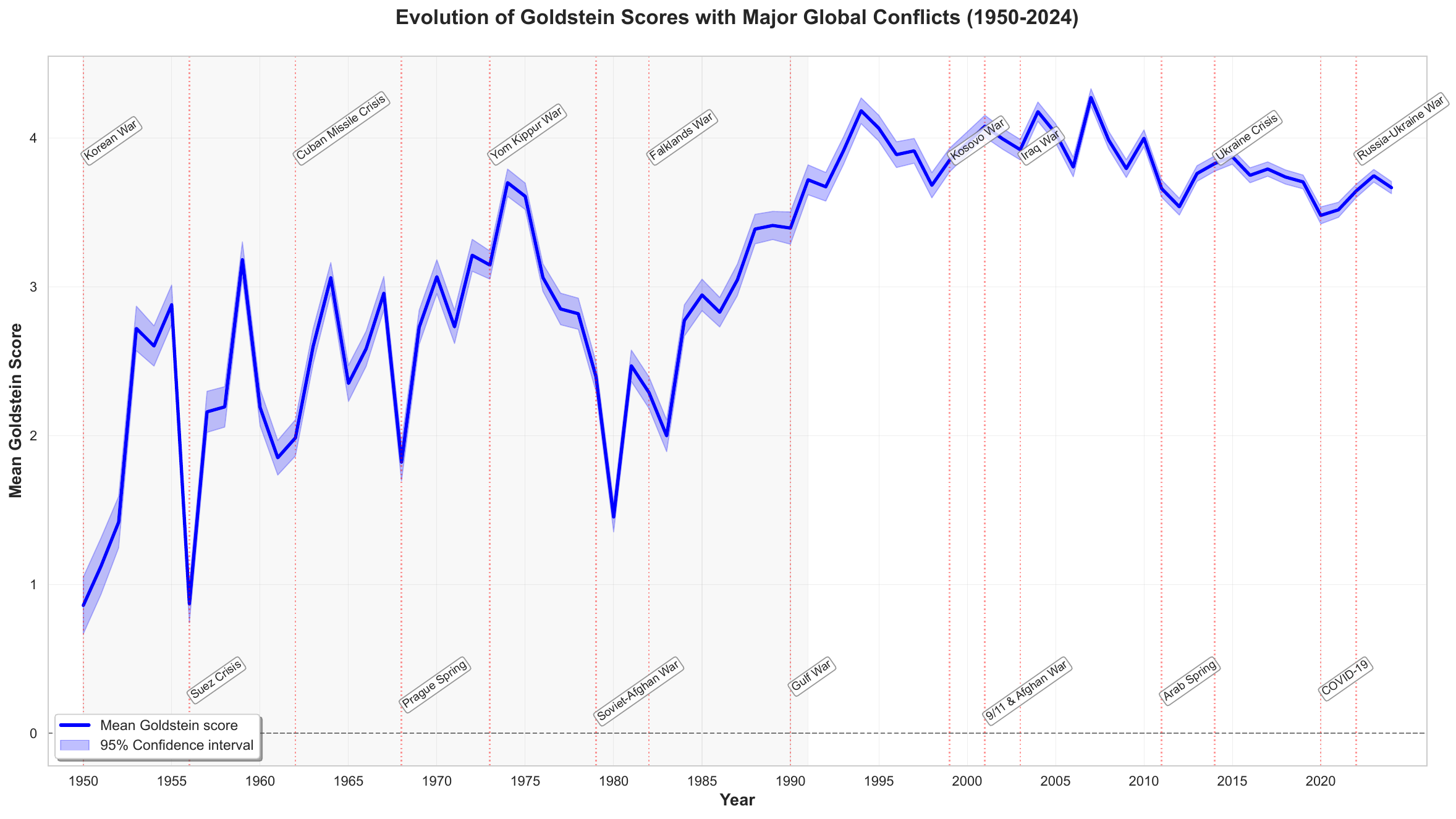}
    \label{fig:global_goldstein_evolution}
    \note{\emph{Notes:} Mean Goldstein scores across all bilateral country pairs with 95\% confidence intervals. Vertical lines indicate major international conflicts. Scores range from $-10$ (maximum conflict) to $+10$ (maximum cooperation). Shaded region denotes the Cold War period (1950--1991).}
\end{figure}

Figure~\ref{fig:global_goldstein_evolution} reveals the secular evolution of global geopolitical relations over seven decades. The Cold War period (1950--1991) exhibits persistently low cooperation levels, with mean scores oscillating between 1 and 3, punctuated by sharp deteriorations during major crises—the Korean War (1950), Suez Crisis (1956), Cuban Missile Crisis (1962), and Soviet-Afghan War (1979). The end of the Cold War marks a structural break: mean scores surge from 2.5 in 1989 to above 3.5 by 1993, subsequently stabilizing at this elevated level through the 2000s. This period of hyper-globalization coincides with historically high cooperation scores and reduced variance, reflecting expanded multilateral institutions and deepening economic integration. Since 2010, however, scores have plateaued and begun declining, with the Arab Spring (2011), Ukraine Crisis (2014), and COVID-19 pandemic (2020) marking inflection points toward renewed fragmentation. The Russia--Ukraine War (2022) drives scores to their lowest level since the 1980s, consistent with renewed fragmentation of the post-Cold War cooperative pattern.

\begin{figure}[H]
    \centering
    \caption{Geopolitical Alignment Scores for Western European and Other States}
    \begin{subfigure}[b]{0.49\textwidth}
        \caption{Western European and Other States}
        \includegraphics[width=\textwidth]{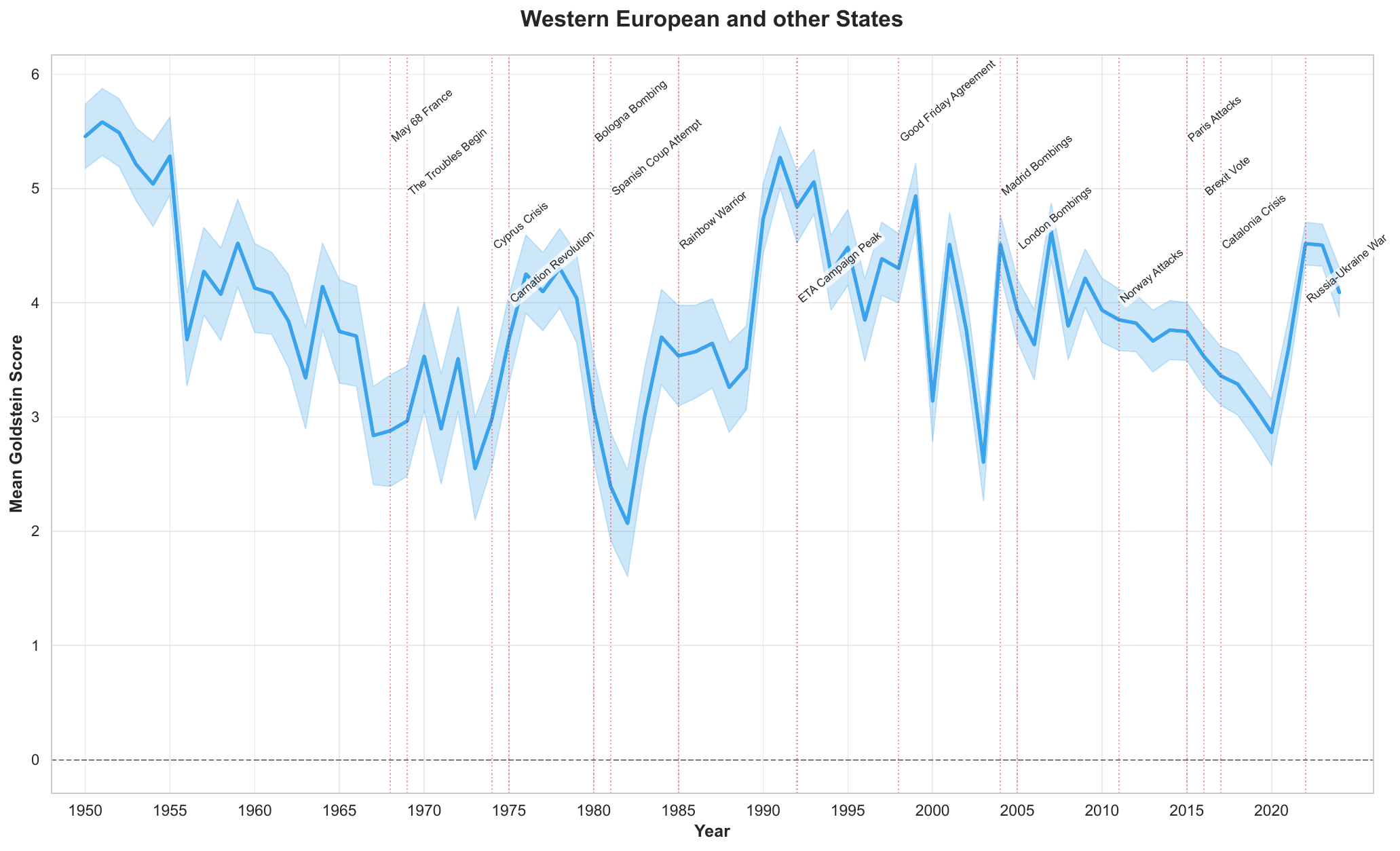}
    \end{subfigure}
    \hfill
    \begin{subfigure}[b]{0.49\textwidth}
        \caption{WEO States vs. Other Regions}
        \includegraphics[width=\textwidth]{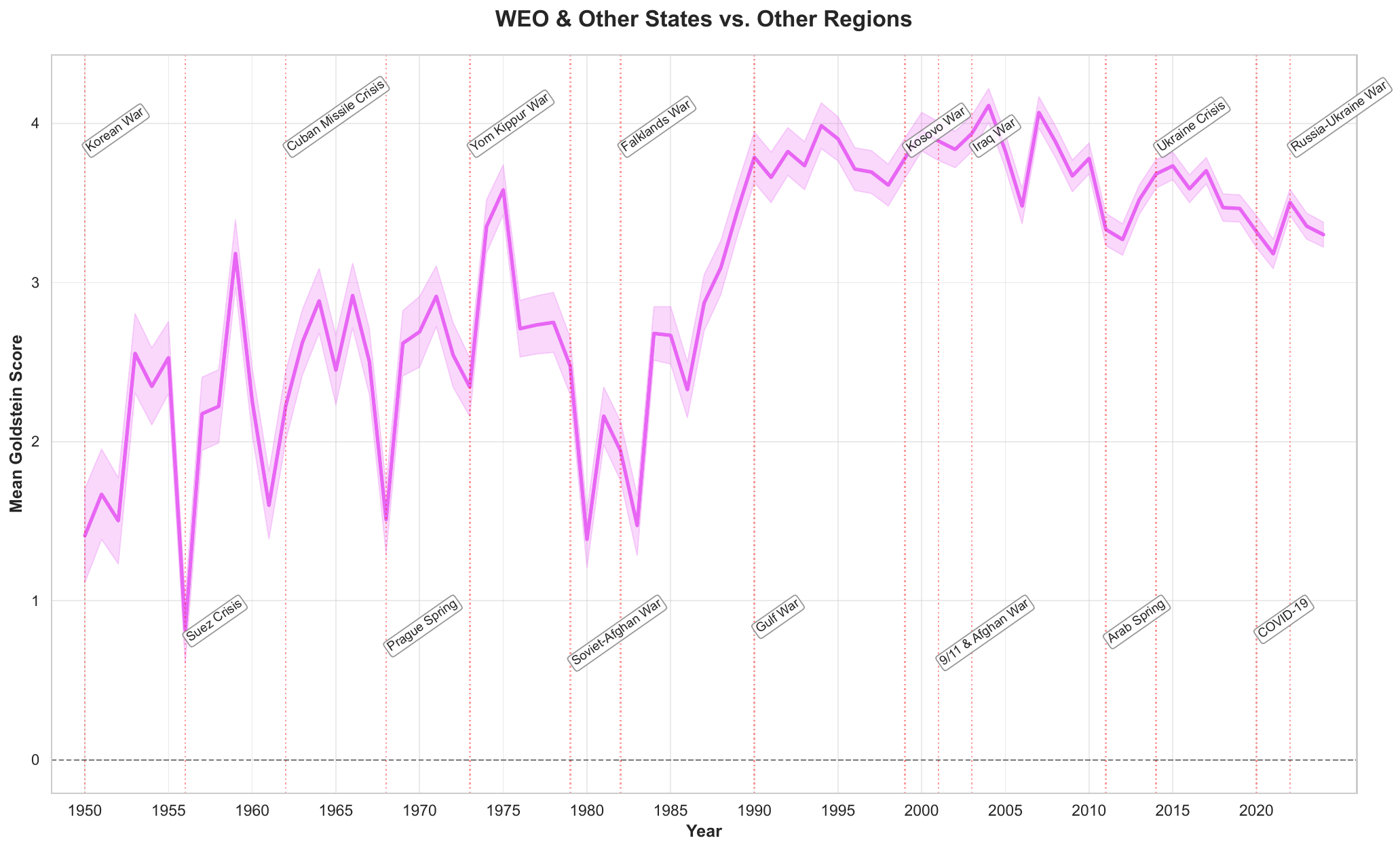}
    \end{subfigure}
    \label{fig:weo_goldstein}
    \note{\emph{Notes:} Panel A shows intra-regional scores among Western European and Other States. Panel B shows inter-regional scores between WEO states and all other regions. Shaded areas represent 95\% confidence intervals.}
\end{figure}

Figure~\ref{fig:weo_goldstein} examines Western European and Other States (WEO), the core of the postwar liberal order. Panel A demonstrates stable intra-regional cooperation, with scores consistently above 3 and reaching peaks near 5 during periods of institutional deepening—the Single European Act (1986), Maastricht Treaty (1992), and Eastern enlargement (2004). Notable disruptions include the Cyprus Crisis (1974), which temporarily depressed scores to 2.5, and the dual shocks of Brexit (2016) and the Russia--Ukraine War (2022), which have driven scores below 3 for the first time since the 1980s. Panel B reveals that WEO states' relationships with other regions broadly track global patterns while maintaining systematically higher baseline cooperation—typically 0.5--1.0 points above the global mean. The divergence since 2015 between stable intra-WEO cooperation and deteriorating WEO-other relations is consistent with selective decoupling: Western states maintain deep integration among themselves while political relationships with the Global South and non-aligned states have deteriorated.

\begin{figure}[H]
    \centering
    \caption{Regional Geopolitical Alignment Scores: Eastern Europe and Asia-Pacific}
    \begin{subfigure}[b]{0.49\textwidth}
        \caption{Eastern European States}
        \includegraphics[width=\textwidth]{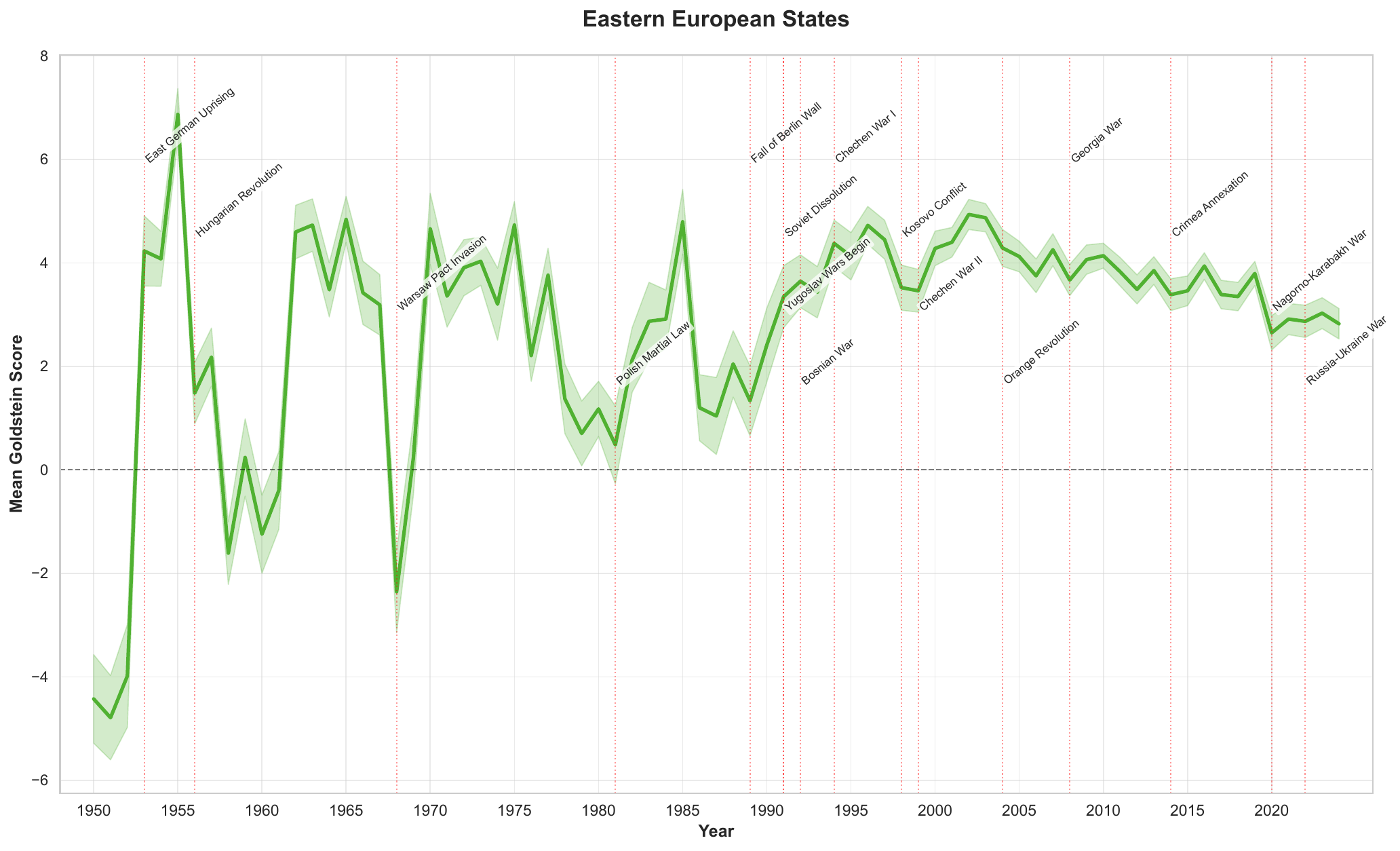}
    \end{subfigure}
    \hfill
    \begin{subfigure}[b]{0.49\textwidth}
        \caption{Asia-Pacific States}
        \includegraphics[width=\textwidth]{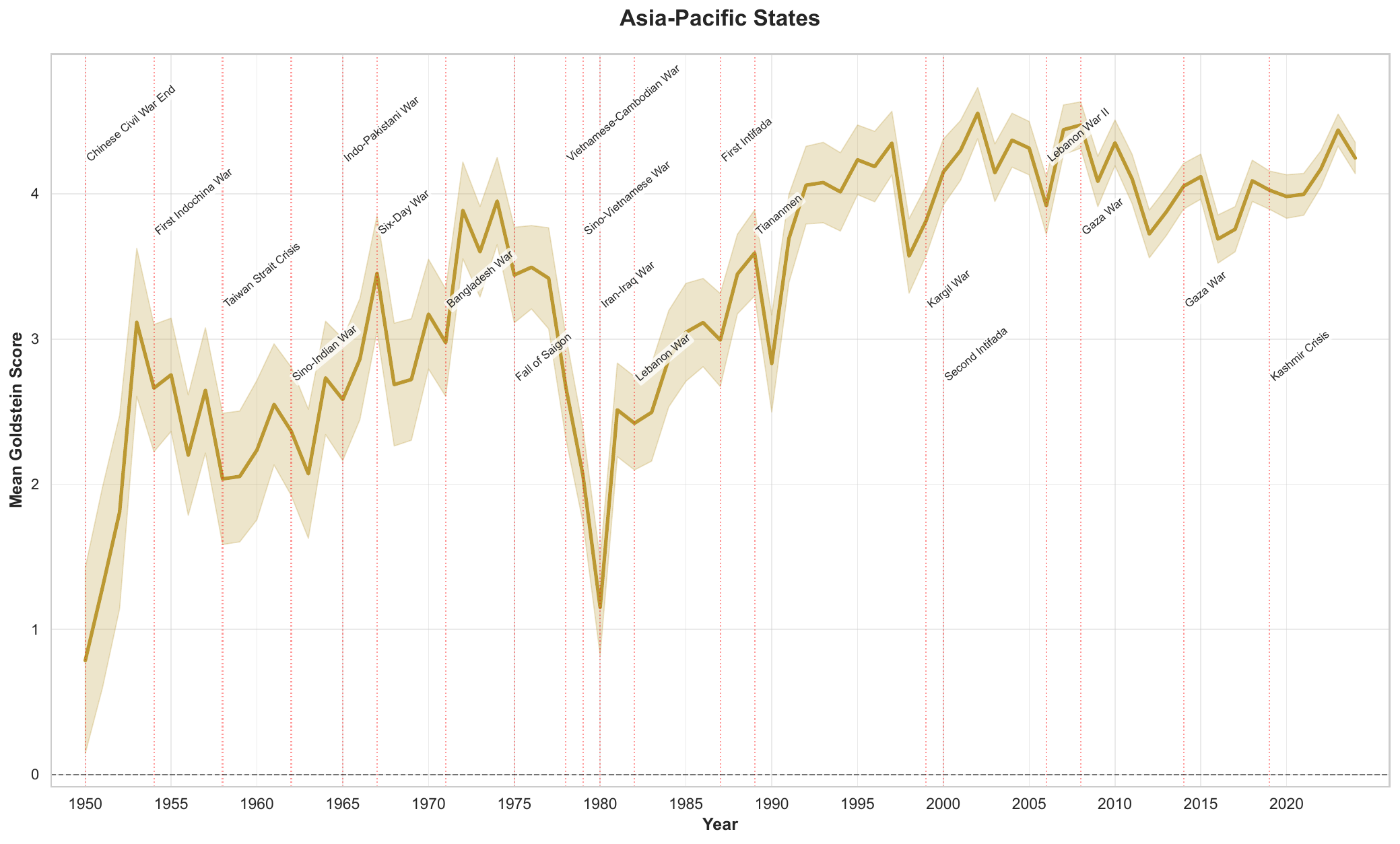}
    \end{subfigure}
    \label{fig:eastern_asia_goldstein}
\end{figure}

Figure~\ref{fig:eastern_asia_goldstein} contrasts two regions with divergent trajectories. Eastern European states (Panel a) exhibit the highest volatility in our sample, with scores swinging from $-5$ during the Hungarian Revolution (1956) to above 7 immediately following German reunification (1990). The Soviet collapse initiates a decade of negative scores as Yugoslavia disintegrates and post-Soviet conflicts erupt. EU accession drives steady improvement from 2000 to 2014, with scores stabilizing around 4. The Crimean annexation (2014) triggers renewed deterioration, while the 2022 invasion produces the sharpest decline in any region, with scores falling toward zero and reversing much of the post-Cold War cooperation gain. Asia-Pacific states (Panel b) display a contrasting pattern of steady improvement from hostile relations in 1950 (Chinese Civil War, Korean War) to consistent cooperation above 4 by 2000. This trajectory reflects sequential conflict resolution—the Vietnam War's end (1975), Sino-Vietnamese normalization (1991), and ASEAN expansion—combined with deepening economic integration. Unlike other regions, Asia-Pacific scores remain stable post-2010 despite U.S.--China tensions, suggesting that economic interdependence continues to moderate political conflicts.

\begin{figure}[H]
    \centering
    \caption{Regional Geopolitical Alignment Scores: Africa and Latin America}
    \begin{subfigure}[b]{0.49\textwidth}
        \caption{African States}
        \includegraphics[width=\textwidth]{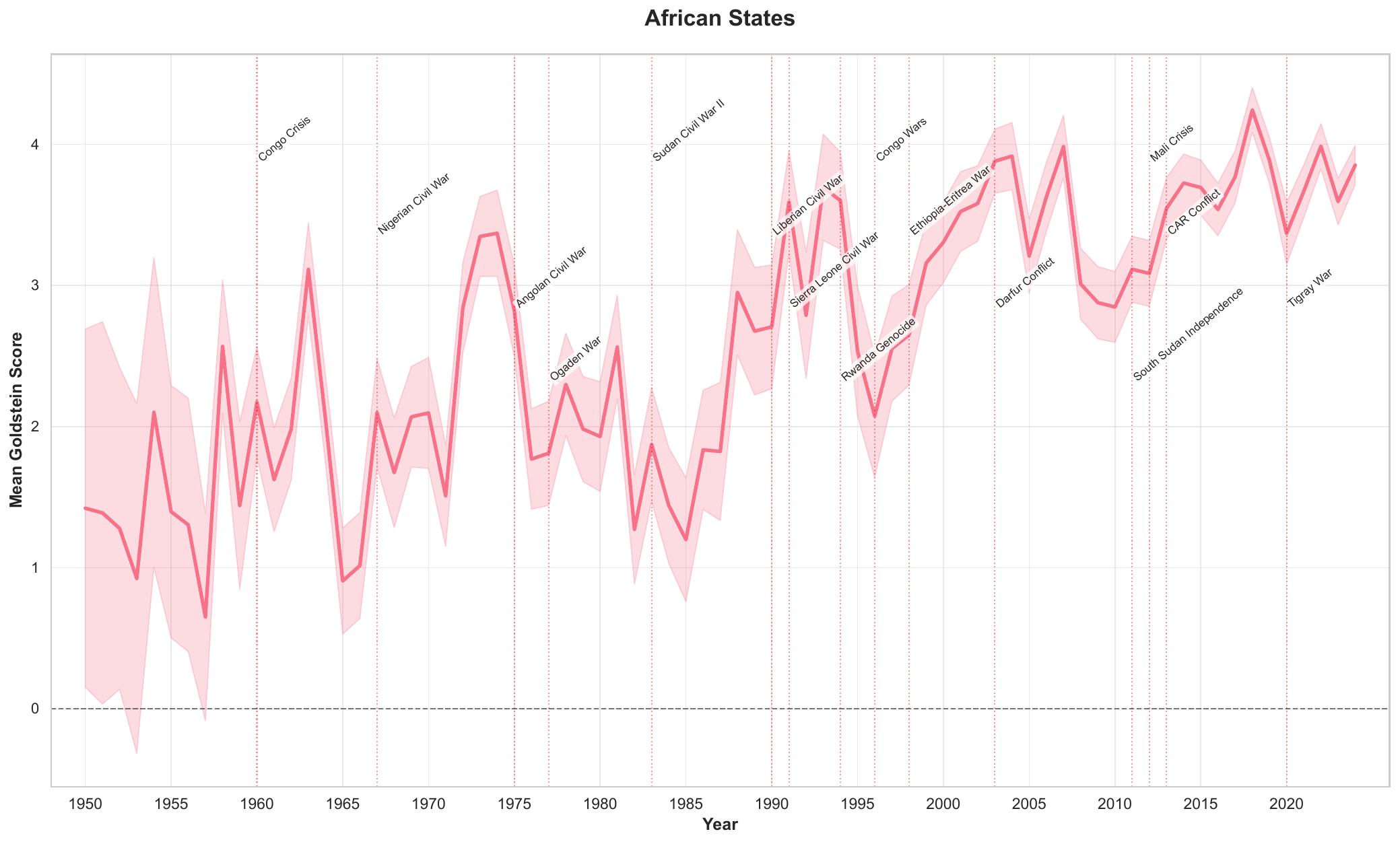}
    \end{subfigure}
    \hfill
    \begin{subfigure}[b]{0.49\textwidth}
        \caption{Latin American States}
        \includegraphics[width=\textwidth]{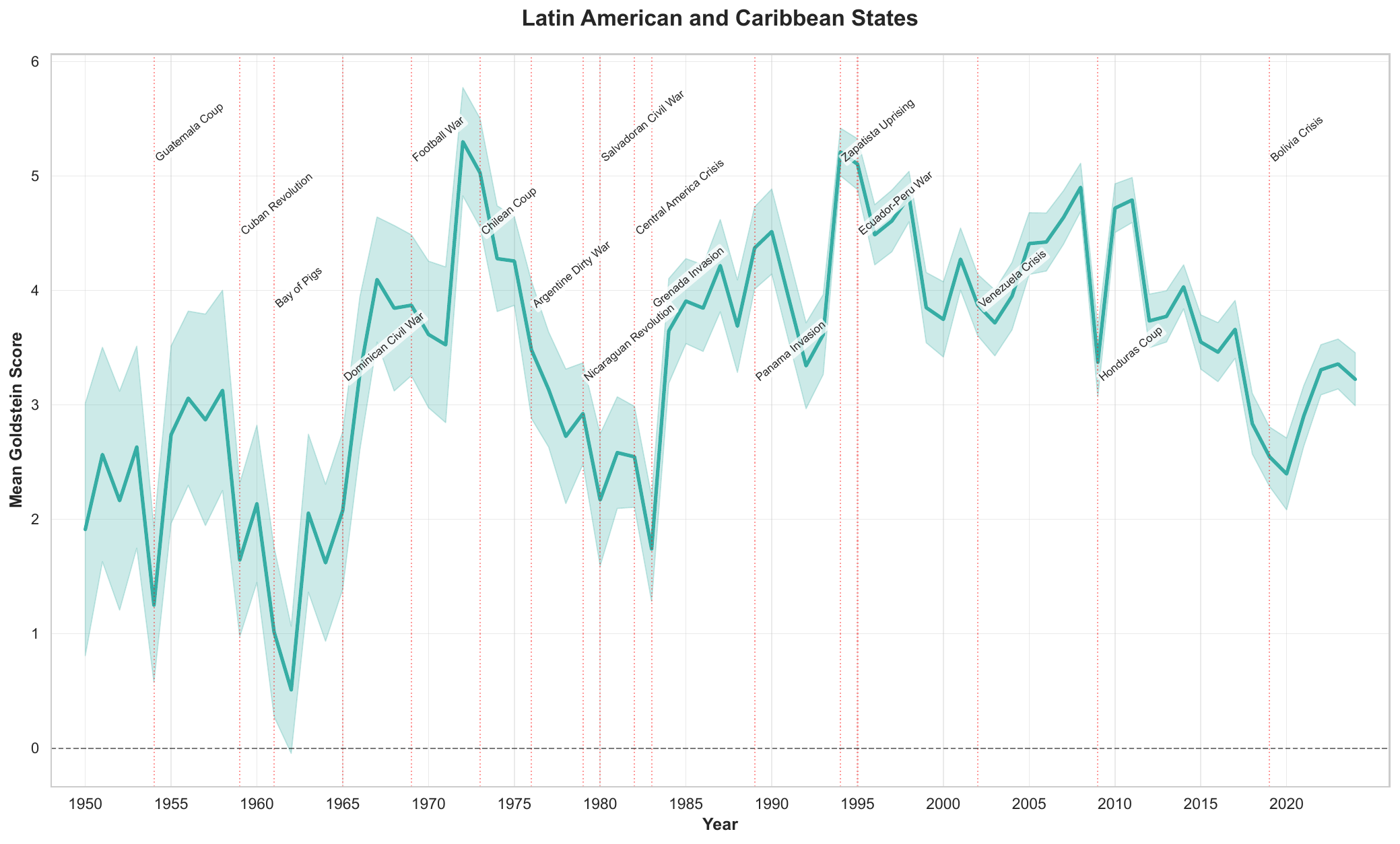}
    \end{subfigure}
    \label{fig:africa_latam_goldstein}
\end{figure}

Figure~\ref{fig:africa_latam_goldstein} documents geopolitical evolution in the Global South. African states (Panel a) show persistent improvement from below 1 during the decolonization conflicts of the 1950s--1960s to above 4 by 2020. The upward trend is punctuated by severe regional crises—the Nigerian Civil War (1967), Angolan Civil War (1975), Rwandan Genocide (1994), and Congo Wars (1996--2003)—each causing temporary score collapses. The post-2000 acceleration reflects both conflict resolution (the conclusion of Sierra Leone and Liberian civil wars) and expanding South-South cooperation, particularly with China. The Tigray War (2020) marks the first major reversal in two decades. Latin American states (Panel b) display a distinct pattern of high volatility around a stable mean of 3--4. Major disruptions correspond to ideological conflicts—the Cuban Revolution (1959), Chilean coup (1973), Central American crises (1980s)—and economic shocks, particularly the debt crisis (1982) that produces the sharpest decline. The post-2000 period shows unusual stability despite political polarization, suggesting that regional economic integration through Mercosur and the Pacific Alliance provides resilience against bilateral political tensions. The decline since 2019 reflects both Venezuela crisis spillovers and COVID-19's disproportionate impact on the region.

\subsection{LLM Prompt Design and Implementation} \label{app:llm_prompt}

This section documents the prompt structure used to instruct the large language model (Gemini 2.5 Pro with Google Search grounding) for compiling and classifying bilateral geopolitical events. For each country-pair-year observation, the LLM receives a structured prompt specifying two countries and a target year, then executes the procedure described below.

\subsubsection{Prompt Evolution}

The prompt design evolved through three versions, each addressing limitations identified in the preceding iteration:

\begin{itemize}
\item \textbf{Version 0 (Prototype):} A minimal prompt requesting bilateral events with basic Goldstein scoring. This version produced inconsistent event granularity and lacked systematic coverage of economic dimensions.

\item \textbf{Version 1 (Topic Taxonomy):} Introduced the six-category event taxonomy (economic, diplomatic, security, legal, multilateral, other) with explicit subcategories. This structured the LLM's search process and substantially improved coverage of non-military dimensions. However, event descriptions remained variable in length and detail.

\item \textbf{Version 2 (Current):} The production prompt used for the full database. It adds mechanism-first reasoning, requiring the LLM to identify the bilateral relationship context before searching for events. It enforces consistent output formatting and includes the complete CAMEO codebook with Goldstein score mappings as reference material.
\end{itemize}

The remainder of this section documents the Version 2 prompt in full.

\subsubsection{Overview and Relationship Assessment Framework} \label{app:prompt_overview}

The prompt instructs the LLM to generate a single JSON object containing all significant political events that characterized the bilateral relationship between two specified countries during a given year. Each event is classified using the CAMEO framework (Section~\ref{app:prompt_cameo}) and scored on the Goldstein scale (Section~\ref{app:prompt_goldstein}).

\paragraph{Relationship Assessment Framework.}
In addition to event-level classification, the LLM assigns an overall relationship assessment for each country-pair-year, selecting one category from a nine-point ordinal scale:

\begin{itemize}
    \item \textbf{State of War / Active Conflict}: Open, large-scale armed conflict.
    \item \textbf{Crisis / Intense Confrontation}: Brink of war; high tension, major disputes, limited clashes.
    \item \textbf{Hostile / Antagonistic Relationship}: Deep animosity; sanctions, diplomatic friction, negative rhetoric.
    \item \textbf{Competitive / Rivalrous Relationship}: Strategic competition with limited cooperation.
    \item \textbf{Limited Contact / Cool Relationship}: Minimal, neutral interaction.
    \item \textbf{Selective Cooperation / Transactional Relationship}: Cooperation on specific mutual interests amid broader competition.
    \item \textbf{Broad Cooperation / Partnership}: Cooperation across many sectors; regular dialogue.
    \item \textbf{Strategic Partnership}: Deep cooperation on major strategic issues; high trust.
    \item \textbf{Alliance}: Formal mutual support treaty (especially military); highest integration.
\end{itemize}

\subsubsection{Analytical Procedure} \label{app:prompt_procedure}

The LLM executes five sequential steps for each country-pair-year observation.

\paragraph{Step 0: Verify Political Entities.}
Before analyzing interactions, the LLM verifies whether the specified countries existed as political entities in the target year. If not, it identifies the primary political entity controlling the relevant territory---for example, the Soviet Union for Russia before December 26, 1991. The verified entity names are used throughout the analysis and reflected in the output.

\paragraph{Step 1: Search and Compile Events.}
The LLM systematically searches for all significant interactions between the verified entities during the target year, drawing on its knowledge base and internet sources. Source prioritization follows: (i) official government sources; (ii) international organizations (UN, WTO, IMF); (iii) major news archives (Reuters, AP, BBC); (iv) academic sources and diplomatic archives. Each event must be verified as occurring in the specified year.

\paragraph{Step 2: Identify Significant Political Events.}
The LLM identifies all significant political events involving direct interaction between the two entities that affect or indicate the state of the bilateral relationship. Each event is classified under one primary category based on its dominant mechanism; when economic tools are employed for political purposes, the event is classified under economic categories. The prompt instructs the LLM to search across six dimensions encompassing 22 subcategories (see Table~\ref{tab:event_categories} for the full taxonomy):

\begin{itemize}
    \item \textit{A. Economic Relations} (A1--A6): Trade policy, financial relations, economic coercion, strategic sectors, integration, and other economic events.
    \item \textit{B. Diplomatic and Political Relations} (B1--B4): Formal diplomacy, high-level interactions, public diplomacy, and cultural exchanges.
    \item \textit{C. Security and Defense} (C1--C3): Military cooperation, security incidents, and intelligence/cyber.
    \item \textit{D. Legal, Territorial, and Movement} (D1--D3): Legal actions, territorial disputes, and movement of people.
    \item \textit{E. Multilateral and Global Governance} (E1--E2): International organizations and global issues.
    \item \textit{F. Other Significant Events} (F1--F7): Historical/symbolic, humanitarian, sports, technology, environment, communications, and other interactions.
\end{itemize}

\paragraph{Step 3: Analyze Each Event.}
For each identified event, the LLM extracts and records the following:

\begin{enumerate}
    \item \textit{Basic information}: Initiator/target country assignment, event name, and detailed event description.
    \item \textit{Temporal details}: The most precise temporal information available---exact date (\texttt{YYYY-MM-DD}), month (\texttt{YYYY-MM}), quarter (\texttt{YYYY-Q\#}), and year---filled hierarchically from most to least precise.
    \item \textit{CAMEO classification}: Quadrant class, root code, and event code, following the codebook in Section~\ref{app:prompt_cameo}.
    \item \textit{Goldstein Scale score}: A numerical score from $-10.0$ to $+10.0$, following the guidelines in Section~\ref{app:prompt_goldstein}.
    \item \textit{International organization involvement}: Whether any international organization played a substantive role (as venue, actor, or instrument), with identification of the specific organizations involved.
    \item \textit{Event category}: Primary category code (A1--F7) from Table~\ref{tab:event_categories}.
    \item \textit{Evaluation summary}: A comprehensive justification explaining the CAMEO classification, Goldstein score, event category assignment, temporal details, and any entity verification issues.
\end{enumerate}

\paragraph{Handling Third Parties.}
When events involve third parties or multilateral contexts, the CAMEO classification and Goldstein score focus on the bilateral relationship between the two focal countries. For example, if both countries jointly cooperate in a multilateral agreement, the event is classified based on their cooperative action toward each other, not the broader multilateral context.

\paragraph{Step 4: Overall Relationship Assessment.}
The LLM selects one overall relationship category for the target year, integrating the pattern, frequency, and intensity of all identified events with historical context. If no significant events are found, the assessment is based on the absence of interaction and historical trends.

\subsubsection{JSON Output Structure} \label{app:prompt_json}

The output is a single JSON object with key \texttt{historical\_political\_events} containing a list of event objects. Each event object includes the following fields:

\begin{itemize}
    \item \texttt{year}: Integer (the target year).
    \item \texttt{country1}, \texttt{country2}: Strings (verified entity names).
    \item \texttt{event\_name}, \texttt{event\_description}: Strings.
    \item \texttt{event\_category}: String (A1--F7) or null.
    \item \texttt{event\_exact\_date}: String (\texttt{YYYY-MM-DD}) or null.
    \item \texttt{event\_month}: String (\texttt{YYYY-MM}) or null.
    \item \texttt{event\_quarter}: String (\texttt{YYYY-Q\#}) or null.
    \item \texttt{event\_year}: Integer.
    \item \texttt{CAMEO\_quad\_class}: String (``Verbal Cooperation,'' ``Material Cooperation,'' ``Verbal Conflict,'' or ``Material Conflict'') or null.
    \item \texttt{CAMEO\_root\_code}: String (two-digit) or null.
    \item \texttt{CAMEO\_event\_code}: String (three-digit) or null.
    \item \texttt{Goldstein\_Scale}: Number ($-10.0$ to $+10.0$) or null.
    \item \texttt{international\_org\_involvement}: Object with subfields \texttt{involved} (Boolean) and \texttt{primary\_organizations} (array of strings or null), or null.
    \item \texttt{relationship}: String (one of the nine relationship categories, uniform across all events in a given year).
    \item \texttt{evaluation\_summary}: String.
\end{itemize}

When no significant events are found, the output contains a single placeholder object with \texttt{event\_name} set to ``No Significant Bilateral Events Found'' and null values for classification fields, accompanied by a context-based relationship assessment.

\subsubsection{CAMEO Classification} \label{app:prompt_cameo}

We employ the Conflict and Mediation Event Observations (CAMEO) framework \citep{schrodt2012cameo} to classify each event along two dimensions: cooperation versus conflict, and verbal versus material. This produces four quadrant classes with hierarchical coding: root codes (two-digit) represent general action categories, and event codes (three-digit) specify precise actions. The LLM is instructed to: (i) identify the core bilateral action; (ii) determine the quadrant class; (iii) select the root code; and (iv) choose the most precise event code. We emphasize accurate classification of economic actions---distinguishing broad sanctions (code 163) from targeted administrative sanctions (code 172)---and mediation events (codes 028, 038, 039, 045, 108, 126, 135, 165). The complete CAMEO codebook with 20 root codes and 143 event codes is provided to the LLM as reference material.

\subsubsection{Goldstein Scale Scoring Guidelines} \label{app:prompt_goldstein}

Each event receives a Goldstein Scale score \citep{goldstein1992} ranging from $-10.0$ (maximum conflict) to $+10.0$ (maximum cooperation). The LLM is instructed to assign scores using the CAMEO event code as the primary reference, with limited contextual adjustments. Table~\ref{tab:goldstein_ranges} presents the scoring guidelines.

\begin{table}[ht]
\centering
\caption{Goldstein Scale Scoring Guidelines}
\label{tab:goldstein_ranges}
\begin{tabular}{cl}
\toprule
\textbf{Range} & \textbf{Description} \\
\midrule
$+8$ to $+10$ & Major concessions, peace agreements, alliance formations \\
$+6$ to $+8$ & Significant aid, material cooperation, major agreements \\
$+2$ to $+5$ & Consultations, cooperative intent, routine diplomatic meetings \\
$0$ to $+2$ & Initial positive gestures, expressions of interest \\
$0$ & Neutral statements or actions \\
$-2$ to $0$ & Mild disagreements, tensions, negative rhetoric \\
$-2$ to $-5$ & Diplomatic protests, rejections, criticism, minor sanctions \\
$-5$ to $-8$ & Threats, coercion, severe sanctions, military mobilizations \\
$-8$ to $-10$ & Military attacks, mass violence, war declarations \\
\bottomrule
\end{tabular}
\note{\emph{Notes:} Guidelines provided to the LLM for Goldstein Scale assignment. The score primarily reflects the typical intensity associated with the CAMEO event code. Contextual adjustments are permitted when specific circumstances clearly indicate higher or lower intensity than typical for the event category---for example, a first meeting after years of hostility may score higher than a routine annual meeting. All adjustments must be justified in the evaluation summary. Source: \citet{goldstein1992}.}
\end{table}

The scoring instructions emphasize three principles. First, the score should primarily reflect the event type identified by the CAMEO event code, ensuring consistency across the database. Second, contextual adjustments are permitted but must remain logically consistent with the CAMEO code's general intensity level---a routine diplomatic meeting (CAMEO 040) should not be scored as highly as a major peace treaty (CAMEO 057). Third, when events involve third parties, the score reflects the impact on the bilateral relationship between the two focal countries, not the broader multilateral context.

\subsubsection{International Organization Involvement} \label{app:prompt_io}

For each event, the LLM records whether international organizations played a substantive role---as a venue for the interaction, a party to the dispute, or an instrument of policy. The prompt specifies a comprehensive list of organizations spanning four tiers:

\begin{itemize}
    \item \textit{UN system}: United Nations (General Assembly, Security Council), ICJ, ICC, WTO, IMF, World Bank, WHO, IAEA, ILO, UNESCO, UNHCR, UNICEF, UNEP.
    \item \textit{Regional organizations}: EU, NATO, OSCE, Council of Europe, African Union, ASEAN, OAS, MERCOSUR, Arab League, GCC, SAARC, CIS, SCO, APEC, Pacific Alliance, CARICOM, ECOWAS, SADC, EAC.
    \item \textit{Economic and development bodies}: OECD, G7, G20, BRICS, OPEC, ADB, AfDB, IDB, EBRD, AIIB, NDB, BIS.
    \item \textit{Specialized organizations}: INTERPOL, IOM, ICAO, IMO, ITU, WIPO, ISO, ICRC, Commonwealth, OIC, Non-Aligned Movement.
\end{itemize}

Passing references to organizations are excluded; only substantive involvement is recorded.

\end{document}